\documentclass[twocolumn, twocolappendix]{aastex63}

\usepackage{amsmath}	
\usepackage{amssymb}	
\usepackage{mathrsfs}

\usepackage[mathlines]{lineno}

\usepackage{bigints}
\usepackage{CJK}
\usepackage{lipsum}

\newcommand{\Msun}{M_\odot}
\newcommand{\Msunyr}{M_\odot~{\rm yr}^{-1}}
\newcommand{\Mh}{M_\mathrm{h}}
\newcommand{\Mbh}{M_\bullet}

\newcommand{\vbsm}{v_\mathrm{bsm}}

\newcommand{\Mdot}{\dot{M}}
\newcommand{\tlife}{\tau}
\newcommand{\fseed}{f_\mathrm{seed}}
\newcommand{\fobsc}{f_\mathrm{obsc}}

\newcommand{\Muv}{M_{1450}}
\newcommand{\Lbol}{L_\mathrm{bol}}

\newcommand{\D}{\mathrm{d}}

\defcitealias{2010AJ....140..546W}{W10}
\defcitealias{2018ApJ...869..150M}{M18}

\shorttitle{BH growth toward $z=$ 6 BHMF \& QLF}
\shortauthors{Li et al.}
\graphicspath{{./}{./figs/}}

\begin{document}
\begin{CJK*}{UTF8}{gbsn}

\title{The Assembly of Black Hole Mass and Luminosity Functions of High-redshift Quasars\\via Multiple Accretion Episodes}

\correspondingauthor{Wenxiu Li; Kohei Inayoshi}
\email{wenxiuli@pku.edu.cn; inayoshi@pku.edu.cn}

\author[0000-0002-1044-4081]{Wenxiu Li}
\affiliation{Department of Astronomy, School of Physics, Peking University, Beijing, 100871, China}
\author[0000-0001-9840-4959]{Kohei Inayoshi}
\affiliation{Kavli Institute for Astronomy and Astrophysics, Peking University, Beijing 100871, China}
\author[0000-0003-2984-6803]{Masafusa Onoue}
\altaffiliation{Kavli Astrophysics Fellow}
\affiliation{Kavli Institute for Astronomy and Astrophysics, Peking University, Beijing 100871, China}
\affiliation{Kavli Institute for the Physics and Mathematics of the Universe (Kavli IPMU, WPI), The University of Tokyo, Chiba 277-8583, Japan}
\author[0000-0003-3467-6079]{Daisuke Toyouchi}
\affiliation{Research Center for the Early Universe (RESCEU), The University of Tokyo, Hongo, 7-3-1, Bunkyo-ku Tokyo, 113-0033, Japan}

\begin{abstract}
The early evolution of the quasar luminosity function (QLF) and black hole mass function (BHMF) encodes 
key information on the physics determining the radiative and accretion processes of supermassive 
black holes (BHs) in high-$z$ quasars.
Although the QLF shape has been constrained by recent observations,
it remains challenging to develop a theoretical model that explains its redshift evolution associated with BH growth self-consistently.
In this study, based on a semi-analytical model for
the BH formation and growth,
we construct the QLF and BHMF of the early BH population that experiences multiple accretion bursts,
in each of which a constant Eddington ratio is assigned following a Schechter distribution function.
Our best-fit model to reproduce the observed QLF and BHMF at $z\simeq 6$
suggests that several episodes of moderate super-Eddington accretion occur
and each of them lasts for $\tau \simeq 20-30$ Myr.
The average duty cycle in super-Eddington phases is $\simeq 15\%$ for massive BHs that reach $\gtrsim 10^8~M_\odot$ by $z\simeq 6$,
which is nearly twice that of the entire population.
We find that the observed Eddington-ratio distribution function is skewed to a log-normal shape 
owing to detection limits of quasar surveys.
The predicted redshift evolution of the QLF and BHMF suggests a rapid decay of their number and mass density 
in a cosmic volume toward $z\gtrsim 6$.
These results will be unveiled by future deep and wide surveys with the James Webb Space Telescope, Roman Space Telescope, and Euclid.
\end{abstract}



\vspace{5mm}
\section{Introduction} \label{sec:intro}
The growth of massive BHs in active galactic nuclei (AGN) is elegantly related to their luminous accretion phases
across the cosmic time \citep{1982MNRAS.200..115S}.
The cosmic BH accretion history and their radiative efficiency are well constrained by 
comparing the mass density of supermassive black holes (SMBHs) in the local universe and the mass accreted onto BHs inferred from
the integration of the QLF based on multi-wavelength observations
\citep[e.g.][]{1971ApJ...170..223C,1992MNRAS.259..725S,2002MNRAS.335..965Y,2004MNRAS.351..169M,2008MNRAS.388.1011M,2004MNRAS.354.1020S,
2009ApJ...690...20S,2014MNRAS.439.2736D,2014ApJ...786..104U,2017A&A...600A..64T,2022ApJ...934...66S}.

Currently the $z\sim 6$ quasar population is well constrained by multiband surveys
\citep{2008AJ....135.1057J,2010AJ....139..906W,2016arXiv161205560C,2018PASJ...70S..35M,2019ApJ...883..183M,2019AJ....157..168D}
along with the QLF \citep[e.g.,][]{2015ApJ...798...28K,2016ApJ...833..222J,2017ApJ...847L..15O,2018ApJ...869..150M}.
Deep spectroscopic observations of high-$z$ quasars enable us to measure the virial BH mass 
with the Mg~{\sc ii} single-epoch method and bring insights of the mass distribution
(e.g., \citealt{2007AJ....134.1150J,2007ApJ...669...32K,2010AJ....140..546W};
\citealt{2018Natur.553..473B,2019ApJ...880...77O,2019ApJ...873...35S}). 
However, extrapolation of the Soltan-Paczy{\'n}ski argument toward higher redshifts of $z\gtrsim$ 6 is still limited 
because of the current capability of high-$z$ quasar observations 
\citep{2013ApJ...768..105M,2016ApJ...829...33Y,2019ApJ...884...30W,2019BAAS...51c.121F}.
Nevertheless, in the era of the James Webb Space Telescope (JWST) and forthcoming facilities,
e.g., the Roman Space Telescope (RST) and Euclid,
infrared imaging and spectroscopic observations will unveil a wealth of information on the
high-$z$ quasar properties and their environments 
\citep{2019BAAS...51c..45R, 2019arXiv190205569A, 2011arXiv1110.3193L}. 
Deep observations of high-$z$ quasars and their host galaxies will shed light on the early BH evolution and help 
answer questions regarding the existence of SMBHs in the early universe \citep{2012Sci...337..544V,2013ASSL..396..293H,2020ARA&A..58...27I}, 
and early development of BH-galaxy coevolution 
\citep[e.g.,][]{2013ApJ...773...44W,2017ApJ...851L...8V,2021ApJ...914...36I,2022ApJ...927..237I,2022MNRAS.511.3751H}.
Future gravitational-wave observations both via space interferometers such as LISA, Tianqin, and Taiji, 
and pulsar timing array experiments \citep{2008MNRAS.390..192S,LISA_2017,Tianqin_2016,SKA_09}
will enable us to probe the abundance of coalescing massive BH binaries at high redshifts, 
setting constraints on the properties of their quasar counterparts
\citep{2006ApJ...637...27K,2009ApJ...700.1952H,2022arXiv220714309P}.

The shapes of the QLF and BHMF contain key information on the physics characterizing the radiative 
and accretion processes of high-$z$ SMBHs, as well as their seeding mechanisms in principle.
Two approaches have been utilized to model those distribution functions.  
The first is semi-analytical modeling, in which BH seed formation, gas accretion, and BH mergers associated 
with the hierarchical growth of the parent dark matter (DM) halos are taken into account in a simplified way. 
This is an effective way of examining the statistical properties of the early BH population and making predictions 
that can be directly compared with the observed QLF and BHMF
\citep[e.g.,][]{1998ApJ...503..505H,2010ApJ...718..231S,2018MNRAS.474.1995R,2018MNRAS.481.3278R,2019MNRAS.486.2336D,
2021MNRAS.500.2146P,2021MNRAS.508.2706Y,2021ApJ...910L..11K,2022MNRAS.511..616T,2022arXiv220714689O}. 
However, due to the limitation of capturing the detailed properties of the relevant physics, 
a large number of model parameters need to be calibrated based on the observed distribution functions 
of lower-$z$ quasar populations \citep[e.g.,][]{2007ApJ...669...45H}
and empirical correlations between the SMBH mass and their host galaxy properties seen
in the local universe \citep{2013ARA&A..51..511K}.
The second approach is more phenomenological, using extrapolation of the low-$z$ QLF evolution with 
fitting function forms \citep[e.g.,][]{2019MNRAS.488.1035K,2020MNRAS.495.3252S,2022ApJ...938...25F}.
This method also allows us to forecast the higher-$z$ QLF and bring insights for future surveys.
Nevertheless, those semi-analytical and phenomenological approaches introduce numerous physical parameters
and make the essential mechanisms reproducing the observed results elusive.
Therefore, it is required to construct a theoretical model characterized by a minimum number of parameters
but capable of dealing with the essential physical processes.

To determine the initial conditions of the early BH assembly,
the formation pathway of seed BHs has been extensively investigated by numerical simulations and semi-analytical studies
\citep{2006MNRAS.370..289B,2009MNRAS.396..343R,2009ApJ...696.1798T,2012MNRAS.422.2051N,2014ApJ...781...60H,
2015MNRAS.448..568H,2018MNRAS.474.3825V,2021MNRAS.506..613S,2023MNRAS.518.1601T,2022MNRAS.516..138B}.
Massive star formation episodes in the early universe is substantially modulated by external environmental effects such as
(i) H$_2$ photo-dissociating irradiation from nearby galaxies
\citep{2001ApJ...546..635O,2002ApJ...569..558O,2003Natur.425..812B,2010MNRAS.402.1249S,2014MNRAS.445..544S,2014MNRAS.445..107V,2016ApJ...832..134C},
(ii) supersonic baryonic streaming motion that delays gas collapse in halos
\citep{2012MNRAS.424.1335F, 2014MNRAS.439.1092T, 2018ApJ...855...17H,2017MNRAS.471.4878S,2018MNRAS.479.4017I},
and (iii) dynamical heating caused by frequent halo mergers and turbulence injection led by cold accretion flows
\citep{2003ApJ...592..645Y,2010Natur.466.1082M,2015ApJ...810...51M,2019Natur.566...85W,2022Natur.607...48L}.
Those effects keep isothermal collapse ($T\sim 8000~{\rm K}$) of massive gas at high accretion rates of $\gtrsim 0.1~\Msunyr$
without vigorous fragmentation \citep{2014MNRAS.445L.109I,2015MNRAS.446.2380B,2016PASA...33...51L},
and allow supermassive stars with $\sim 10^5~\Msun$ (presumably seed BHs) to form at the halo centers 
\citep{2013ApJ...778..178H,2013A&A...558A..59S,2019PASA...36...27W,2023MNRAS.518.1601T}.
Although the conditions required for heavy seed formation are thought to be too stringent to be realized in the typical regions of the early universe
\citep{2008MNRAS.391.1961D,2009ApJ...695.1430A,2015MNRAS.450.4350I},
recent studies by \citet{2021MNRAS.503.5046L} and \citet{2021ApJ...917...60L} found that the formation of heavy seeds
accelerates in rare, overdense regions of the universe, where non-linear galaxy clustering boosts the irradiation intensity
and frequency of halo mergers.
In such massive halos, intense cold streams through cosmic webs and supersonic turbulent motion promote the formation
of massive gas clouds that collapse into supermassive stars and seed BHs \citep{2009Natur.457..451D,2012MNRAS.422.2539I,2022Natur.607...48L}.
The mass distribution of seed BHs is expected to be substantially top-heavy, extending the upper mass to $\gtrsim 10^{4-5}~\Msun$
\citep{2021ApJ...917...60L, 2023MNRAS.518.1601T}.
Metal enrichment of the progenitor halos due to internal and external star formation activity would regulate the formation of 
seed BHs and alter the characteristic of the mass distribution function (see also a quantitative argument on the low efficiency of enrichment in 
\citealt{2021ApJ...917...60L}).
Further studies with hydrodynamic simulations are required to improve our understanding of the efficiency of metal enrichment in quasar progenitor halos
\citep[e.g.,][]{2018MNRAS.475.4378C}.


In this paper, we propose a theoretical model for the redshift-dependent QLF and BHMF at $z\gtrsim 6$,
applying the semi-analytic seed formation model from \cite{2021ApJ...917...60L} to study the BH growth in the early universe.
We assume that the early BH population experiences multiple accretion bursts, in each of which a constant Eddington ratio is assigned 
following a Schechter distribution function. 
We conduct the Markov Chain Monte Carlo (MCMC) fitting procedure to optimize the BH growth parameters 
so that the observed QLF and BHMF at $z\simeq 6$ are simultaneously reproduced. 
Our best-fit model suggests several episodes of modest super-Eddington accretion and requires
individual burst durations of $\simeq 20-30$ Myr to avoid over(under)-production of BHs at the high-mass end.
We also find that the observed Eddington-ratio distribution function is skewed to a log-normal shape from the intrinsic Schechter-like function
owing to detection limits of quasar observations.
We further discuss the redshift evolution of the QLF and BHMF at $z\gtrsim 6$,
and give implications for future observations.
Those results will be tested by future deep and wide-field surveys with JWST, RST, and Euclid,
by conducting spectroscopic measurements of individual BH masses and constructing high-$z$ QLFs.

This paper is organized as follows. 
In Section~\ref{sec:method}, we describe the semi-analytical model for BH seeding and subsequent growth via accretion,
and explain the MCMC fitting procedure to constrain the theoretical model by the high-$z$ quasar observations.
In Section~\ref{sec:fitting_result}, we present the fitting result for the model parameters reproducing the observational data,
and discuss the physics determining the BH growth.
In Section~\ref{sec:cosm}, we discuss the cosmological evolution of the BHMF and QLF suggested by our best-fit model,
as well as the detection number of those BHs at $z\sim 6$--$10$ with Euclid and RST.
In Section~\ref{sec:discussion}, we present the evolutionary tracks of individual BHs and their statistical properties at $z\gtrsim 6$,
based on the growth model calibrated above.
We finally summarize our findings in Section~\ref{sec:sum}.
Throughout this work, we apply the cosmological parameters from \cite{2016A&A...594A..13P},
i.e., $\Omega_{\mathrm{m}}=0.307,~\Omega_{\Lambda}=0.693,~
\Omega_{\mathrm{b}}=0.0486$, and $H_0=67.7 \mathrm{~km} \mathrm{~s}^{-1} \mathrm{Mpc}^{-1}$.
All magnitudes quoted in this paper are in the AB system.

\vspace{2mm}
\section{METHODOLOGY}\label{sec:method}

\vspace{2mm}
\subsection{Seed BH formation}\label{sec:seed}
The QLFs at high redshifts are determined by the BH seeding mechanism and subsequent growth. 
We first describe our model regarding the formation of BH seeds in progenitor DM halos 
that end up in high-$z$ quasar host galaxies with halo mass of $\Mh \gtrsim 10^{11}~\Msun$.
The mass range of those DM halos for $z\sim 6$ quasars is motivated by the halo mass measurement, 
where the rotation velocity of gas based on the [C~{\sc ii}] 158 $\micron$ line width is assumed to be the circular velocity of the halo
\citep{2002ApJ...578...90F,2013ApJ...773...44W,2019ApJ...872L..29S}.
In our model, we consider three parent halos with $\Mh = 10^{11}$, $10^{12}$, and $10^{13} ~\Msun$ at $z=$ 6
and generate $N_{\rm tot} (= 10^4)$ merger trees for each parent halo mass backward in time using the {\tt GALFORM} 
semi-analytic algorithm based on the extended Press-Schechter formalism 
\citep{1974ApJ...187..425P,2000MNRAS.319..168C,2008MNRAS.383..557P}.
For each tree, we adopt a minimum DM halo mass of $M_{\rm h,min}=10^5~\Msun$, which is small enough to capture 
the earliest star formation episodes via H$_2$ line cooling at the high-$z$ universe \citep{1996ApJ...464..523H,1997ApJ...474....1T}.

We consider seed BH formation in the main progenitors of quasar host galaxies, following a semi-analytical model established by \citet{2021ApJ...917...60L}.
In the highly-biased, overdense regions of the universe, those progenitor halos are likely irradiated by intense H$_2$-photodissociating radiation 
from nearby star-forming galaxies and heat the interior gas by successive mergers. 
The two effects counteracting H$_2$ formation and cooling prevent gas collapse and delay prior star formation \citep[e.g.,][]{2014MNRAS.445..107V,2019Natur.566...85W}.
Under such peculiar circumstances, massive clouds collapse to the halo centers at high accretion rates and leave massive stars (presumably seed BHs) behind.
Our model takes into account various key physical processes on star formation such as halo merger heating, radiative cooling, and
(photo-)chemical reaction networks.
The time evolution of the gas dynamics and thermal states is calculated in a self-consistent way with the halo assembly history.
We calculate the time-dependent H$_2$ photo-dissociating radiation flux (${\rm H}_2+\gamma \rightarrow 2~{\rm H}$) following \citet{2014MNRAS.442.2036D}.
The flux is measured at one free-fall time, $t_{\rm ff} \simeq 32 [\left(1+z\right)/21]^{-3/2}~{\rm Myr}$, of the individual source halos after their single star bursts.
The value of $t_{\rm ff}$ is typically longer than the lifetime of massive stars that dominate the production of UV radiation.
Therefore, this treatment underestimates the photo-dissociating flux since the assumption of an old stellar population provides less UV radiation than smoothly-proceeding star formation \citep{2021MNRAS.503.5046L}.
In this sense, our model gives a conservative estimate of the formation efficiency of heavy seed BHs.
In addition, we take account of H$^-$ photo-detachment (${\rm H}^-+\gamma \rightarrow {\rm H}+{\rm e}^-$) 
caused by irradiation from nearby galaxies.
This effect suppresses H$_2$ formation through electron-catalysed reactions (${\rm H}+{\rm e}^- \rightarrow {\rm H}^- + \gamma$; 
${\rm H}^- + {\rm H} \rightarrow {\rm H}_2 + {\rm e}^-$), and thus delays gravitational collapse of gas clouds at the halo centers 
\citep{2001ApJ...546..635O,2010MNRAS.402.1249S}.
In this paper, we approximate the galaxy spectrum to be a black-body spectrum with $T_{\rm rad}=2\times 10^4~{\rm K}$ \citep{2014MNRAS.445..544S,2015MNRAS.450.4350I}.

In addition, we consider baryonic streaming motion relative to DM produced in the epoch of cosmic recombination at $z_{\rm rec} \simeq 1100$.
This effect also delays gas collapse and star formation in protogalaxies through injection of kinetic energy into gas
\citep[e.g.,][]{2012MNRAS.424.1335F, 2014MNRAS.439.1092T, 2017Sci...357.1375H,2019MNRAS.484.3510S,2021ApJ...917...60L}.
The amplitude of the streaming velocity is set to $\vbsm = 0$ and $\vbsm=1~\sigma_{\rm bsm}$, where $\sigma_{\rm bsm}=30~{\rm km~s}^{-1} (1+z)/(1+z_{\rm rec})$
is the root-mean-square speed at a redshift of $z$ \citep{2010PhRvD..82h3520T}.
We calculate the effective sound speed of gas as $c_{\rm eff} = \{c_{\rm s}^2+v_{\rm tur}^2/3+\left(\alpha_0 \vbsm \right)^2 \}^{1/2}$ at the center of each progenitor halo,
where $c_{\rm s}$ is the thermal sound speed, $v_{\rm tur}$ is the turbulent velocity of gas (i.e., the specific kinetic energy accumulated through halo mergers), and
the coefficient is set to $\alpha_0=1$ \footnote[1]{The value of $\alpha_0=1.0$ is motivated by cosmological simulations of massive primordial stars
under fast streaming velocities \citep{2017Sci...357.1375H,2019MNRAS.484.3510S}.
Note that our previous study in \citet{2021ApJ...917...60L} adopted $\alpha_0=4.7$ \citep{2018ApJ...855...17H}.}.
When the gas core becomes gravitationally unstable, the evolution of the gas density profile is well described by 
the Penston-Larson self-similar solution \citep{1969MNRAS.144..425P,1969MNRAS.145..271L}. 
The mass accretion rate from the envelope can be written as $\dot{M}=c_{\rm eff}^3/G$,
where $G$ is the gravitational constant.

At the vicinity of an accreting protostar, an accretion disk forms due to angular momentum of the inflowing material from large scales
\citep{2010MNRAS.403...45S,2011Sci...331.1040C,2011ApJ...737...75G}.
The episodic nature of disk accretion decreases the time-averaged accretion rate onto the protostar as $\Mdot_{\star}=\eta \Mdot$, 
where we adopt $\eta=0.3$ as the conversion efficiency owing to angular momentum of the accreting flow \citep{2016MNRAS.459.1137S, 2023MNRAS.518.1601T}. 
The size evolution of an accreting protostar and the radiative feedback strength depend sensitively on whether
the accretion rate becomes higher than a critical value of $\dot{M}_{\mathrm{crit}}=0.04~\Msunyr$ 
\citep{2001ApJ...561L..55O,2013ApJ...778..178H,2013A&A...558A..59S,2015MNRAS.452..755S,2018MNRAS.474.2757H}.
For $\dot{M}_\star > \Mdot_\mathrm{crit}$, the stellar envelope inflates owing to rapid entropy injection through accreting matter,
keeping its surface temperature as low as $T_\mathrm{eff} \simeq 5000$ K. 
Since stellar UV radiation is hardly emitted from the cold surface, the accretion flow efficiently feeds the central star without being 
impeded by radiative feedback.
As a result, the stellar mass reaches $\sim 10^5~\Msun$ but is limited by the general-relativistic instability to
\begin{equation}
 M_{\star, \mathrm{GR}} \simeq\left[0.83~\log \left(\frac{\dot{M}_{\star}}{\Msunyr}\right)+2.48\right] \times 10^{5} ~\Msun,
\end{equation}
in the range of $\dot{M}_\star\gtrsim 0.1~\Msunyr$ \citep{2016PhRvD..94b1501S,2017ApJ...842L...6W,2019PASA...36...27W}.
On the other hand, for $\dot{M}_\star \leq \Mdot_\mathrm{crit}$, the protostar begins to contract by loosing its thermal energy via radiative diffusion
and increases its surface temperature to $T_\mathrm{eff} \sim 10^5$ K.
Hence, intense stellar UV radiation heats the disk surface and launches mass outflows, preventing mass supply from the disk to the star
\citep{2008ApJ...681..771M,2011Sci...334.1250H}. 
In this case, the stellar mass is determined by the balance between mass accretion and mass loss via photoevaporation \citep{2013ApJ...773..155T}.
The feedback-regulated mass can be approximated as
\footnote[2]{A recent study by \citet{2023MNRAS.518.1601T} provides a more sophisticated prescription of the feedback-regulated stellar mass based on their radiation hydrodynamical simulations.}
\begin{equation}
M_{\star, \mathrm{fb}} \simeq 2.9 \times 10^{3} \Msun\left(\frac{\dot{M}_{\star}}{0.01 M_{\odot} \mathrm{yr}^{-1}}\right),
\end{equation}
see more details in \cite{2021ApJ...917...60L}.
For an intermediate accretion rate between $\Mdot_\mathrm{crit}$ and $0.1~\Msunyr$, 
we perform logarithmic interpolation in the $\Mdot_\star - M_\star$ plane to smoothly connect 
$M_{\star, \mathrm{fb}}$ and $M_{\star, \mathrm{GR}}$ at the boundaries.
At the end of the stellar lifetime, those massive primordial stars result in BHs without significant mass loss 
via stellar winds \citep{2003ApJ...591..288H,2015MNRAS.451.4086S}.
One important caveat is the effect of stellar rotation on the structure evolution.
Namely, the mass loss rate of a rotating massive star via winds would be enhanced by the centrifugal force on the surface
or suppressed by efficient mixing of the interior structure by meridional circulation \citep[see][references therein]{2008A&A...489..685E,2012A&A...542A.113Y}.
Although those effects in the high mass regime are poorly understood yet, full general-relativistic simulations of the gravitational collapse of a rotating supermassive star 
show that most of the stellar mass is eventually swallowed by the newly born BH, ejecting only $\sim 10\%$ of the mass \citep{2002ApJ...572L..39S,2016PhRvD..94b1501S}.
Therefore, we assume that the mass of a remnant BH is equal to that of its stellar progenitor.

\begin{figure*}
\centering
\includegraphics[width=140mm]{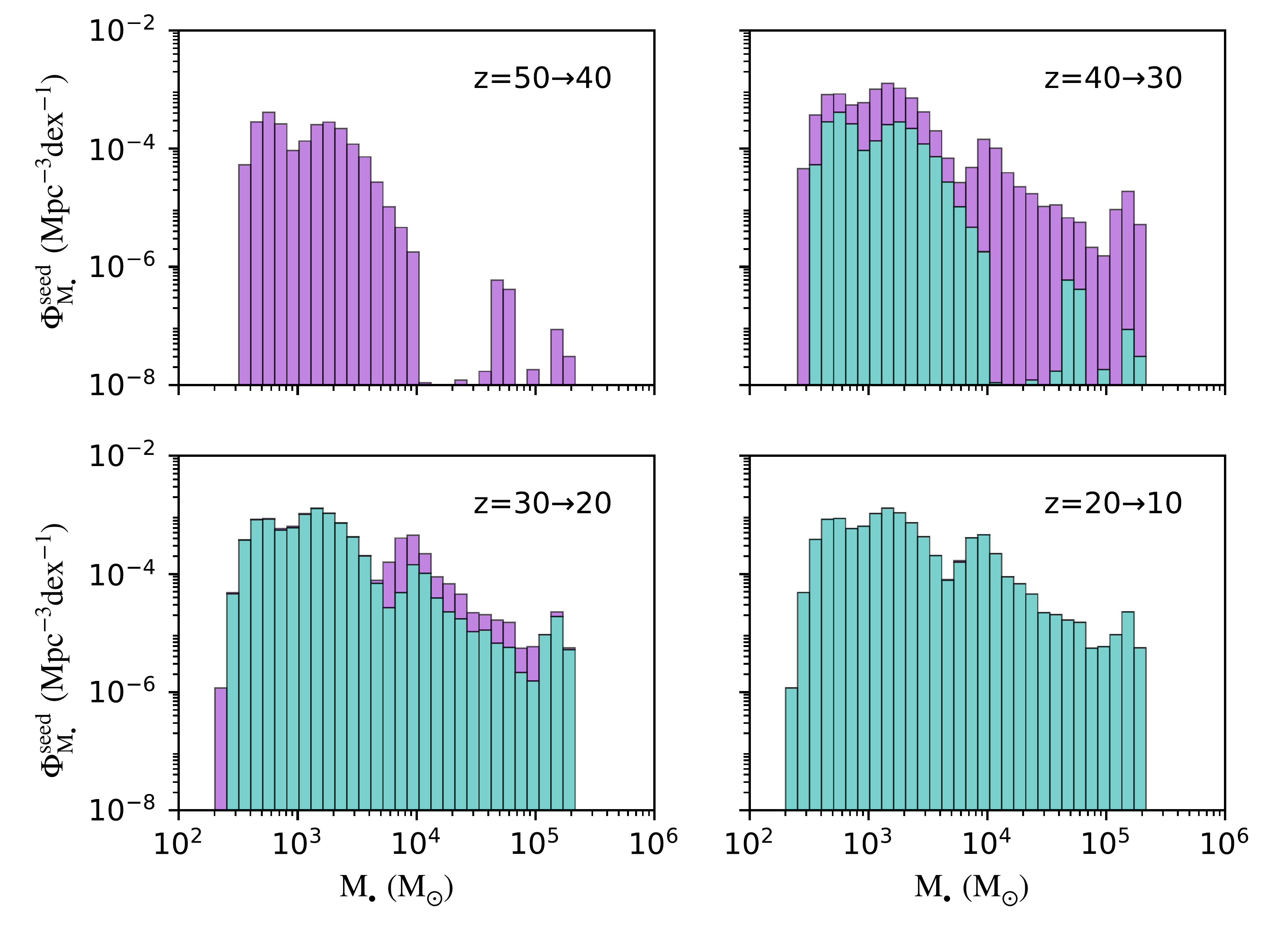}
\caption{
Mass distribution of seed BHs formed in quasar host galaxies at different redshifts ($10\leq z \leq 50$).
In each panel, the contribution from newly-born BHs in the redshift interval is highlighted in magenta,
and the cumulative number distribution of BHs formed by prior episodes is shown in green.
The majority of seed BHs form in the epoch of $20\lesssim z \lesssim 40$,
and the distribution is extended to $M_\bullet \gtrsim 10^5~\Msun$, 
reflecting the halo assembly and environmental effects discussed in Section~\ref{sec:seed}.
}
\label{fig:seedmf}
\vspace{5mm}
\end{figure*}

Let us consider that a seed BH with a mass of $M_{\bullet,i}$ formed in the $i$-th halo merger tree at the cosmic age 
$t_i$ ($1\leq i \leq N_{\rm tot}=10^4$),
and each of them contributes to the probability distribution function (PDF) of seed mass at a time $t$ as
\begin{equation}
  p^{\rm seed}_i (\Mbh,t)\equiv \frac{\D^2 P^{\rm seed}_i}{\D \Mbh ~\D t}= \frac{ \delta(\Mbh-M_{\bullet,i})~ \delta(t-t_i)}{N_{\rm tot}}.
\end{equation}
Based on the cumulative seed-mass distribution of $\D P^{\rm seed}_i/ \D \Mbh$
(i.e., the integral of $p^{\rm seed}_i$ over time),
we construct the mass function from all $N_{\rm tot}$ seeds formed in one parent halo
with a mass of $\Mh$, where gas has a coherent streaming velocity of $\vbsm$ as a function of time (and redshift).
The absolute abundance of those seed BHs is calculated with the number density of the parent halo in a comoving volume at $z=6$ 
from the Sheth-Tormen mass function \citep{2001MNRAS.323....1S},
\begin{align}
  n_{\Mh}= \int_{\Mh}^{10\Mh}  \frac{\D n_{\mathrm{ST}}} {\D \Mh'} \D \Mh', 
\end{align}
where we adopt $\Mh = 10^{11}$, $10^{12}$, and $10^{13}~\Msun$, with number densities of
$n_{\Mh} = 9.9\times 10^{-4}$, $6.2\times 10^{-6}$, and $8.9\times 10^{-10}~ \text{Mpc}^{-3}$, respectively.
Combining the PDF and number density normalization, we obtain the seed BH mass function cumulative at time $t$ as
\begin{align}
\Phi_{\Mbh}^{\rm seed}\left(t\right) \equiv \sum_{\Mh, \vbsm} n_{\Mh} f_{\vbsm} 
\cdot \sum_{i}^{N_{\rm tot}} \frac{\D P^{\rm seed}_i}{\D \log \Mbh}\left(t\right) \Big| _{\Mh, \vbsm},
\end{align}
where $f_{\vbsm}$ is the volume fraction of the universe with a streaming velocity.
The value is calculated as $f_{\vbsm} = 0.6$ and $0.4$ for $\vbsm = 0$ and $\vbsm=1~\sigma_{\rm bsm}$, respectively.

Fig.~\ref{fig:seedmf} presents how the seed mass function is developed in each redshift interval.
The majority of seed BHs form in the epoch of $20\lesssim z \lesssim 40$, and the distribution function is 
extended to $M_\bullet \gtrsim 10^5~\Msun$, reflecting the halo assembly and environmental effects.
Note that in this study, we consider massive seed BHs formed in low-metallicity environments,
but neglect a potential contribution from BHs left behind the second-generation star formation in metal-enriched environments.
The latter ones would dominate in number at the low-mass end of the BHMF at $z\sim 6$,
but have little impact on the bulk properties of the observed QLF and BHMF \citep[see][]{2022MNRAS.511..616T}.

\vspace{2mm}
\subsection{BH mass growth}
\label{sec:model}
With BH seeds planted, we study the evolution of the BH mass distribution until $z=6$. 
The growth model of each BH is characterized by a minimal number of free parameters, 
giving the accretion rate by
\begin{equation}
  \label{eq:mdot}
  \Mdot_\bullet = \lambda f(\Mbh) \Mdot_\mathrm{Edd} ,
\end{equation}
where $\lambda$ is the ratio of the quasar bolometric luminosity to its Eddington luminosity $L_{\rm Edd}$,
and $\Mdot_\mathrm{Edd} \equiv L_{\mathrm{Edd}}/\eta_0 c^2$ is the Eddington accretion rate with a radiative efficiency of $\eta_0=0.1$ \citep{1973A&A....24..337S},
which is consistent with the value obtained from the Soltan-Paczy{\'n}ski argument \citep[e.g.,][]{2002MNRAS.335..965Y,2010ApJ...725..388C}.
Based on studies on the mass dependence of the radiative efficiency for AGNs at $z\lesssim$ 3 
\citep{2008MNRAS.390..561C,2012ApJ...749..187L,2014ApJ...786..104U}, 
we introduce a similar functional form of
\begin{equation}
\label{eq:f_M}
f(\Mbh) = \frac{2}{1+\left(\Mbh /M_{\bullet,c} \right)^\delta}, 
\end{equation}
where $M_{\bullet,c}=10^8~\Msun$ is adopted
\footnote[3]{
A series of radiation hydrodynamical simulations for BH accretion
suggest that the gas supplying rate from galactic scales into the nuclear scale is not high enough to sustain 
super-Eddington accretion when the BH mass is higher than the characteristic value \citep{2021ApJ...907...74T}.}.
With a positive value of $\delta$, the BH growth at the high mass end is suppressed,
while the growth speed for less massive BHs is accelerated.
In the limit of $\delta \ll 1$, where $f(\Mbh) \simeq 1$ is nearly independent of $M_{\bullet,c}$,
the model in Eq.~(\ref{eq:f_M}) reproduces an exponential growth with an $e$-folding timescale of
$t_{\rm S} =  \Mbh/\Mdot_{\mathrm{Edd}} \approx 45$ Myr (the so-called Salpeter timescale; \citealt{1964ApJ...140..796S}).
Note that we do not explicitly consider the energy loss with radiation in Eq.~(\ref{eq:mdot}), neglecting a factor of $1-\eta_0=0.9$.
This small deviation from unity can be absorbed by the uncertainty due to the functional form of $f(M_\bullet)$ and does not bring 
a significant impact on the results discussed below.

\begin{figure*}
\centering
\includegraphics[width=170mm]{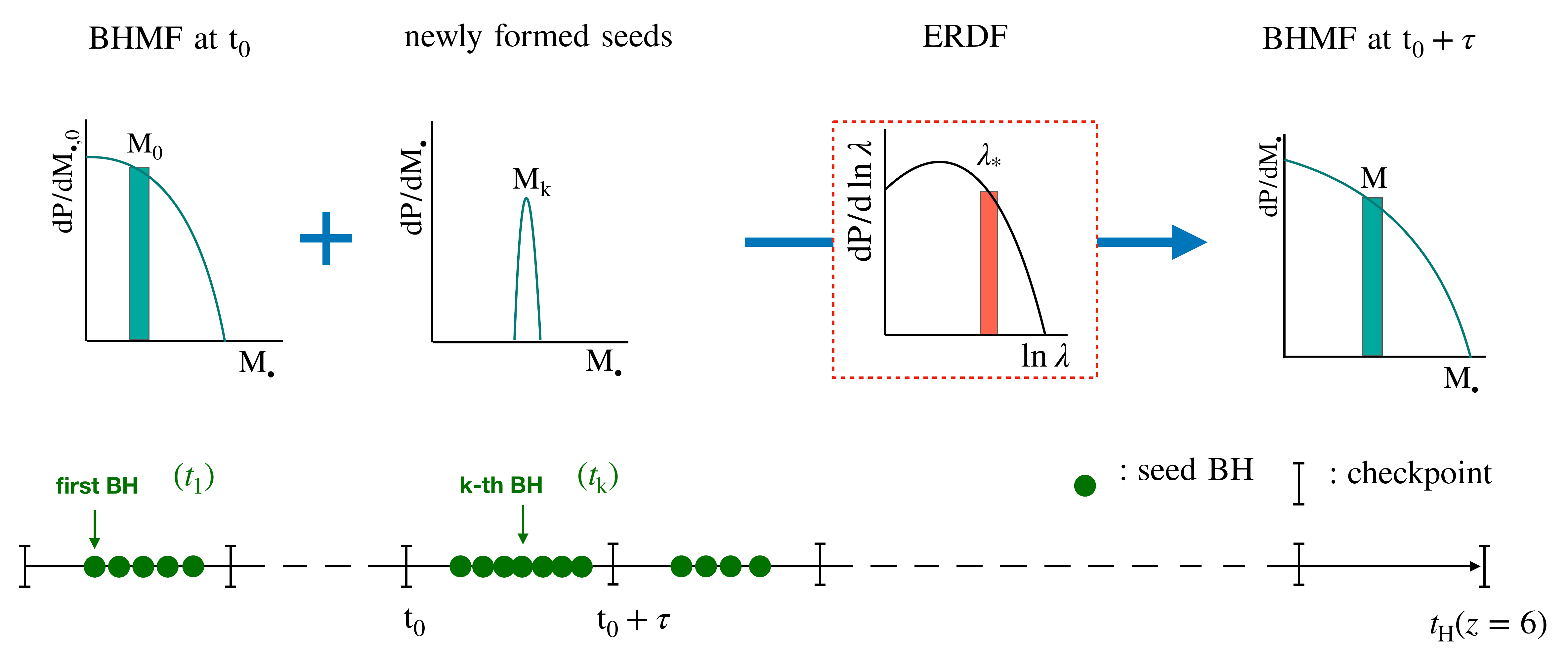}
\caption{
Schematic illustration of the numerical procedure for calculating the BH mass function
between a time interval with a duration of $\tau$.
The time evolution of the distribution is calculated by adding newly-forming seed BHs and 
taking into account growth of existing BHs via mass accretion.
For a given BH population at a checkpoint at $t=T_0$, their growth is calculated with Eqs.~(\ref{eq:dpdm}) and (\ref{eq:mi2m}),
where the ERDF is assumed to follow Eq.~(\ref{eq:Pl}).
For a newly formed seed BH with $M_{\rm k}$ during the interval (i.e., $T_0 \leq t_k < T_0+\tau$), 
we evolve the BH mass in $\Delta t = T_0+\tau -t_k$ and add it to the mass distribution of the existing BHs.
}
\label{fig:scheme}
\vspace{8mm}
\end{figure*}

Quasar activity is thought to take place episodically with accretion bursts triggered by gas inflows to the galactic nuclei
\citep{2005Natur.433..604D,2005ApJ...630..705H}, 
and the BH feeding rate declines by self-regulating feedback processes \citep[e.g.,][]{2008ApJ...686..815Y,2011ApJ...737...26N} 
or gas consumption \citep{1991MNRAS.248..754P,2005ApJ...634..901Y,2007MNRAS.377L..25K}. 
This ``flickering'' pattern of individual quasar luminosity evolution can be directly translated into the diversity 
in the Eddington ratios of quasar samples. 
We here suppose that the Eddington ratio distribution function $\D P/ \D\ln\lambda$ (ERDF) is characterized
with a Schechter function and the distribution function is independent of redshift 
\citep{2006ApJ...639..700H,2009ApJ...698.1550H}.
For unobscured AGNs at low redshifts, the ERDF is characterized with a Schechter function
\citep{2015MNRAS.447.2085S,2016ApJ...826...12J,2018MNRAS.474.1225A}.
Motivated by those facts, we give the ERDF by a Schechter function with two free parameters $\lambda_0$ and $\alpha$ as
\begin{equation}
  \label{eq:Pl}
  \frac{\D P}{ \D \ln \lambda} \propto
  \left(\frac{\lambda} {\lambda_0} \right)^\alpha \exp{\left(-\frac{\lambda}{\lambda_0}\right)}.
\end{equation}
The normalization is set so that the integral of this function over $\lambda_\mathrm{min}(=0.01) \leq \lambda < \infty$ is unity.
The minimum Eddington ratio $\lambda_\mathrm{min}$ is adopted from the simulated evolution of quasar luminosities 
that decrease mildly toward $\lambda \sim 0.1-0.01$ with $\lesssim$ 100 Myr after the peak activity \citep{2011ApJ...737...26N}.
Moreover, a turnover in the ERDF at the lowest value of $\lambda\lesssim 0.01-0.001$ is suggested by observations of X-ray selected AGNs \citep{2018MNRAS.474.1225A}.

To characterize the episodic accretion patterns of individual BHs, 
we introduce a time duration of $\tlife$, during which a single value of $\lambda$ is assigned to an accreting BH
following the ERDF of Eq.~(\ref{eq:Pl}).
In this way, we give a different growth speed of a BH in every period with a duration of $\tlife$ until $z=6$ 
(the corresponding cosmic age is $t_{\rm H}\simeq 913$ Myr).
This treatment enables us to capture the nature of accretion bursts and prohibit an unrealistically 
long-lasting rapid growth phase of a BH.
This phenomenological method avoids numerous uncertainties in modeling of the galaxy assembly, 
gas feeding, and BH feedback processes in the quasar progenitor halos,
which are implemented as sub-grid models in previous cosmological simulations \citep{2017MNRAS.467.4243D,2018MNRAS.473.4003B,2019MNRAS.488.4004L}.
These studies found that the growth of massive BH seeds toward SMBHs is characterized by the intermittent patterns with quiescent and accretion phases.
Instead of treating these ingredients explicitly in our semi-analytical model, we extract 
an averaged but fundamental timescale that governs mutual correlation between the QLF and BHMF,
based on direct comparison to those distribution functions for the observed high-$z$ quasar population.

Thus far, we assume that {\it all} the seed BHs formed in parent halos participate in the assembly of SMBHs and 
end up in quasar host galaxies by $z\simeq 6$.
To relax this stringent assumption, we introduce the seeding fraction of $\fseed (\leq 1)$.
As discussed in \citet{2009ApJ...696.1798T}, a small value of $\fseed <1$ is required to avoid overproduction of SMBHs at $z\sim 6$,
depending on BH seeding and growth mechanisms in semi-analytical models.
Since the value has been poorly constrained both by theoretical and observational studies, we treat it as a free parameter.
We take three different values of $\fseed = 1$, $0.1$, and $0.01$, and explore its effect on shaping the QLF and BHMF.
Note that the seed mass function shown in Fig.~\ref{fig:seedmf} is given with $\fseed = 1$.

\vspace{2mm}
\subsection{BH mass function}\label{sec:MF}

We describe how to calculate the time evolution of the BH mass function, 
combining the semi-analytical growth model and ERDF (see Eqs.~\ref{eq:mdot} and \ref{eq:Pl}).
The schematic illustration of our numerical procedure is shown in Fig.~\ref{fig:scheme}. 
We first set a series of checkpoints with an interval of $\tlife$, 
tracing back from $t_{\rm H}(z=6)$ to just prior to the first seed formation time $t_1$, where $t_i$ is the $i$-th seed forming time
as defined in Section~\ref{sec:seed} ($1\leq i\leq N_{\rm tot}$).
Note that the number of the checkpoints is given by $\lceil (t_{\rm H}-t_1)/\tau \rceil+1$, where $\lceil x\rceil \equiv {\rm min}\{n \in \mathbb{Z}~ |~x\leq n\}$.

The time evolution of the BH mass distribution is calculated from the first checkpoint to $t_{\rm H}(z=6)$ by adding newly-forming seed BHs
and taking into account growth of existing BHs via mass accretion.
For a given mass distribution function $\D P/\D  M_{\bullet, 0}$ at a checkpoint of $t=T_0$, 
we give its evolution at $t=T_0+\Delta t$ as
\begin{equation}
  \label{eq:dpdm}
  \frac{\D P}{\D \Mbh} = \int
   \frac{\D P}{\D \ln \lambda}\Big|_{\lambda_\ast} \cdot
   \frac{\D \ln \lambda_\ast}{\D  \Mbh}\Big|_{\Delta t, M_{\bullet,0}} \cdot
  \frac{\D P}{\D M_{\bullet, 0}}~\D M_{\bullet, 0},
\end{equation}
where $\lambda_\ast$ is the Eddington ratio required for a BH with $M_{\bullet,0}$ to grow up to $\Mbh$ in $\Delta t$,
and is calculated analytically from Eq.~(\ref{eq:mdot}) by
\begin{equation}
  \label{eq:mi2m}
\frac{2 \lambda_\ast \Delta t}{t_{\rm S}}=
  \ln \left(\frac{\Mbh} {M_{\bullet,i}}\right) + \frac{1}{\delta} 
  \left[ \left(\frac{\Mbh}{M_{\bullet,c}}\right)^\delta - \left(\frac{M_{\bullet, 0}}{M_{\bullet,c}}\right)^\delta \right],
\end{equation}
and the derivative of ($\D \ln \lambda_\ast / \D  \Mbh )|_{\Delta t, M_{\bullet,0}}$ is also determined.
For existing BHs, we calculate the evolution of the mass distribution by setting $\Delta t = \tlife$.
For a seed BH formed during this cycle (i.e., $T_0\leq t_k < T_0+\tlife$), 
we evolve $\D P^{\rm seed}_k/\D \Mbh$ by setting $\Delta t = T_0+\tlife -t_k$ and add it to the mass distribution 
of the existing BHs at $t=T_0+\tlife$.
Combining the PDF and number density normalization, the BHMF is given by
\begin{equation}
  \Phi_{\Mbh} 
  =\sum_{\Mh, \vbsm} n_{\Mh} f_{\vbsm} {\fseed} \cdot \frac{\D P}{\D \log \Mbh}\Big|_{\Mh, \vbsm}.
 \end{equation}

To follow the time evolution of the mass function, we set up 
logarithmically spaced mass grids at $10^2 \leq \Mbh /\Msun \leq 10^{12}$.
The number of the grid points is set to $N_{\rm bin}=800$, so that the convergence of the numerical result is ensured.
Our result of $\Phi_{\Mbh}$ is consistent with that obtained from the direct sampling method, where 
we consider the growth of $10^7$ individual BHs rather than the evolution of smooth analytical mass distribution functions.
Our method reduces the statistical errors seen in the direct sampling method 
and thus allows us to extend the BHMF and QLF to the higher mass and brighter end
in a reasonable computational time.

\vspace{2mm}
\subsection{Obscuration-corrected QLF}\label{sec:LF}

Using the BHMF and its evolution, we construct the QLF, which can be directly probed by high-$z$ quasar observations.
For an accreting BH with a bolometric luminosity of $\Lbol=\lambda L_\mathrm{Edd}$, 
its rest-frame ultraviolet (UV) absolute magnitude of $\Muv$ at 1450 $\mathrm{\AA}$ is estimated as
\begin{equation}
  \label{eq:M1450}
  \Muv= -21.0-2.5 \log  \left(\frac{\Lbol}{10^{45}~\mathrm{erg~s}^{-1}} \right) ~[\rm{mag}],
\end{equation}
where a bolometric correction factor of $f^{\rm bol}_{1450}=4.4$ for the monochromatic UV band is adopted
\citep{2006ApJS..166..470R}.
We construct the QLF from the BHMF, combining this conversion factor and ERDF.
However, we note that the shape and normalization of the {\it intrinsic} QLF is not necessarily identical to the {\it observed} QLF
because quasar surveys with rest-frame UV-to-optical bands are not sensitive to the obscured quasar population,
i.e., the observed number of fainter quasars is largely reduced by the obscuration effect
\citep{2003ApJ...598..886U,2007A&A...463...79G,2008A&A...490..905H,2014ApJ...786..104U,2014MNRAS.437.3550M}.

A widely-accepted mechanism causing quasar obscuration is extinction by dusty tori and dense gas clouds 
in the circumnuclear region, hence tightly related to gas fueling and feedback processes from accreting BHs 
\citep[see][for a review]{2018ARA&A..56..625H}.
In addition, dense gas clouds at larger galactic scales may have a significant contribution to quasar obscuration at 
higher redshifts \citep{2020MNRAS.495.2135N}.
However, the physical origin and conditions for high-$z$ quasar obscuration have been poorly understood yet.
Observationally, the obscured fraction of AGNs with hydrogen column density of $N_{\rm H}>10^{22}~{\rm cm}^{-2}$
can be estimated from spectral analyses in the X-ray band
\citep[e.g.,][]{2003ApJ...598..886U,2007A&A...463...79G,2008A&A...490..905H}. 
The obscured fraction is nearly $\gtrsim 80\%$ at $L_{\rm X}<10^{43}~{\rm erg~s}^{-1}$ and 
decreases with the X-ray luminosity \citep{2014ApJ...786..104U,2014MNRAS.437.3550M}.
The exact function shape still remains under debate, but the overall dependence on the X-ray luminosity holds.
Moreover, the obscured fraction is found to increase gradually toward higher redshifts and is saturated at $z\gtrsim 2-3$
\citep{2008A&A...490..905H,2014ApJ...786..104U,2014MNRAS.437.3550M,2018MNRAS.473.2378V,2022A&A...666A..17G}.

In this study, we adopt the obscuration fraction $\fobsc$ based on X-ray observations \citep{2014ApJ...786..104U}:
\begin{eqnarray}
\fobsc && = \text{min}  \left[ \psi_{\rm max}, \psi \right], \\[5pt]
\psi &&= \text{max}\left[\psi_0  - \beta \log (L_{\rm X}/L_{\rm X,0}), \psi_{\rm min}     \right],
\end{eqnarray}
where the parameters are written as $\psi_{\rm max}=0.84$, $\psi_{\rm min}=0.2$, $\psi_{0}=0.73$, $\beta=0.24$, and $L_{\rm X,0}=10^{43.75}~{\rm erg~s}^{-1}$.
%
Here, we use the bolometric correction factor of the hard X-ray band:
\begin{equation}
  f^{\rm bol}_{\rm X} \equiv \frac{\Lbol}{L_{\rm X}}=
  a\left[1+\left(\frac{\log \left(\Lbol / L_\odot\right)}{b}\right)^{c}\right],
\end{equation}
where $a=10.96$, $b = 11.93$, and $c = 17.79$ \citep[see Eq.~2 in][]{2020A&A...636A..73D}. 
It is worth noting that the functional form of the obscuration fraction
is calibrated with X-ray selected AGNs at $z\lesssim 5$,
and the value for high-$z$ quasars is still under investigation.
We leave more comprehensive arguments about this effect for future work.

After taking into account the obscuration effect, 
the observed QLF for a given BHMF and ERDF is calculated as
\begin{equation}
\label{eq:dn_dM1450}
\Phi_{\Muv} 
 = (1 -\fobsc)  
\int \frac{\D P}{\D \ln \lambda}\Big|_{\tilde{\lambda}}  \cdot
\frac{\D \ln \tilde{\lambda}}{\D \Muv} \cdot
 \Phi_{\Mbh} \D \log \Mbh,
\end{equation}
where the values of $\tilde{\lambda}=\lambda(\Muv, \Mbh)$ and $\D \ln \tilde{\lambda}/\D \Muv$ are calculated analytically from Eq.~(\ref{eq:M1450}).
Hence, the observed QLF can be produced simultaneously with the BHMF from a set of model parameters.

\vspace{2mm}
\subsection{MCMC fitting}\label{sec:fitting}

In this section, we describe the MCMC fitting procedure used to optimize 
the BH growth parameters so that the observed BHMF and QLF are consistently reproduced. 
As discussed in Section~\ref{sec:MF}, we consider five parameters of $\tlife$, $\delta$, $\alpha$, $\lambda_0$, and $\fseed$.
We calculate the best-fit values of the first four parameters using the {\tt emcee} Python package for the MCMC sampling \citep{2013PASP..125..306F},
fixing $\fseed$ to a single value.
We vary four parameters instead of five in the fitting to guarantee that the Markov chain can be stabilized. 
The fitting is carried out by sampling 100 walkers with 5000 steps.
We verify the convergence of the posterior distribution results by doubling the chain length.

%
In this analysis, we constrain those four parameters to be within certain physically acceptable ranges, 
assuming a prior probability function $P^{\rm prior}_a$ for each quantity $a=\tlife$, $\delta$, $\lambda_0$, and $\alpha$.
We use a uniform prior on $\tlife$ with a limit of $10~{\rm Myr} \leq \tlife \leq 200~{\rm Myr}$,
motivated by the constraints on the quasar lifetime based on various observations
\citep[e.g.,][]{2004cbhg.symp..169M}.
The parameter of $\delta$ that characterizes a non-exponential BH growth 
is constrained at $-4 \leq \log \delta \leq -0.3$
\footnote[4]{ We first set a uniform prior at $0\leq \delta \leq 1$, 
and find the posterior distribution assembles to $\delta \lesssim 0.1$.
We then switch to impose the $\log \delta$ prior to improve the fitting performance at small $\delta$ values.}.
The prior function is uniform at $\log \delta \geq -3$ and is imposed to decay with an exponential cutoff at $\log \delta < -3$,
which is small enough for the model to be approximately mass independent, i.e., $f(M_\bullet)\simeq1$.
For the characteristic Eddington ratio $\lambda_0$ in the Schechter-type ERDF, 
we impose a Gaussian prior function with a mean value of $\mu_{\lambda_0}=0.6$ and a dispersion of $\sigma_{\lambda_0}=0.4$.
The choice is motivated by the brightest quasar samples at $z=$ 6, whose ERDF is peaked around $\lambda \sim 0.6$ 
after luminosity bias correction \citep[e.g.,][]{2010AJ....140..546W}.
For the low-$\lambda$ slope of $\alpha$ in the ERDF, we assign a Gaussian prior function with a mean value 
$\mu_{\alpha}=0$ and a dispersion of $\sigma_{\alpha}=0.3$.
Note that the value of $\alpha$ remains poorly constrained by the current high-$z$ quasar observations,
but the low-$z$ quasar samples support $-0.3 \lesssim \alpha \lesssim 0.3$ 
\citep[e.g., see the mass-integrated ERDF for comparison; the right panel of Figure 21 in][]{2015MNRAS.447.2085S}.

The observed BHMF data is taken from \citet{2010AJ....140..546W} (hereafter, \citetalias{2010AJ....140..546W}), where 
the virial BH mass measurements for 17 bright quasars at $z\simeq $ 6 are used.
Although the BHMF spans over $10^7 < \Mbh/\Msun <10^{10}$, the best constraint is given around 
$\Mbh \simeq 10^{8.5}~\Msun$
and thus the statistical error size becomes larger both at the lower and higher mass range.
We set up 10 logarithmically spaced mass bins and adopt errors with sizes increasing quadratically as 
$\Phi_{\Mbh}^{\rm err} = \pm \{0.2 +   [\log (\Mbh/10^{8.5}~\Msun)]^2/3 \}~{\rm dex}$,
covering the same mass range as in their bootstrap resamples (see Figure 8 of \citetalias{2010AJ....140..546W}). 
The observed (unobscured) QLF is taken from \citet{2018ApJ...869..150M} (hereafter \citetalias{2018ApJ...869..150M}) 
over a wide range of UV magnitude at $-30 < \Muv <-22$ mag (see their Table 4 and Figure 13). 
The magnitude bins and the error sizes are given consistently with \citetalias{2018ApJ...869..150M}.

In the MCMC fitting, the four parameters are sampled to generate the synthetic BHMF and QLF dataset,
which are compared to the observational dataset. 
The probability is evaluated by a $\chi^2$-type value defined by
%
\begin{align}
  \chi^2 = \sum_{x,i}
  \frac{\left(\log{\Phi^{\rm mod}_{x,i}} - \log{\Phi^{\rm obs}_{x,i}}\right)^2}{(\log{\Phi^{\rm err}_{x,i}})^2}
  - 2 \sum_{a} \ln P^{\rm prior}_a.
\end{align}
The first term on the right hand side is the classical $\chi^2$ value that measures the dispersion of the observational data from the model prediction,
where $\Phi^{\rm mod}_{x,i}$ and $\Phi^{\rm obs}_{x,i}$ represent the modeled and observational value 
on the \textit{i-}th bin of the BHMF ($x=\Mbh$) and QLF ($x=\Muv$), and $\Phi^{\rm err}_{x,i}$ is the corresponding error.
Adopting 10 mass bins and 12 magnitude bins, we treat the errors from comparison both in the BHMF and QLF
with nearly equal weight.
The second term denotes the deviation of the parameter set from the assumed prior distribution functions.
The posterior distribution of the MCMC fitting is stabilized around the minimum value of $\chi^2$ (highest probability), 
with the models best reproducing the $z=$ 6 BHMF and QLF.
%

\vspace{2mm}
\section{Results}\label{sec:fitting_result}

\begin{figure*}
\centering
\includegraphics[width=85mm]{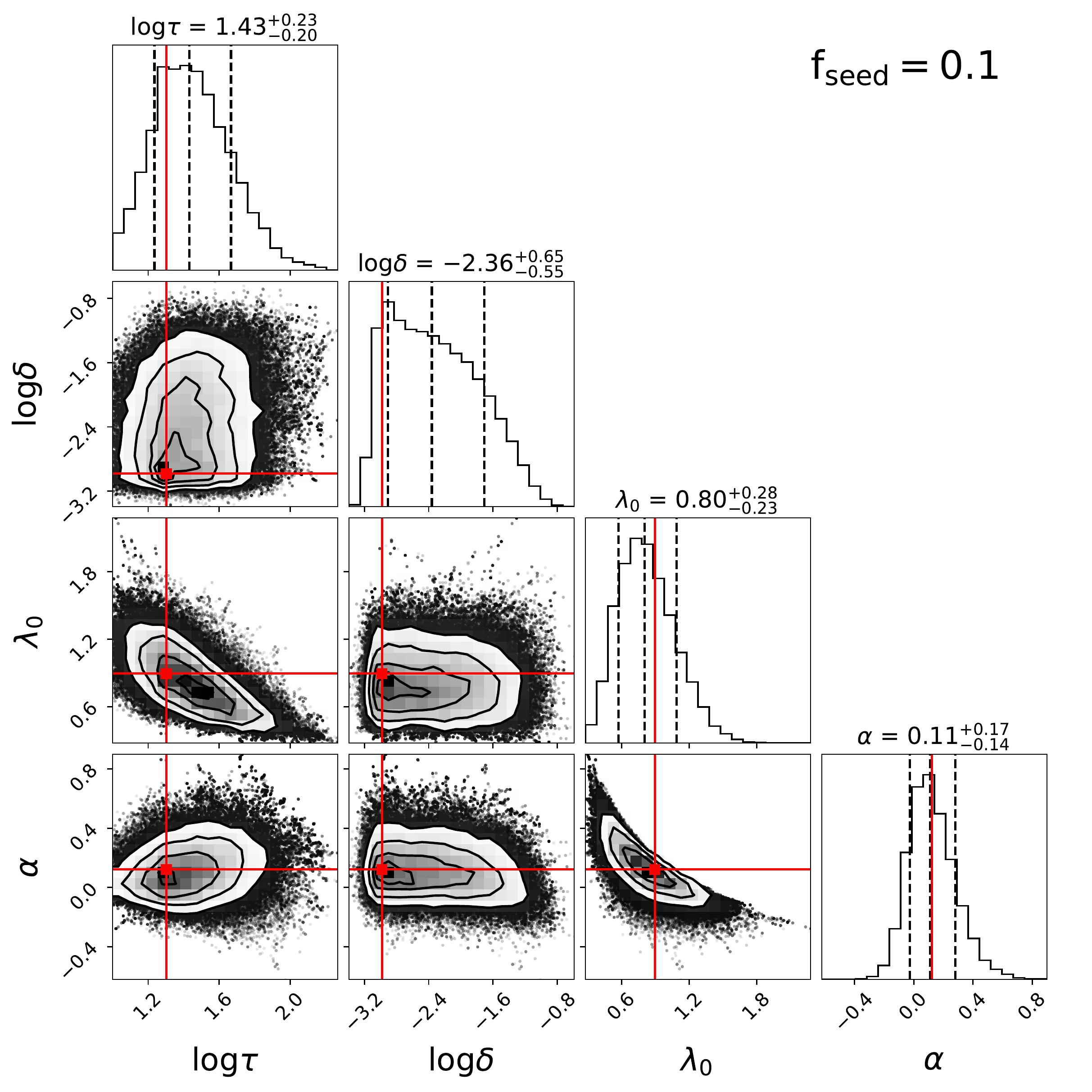}\hspace{3mm}
\includegraphics[width=85mm]{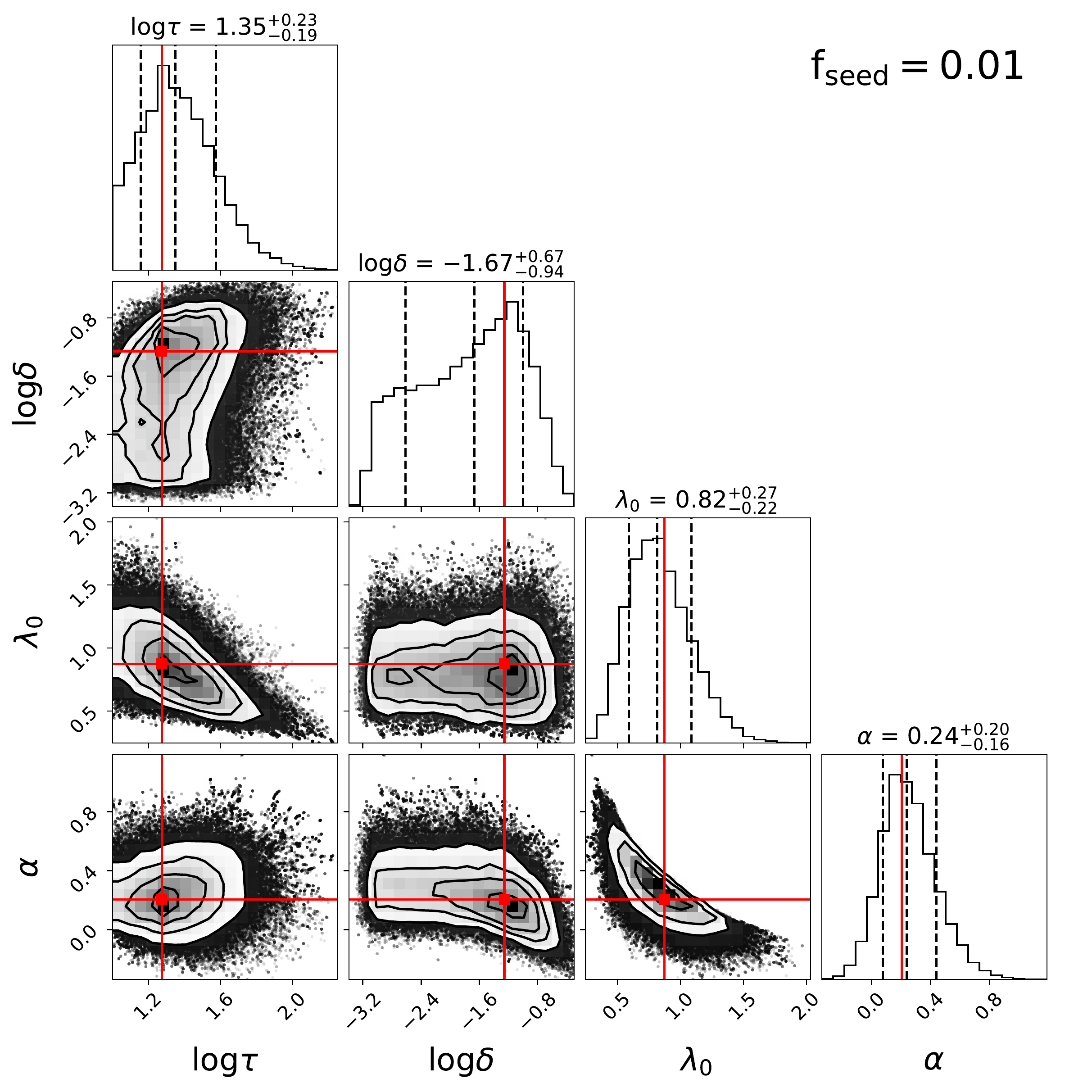}
\caption{
Two dimensional posterior distribution of the four model parameters with $\fseed=0.1$ (left) and 0.01 (right), 
along with the marginalized one dimensional projection.
The three vertical dashed lines in the one-dimensional posterior distribution of each parameter correspond to 
16\%, 50\%, and 84\% quantiles in the cumulative distribution, respectively. 
The best-fit values shown with the red lines are listed in Table~\ref{tab:best_fits}.
}
\label{fig:contour}
\vspace{5mm}
\end{figure*}

We discuss the MCMC fitting results and compare the BHMF and QLF with their observed distribution functions.
The best-fit model parameters together with the $\chi^2$ values for the three cases with $\fseed$ are summarized in Table~\ref{tab:best_fits}.
We note that the ideal case with $\fseed=1.0$, where environmental effects preventing BHs from growing is neglected, yields a $\chi^2$ value 
nearly twice larger than those in the other two cases.
Therefore, in the following, we focus only on the results with $\fseed=0.1$ and $0.01$ in comparison with observations.
In Fig.~\ref{fig:contour}, we visualize the fitting results of the model parameters in two-dimensional 
posterior distribution for the case of $\fseed= 0.1$ (left) and $0.01$ (right), respectively.
The three vertical dashed lines in the one-dimensional posterior distribution of each parameter
correspond to 16\%, 50\%, and 84\% quantiles in the cumulative distribution, respectively.
The best-fit values shown with the red lines are listed in Table~\ref{tab:best_fits},
and are consistent with the peak values in the one-dimensional distribution.

The best-fit solutions yield a small value of $\delta < 0.1$.
This result favors the nearly exponential BH growth model (Eqs.~\ref{eq:mdot} and \ref{eq:f_M}).
The typical duration of quasar activity is required to be $\tlife \simeq 20-30$ Myr for both cases.
This suggests that accreting seed BHs vary their growth speeds and their Eddington ratios 
$\simeq 30-40$ times toward the cosmic time at $z\simeq 6$.
The timescale of $\tlife$ is also related to the quasar lifetime of $t_{\rm Q}\sim 1-10$ Myr, 
which is estimated by the measurement of the physical extents of hydrogen Ly$\alpha$ proximity zones 
observed in the rest-frame UV spectra \citep[e.g.,][]{2018ApJ...867...30E,2019ApJ...884L..19D}.
The characteristic Eddington ratio $\lambda_0$ needs to be close to unity, otherwise the highest luminosity and BH mass
in the model would differ from those in observations.
As shown in Fig.~\ref{fig:contour}, the power-law index $\alpha$ in the ERDF shows an anti-correlation with the characteristic Eddington ratio $\lambda_0$.
With a higher $\lambda_0$ (i.e., faster BH growth), the ERDF needs to be skewed toward the lower $\lambda$ regime with a smaller value of $\alpha$ (i.e., a larger fraction of inactive BHs)
to be consistent with the observed QLF and BHMF.

The best-fit parameters for the two cases reproduce the bulk properties of the BHMF and QLF at $z\sim 6$,
despite the ten-fold difference in the seeding fraction.
With the fitted values for the ERDF, the fraction of super-Eddington accreting BHs is estimated as 
$P(\lambda \geq 1)=0.064$ and $0.075$ for $\fseed=0.1$ and $0.01$, respectively.
While the probability of rapid accretion in each cycle differs slightly between the two cases,
multiple accretion episodes enlarge the difference and accelerate BH growth for the lower seeding fraction.
Moreover, the case with $\fseed=0.01$ requires a higher value of $\delta (\simeq 0.055)$,
promoting the growth of less massive BHs with $M_\bullet <10^8~\Msun$.
Therefore, a larger fraction of seed BHs are delivered into the SMBH regime,
and the ten-fold difference in $\fseed$ is nearly compensated.

\begin{figure*}
\centering
\includegraphics[width=85mm]{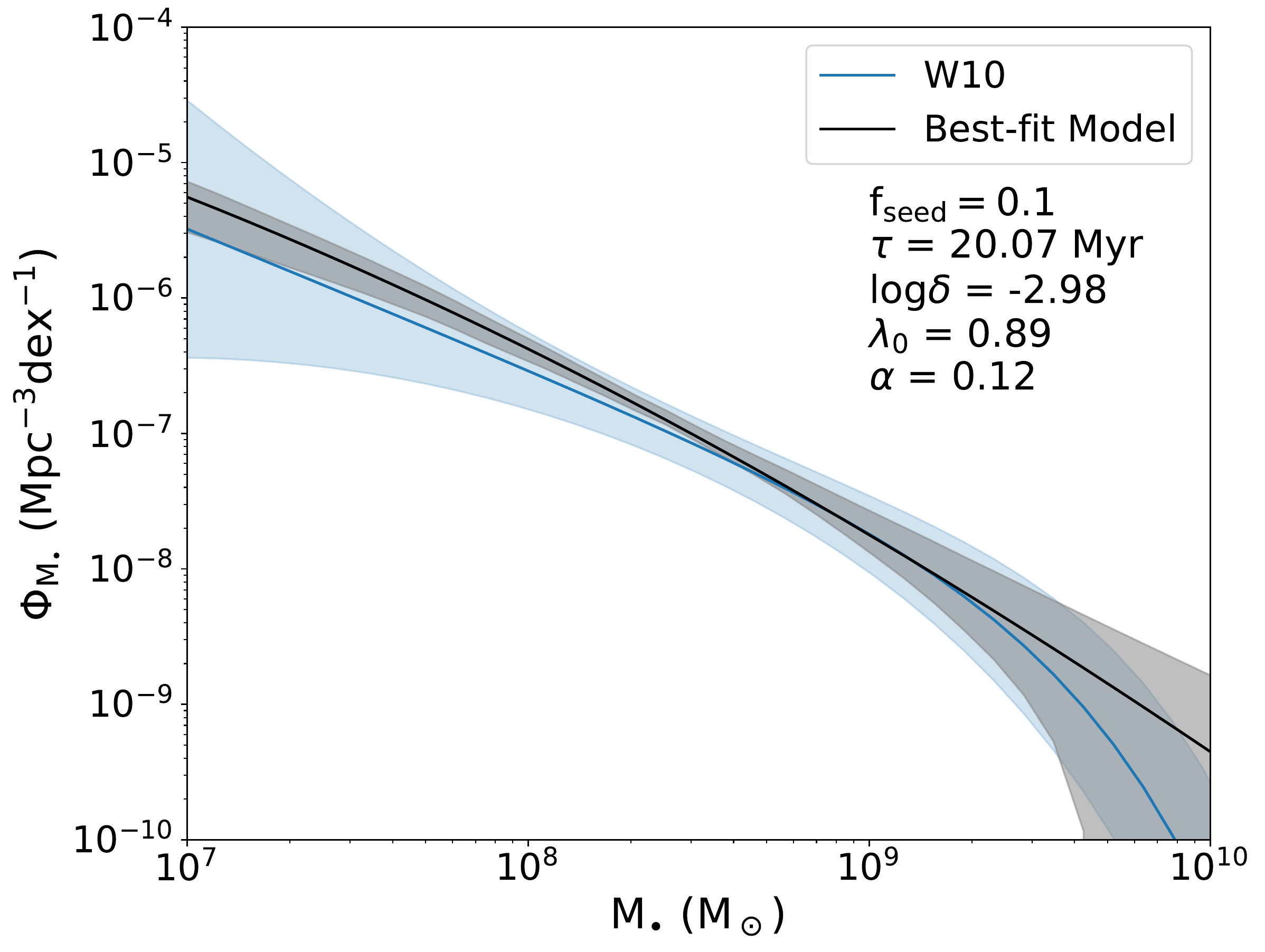}\hspace{3mm}
\includegraphics[width=85mm]{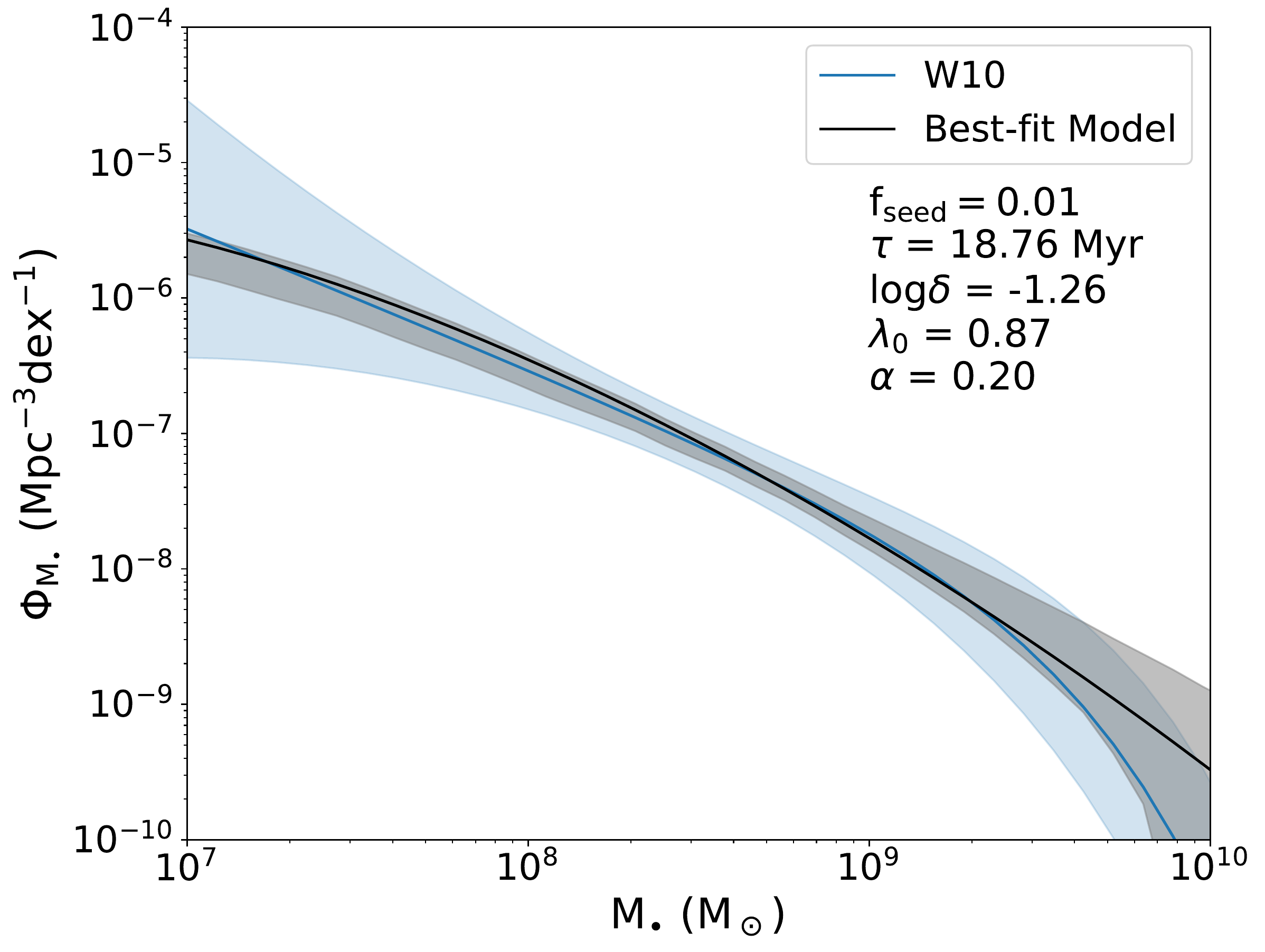}
\caption{
The BH mass function at $z=6$ with the best-fit parameters (black curve) and the $1\sigma$ statistical error (grey shaded region) 
for the cases with $\fseed=0.1$ (left) and $0.01$ (right). 
For comparison, the BHMF constructed by \citetalias{2010AJ....140..546W} and their statistical errors (see text) are shown with the cyan curve and shaded region.
}
\label{fig:fitmf}
\vspace{2mm}
\end{figure*}

In Fig.~\ref{fig:fitmf}, we show the BHMFs at $z=$ 6 reproduced by the best-fit parameters for 
$\fseed = 0.1$ (left) and $0.01$ (right) as well as
the $1\sigma$ standard deviation calculated from random parameter sampling in the Markov chain.
For comparison, we present the BHMF inferred by \citetalias{2010AJ....140..546W} along with the statistical errors.
Note that \citetalias{2010AJ....140..546W} constructed the BHMF using quasar samples with 
$10^8~\Msun \lesssim \Mbh \lesssim 3\times 10^9~\Msun$.
Overall, our best-fit BHMF model agrees with their constraints in this mass range.
The power-law index of the BHMF ($\gamma_{M_\bullet} \equiv \D \ln \Phi_{M_\bullet}/\D \ln M_\bullet$) is as steep as $\gamma_{M_\bullet}\lesssim -1$
at $M_\bullet \gtrsim 10^7~(3\times 10^7)~\Msun$ for $\fseed = 0.1 ~(0.01)$.
Thus, the total mass budget of the entire BH population is dominated by BHs with the characteristic mass (see also Section~\ref{sec:cosm}). 
The discrepancy between the model and observational data enlarges at the high- and low-mass end,
reflecting poor constraints on the BHMF from the current observations.

\begin{deluxetable}{cccccccc}
\tablecolumns{1}
\renewcommand\thetable{1} 
\tablewidth{0pt}
\tablecaption{Best-fit parameters for BH growth\label{tab:best_fits}} 
\tablehead{
  \colhead{$\fseed$}  && \colhead{$\tau$ (Myr)} & \colhead{$\log\delta$} &  \colhead{$\lambda_0$}  & \colhead{$\alpha$}  & \colhead{$\chi^2$}
}
\startdata
1.0    &&  23.13 &    -2.97    &   0.96    &   -0.06    &    15.77   \\
0.1    &&  20.07 &    -2.98    &   0.89    &    0.12    &    8.27    \\
0.01   &&  18.76 &    -1.26    &   0.87    &    0.20    &    6.21    \\
\enddata
\tablecomments{
The best-fit parameters optimized by MCMC sampling for the three different cases of $\fseed$.
The fitting performance is quantified with the relative magnitudes of the $\chi^2$ value.}
\end{deluxetable}

\begin{figure*}
\centering
\includegraphics[width=85mm]{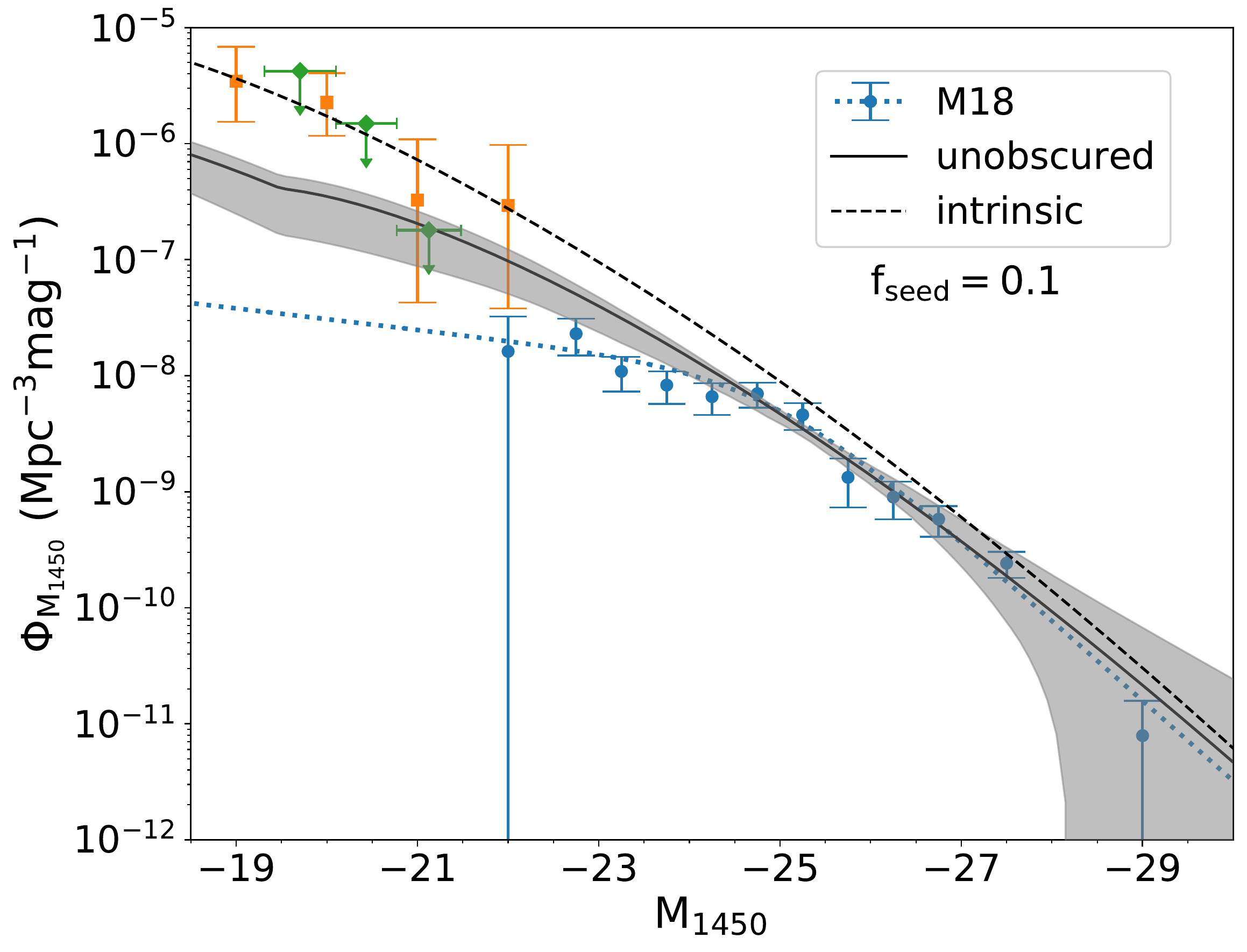}\hspace{3mm}
\includegraphics[width=85mm]{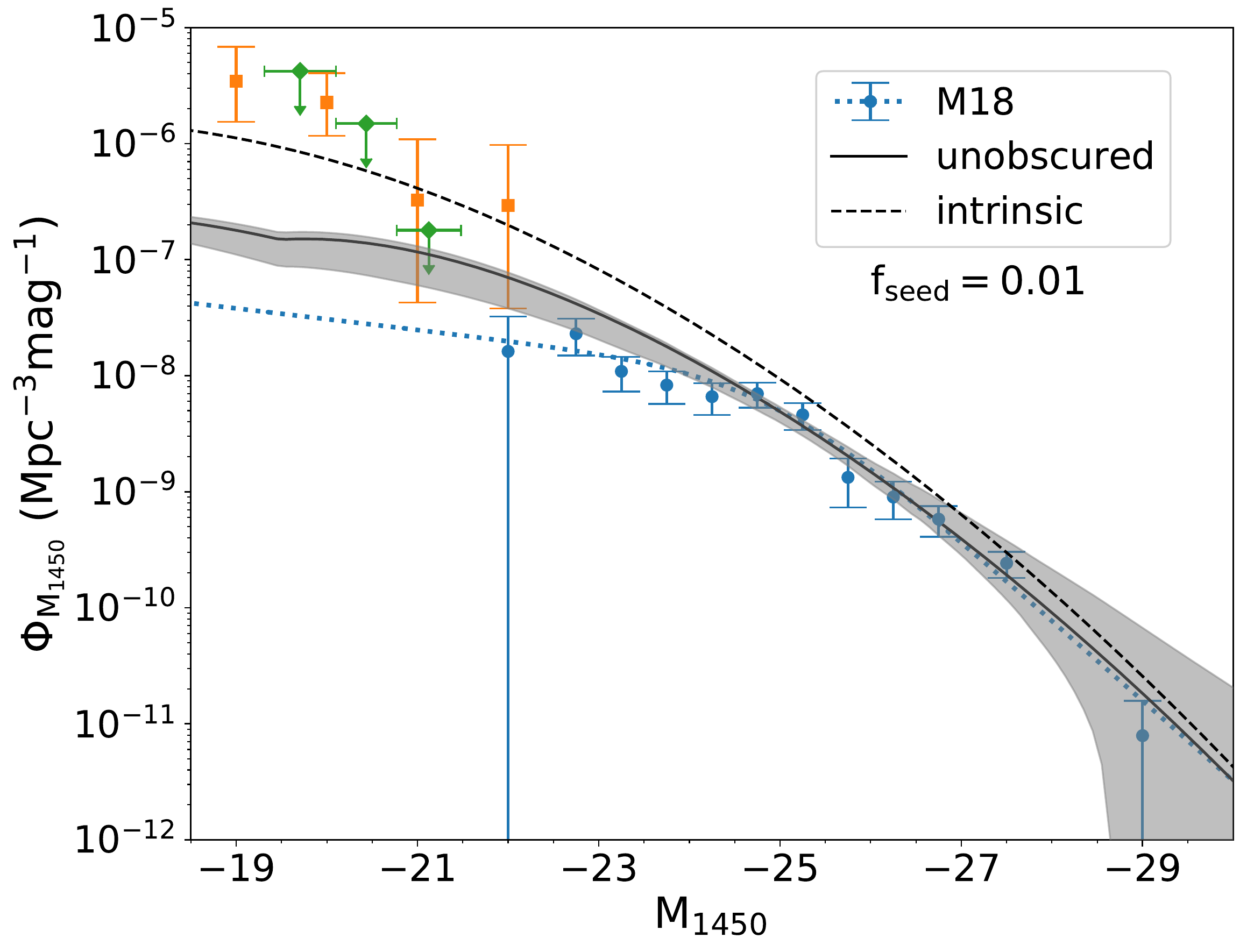}
\caption{
The quasar luminosity function at $z=6$ with the best-fit parameters (black curve) and the $1\sigma$ statistical error (grey shaded region) 
for the cases with $\fseed=0.1$ (left) and $0.01$ (right).
The solid and dashed curves are the unobscured and intrinsic QLF, respectively.
We overlay the observational data for the unobscured QLF taken from \citetalias{2018ApJ...869..150M} in blue and their parametric function with a double power-law function (dotted curve).
Upper bounds of the number density of faint quasars are given by \citet{2022NatAs...6..850J}, based on the Hubble Space Telescope imaging observations (green).
The QLF data based on X-ray selected quasars at $z\sim 5.55$ are overlaid in orange \citep{2019ApJ...884...19G}, for which the normalization is scaled to $z=6$
by a factor of $10^{-0.72\Delta z}$ \citep{2016ApJ...833..222J}, where $\Delta z=0.45$.
}
\label{fig:fitlf}
\vspace{5mm}
\end{figure*}

\begin{figure*}
\centering
\includegraphics[width=85mm]{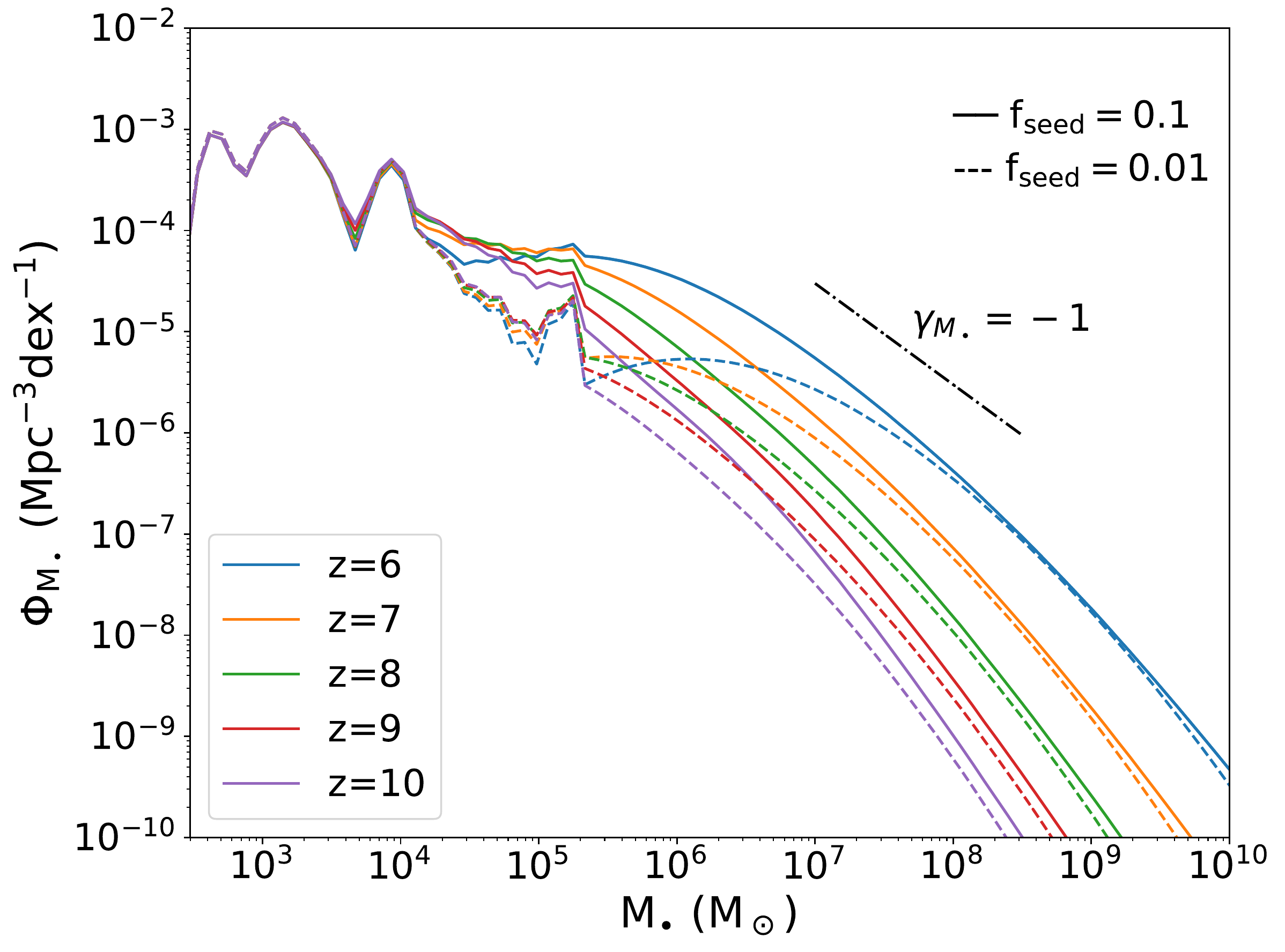}\hspace{5mm}
\includegraphics[width=81mm]{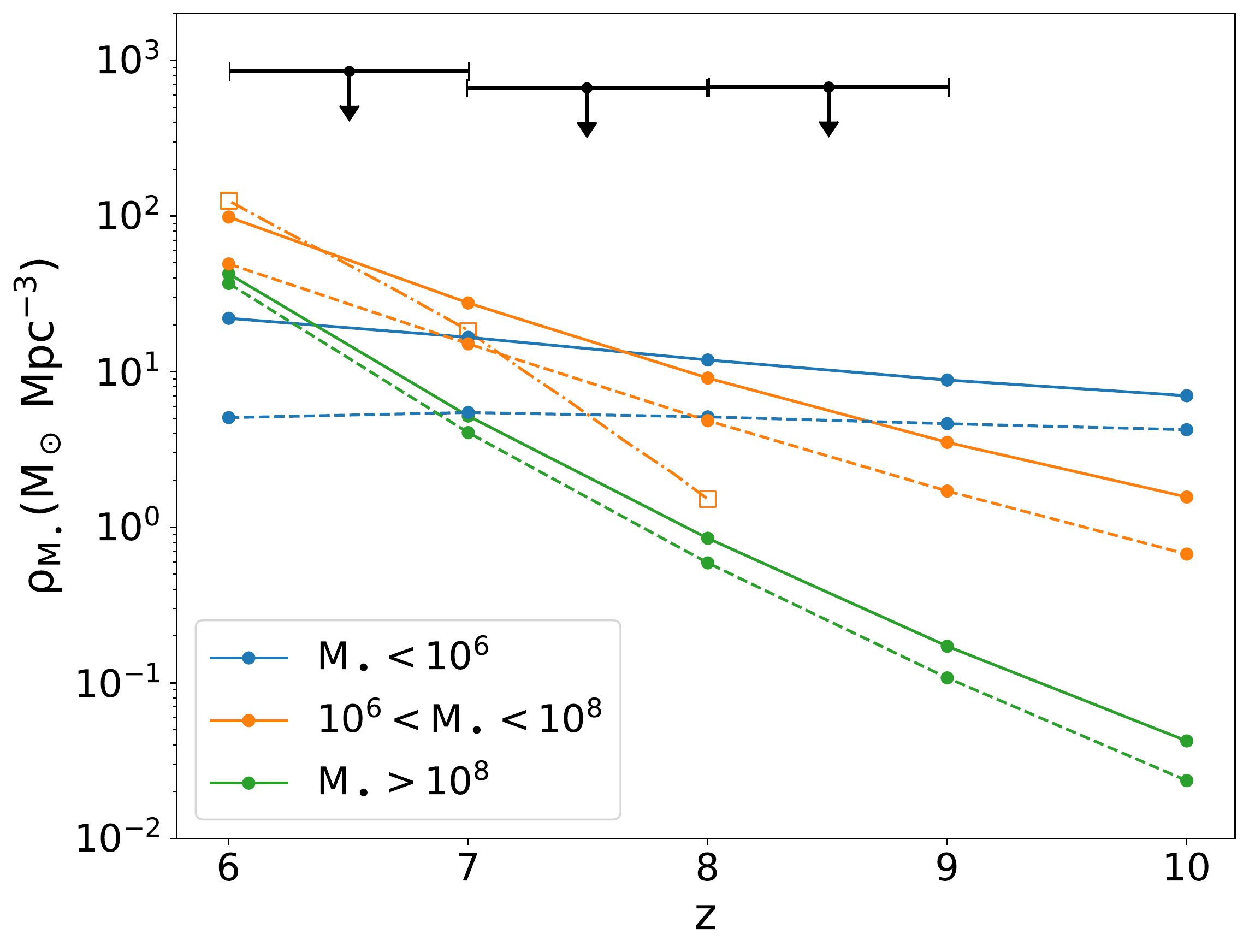}
\caption{
{\it Left}: The predicted BH mass functions at $z=6$--$10$ for the case with $\fseed=0.1$ (solid) and $0.01$ (dashed), including the seed population.
{\it Right}: the cumulative mass density for BHs in a comoving volume at three different mass ranges: $M_\bullet \leq 10^6~\Msun$ (blue), $10^6~\Msun < M_\bullet < 10^8~\Msun$ (orange),
and $M_\bullet \geq 10^8~\Msun$ (green).
The dashed-dotted curve shows the results from the cosmological simulations by \citet{2022MNRAS.513..670N}.
The black arrows present upper bounds of the mass accreted onto BHs obtained from measurements of unresolved cosmic X-ray background radiation \citep{2013ApJ...778..130T}.
}
\label{fig:BHMF_rhoz}
\vspace{5mm}
\end{figure*}

Fig.~\ref{fig:fitlf} presents the QLFs at $z=6$ reproduced by the best-fit parameters as well as the $1\sigma$ spreads
for $\fseed = 0.1$ (left) and $0.01$ (right).
The solid and dashed curve are the unobscured and intrinsic QLF, respectively.
The latter includes both obscured and unobscured quasars.
For comparison, we overlay the observational data taken from \citetalias{2018ApJ...869..150M} in blue
and their parametric QLF with a double power-law function (dotted curve).
Our best-fit unobscured QLF is consistent with the observed one at $-29~{\rm mag} \lesssim \Muv \lesssim -25~{\rm mag}$,
but overproduces fainter quasars at $\Muv\gtrsim -24~{\rm mag}$.
Similarly to the BHMF, the difference between the model and observational data at the faint end becomes smaller with
the lower seeding fraction.
Since the progenitors of those faint quasars originate from lower mass seed BHs with $\fseed =0.01$ (see discussion above),
the BH population on the peak of the mass distribution naturally produces a flatter slope of the QLF at the faint end.
Note that while our best-fit BHMF model over-produces the BH abundance in the high mass end at $\Mbh>10^9~\Msun$
(although consistent within the 1-sigma error), the discrepancy becomes moderate in the brightest end of the QLF because low-$\lambda$ BHs are more abundant with the best-fit EDRF
\footnote[5]{The recent work by \citet{2022MNRAS.517.2659W} updates the construction of the $z=6$ BHMF and shows a higher abundance at the high-mass end, consistent with our results.}.
In addition, we show upper bounds of the faint quasar abundance at $\Muv\gtrsim -22~{\rm mag}$ \citep[green;][]{2022NatAs...6..850J}
based on a search for point sources in the survey fields of the Hubble Space Telescope.
Our unobscured QLF lies within the constraints.

Quasar observations in X-rays also provide 
useful constraints on the intrinsic population because of less obscuration in X-rays.
Based on the X-ray selected faint quasars, \cite{2019ApJ...884...19G} reported a number density of 
faint quasars at $\Muv \gtrsim -22$ mag.
Here, we show their $z=5.55$ luminosity function in orange,
for which the normalization is scaled to $z=6$ by a factor of
$10^{-0.72\Delta z}$ \citep{2016ApJ...833..222J}, where $\Delta z=0.45$,
These values are substantially higher than those expected from extrapolation of the rest-UV based QLF by 
\citetalias{2018ApJ...869..150M} (dotted curve) down to the faint end.
In contrast, our best-fitted intrinsic QLF (dashed curve) is broadly consistent with the faint-end of the X-ray based QLF within the errors.

From another point of view, the discrepancy between the UV and X-ray QLFs at the faint end would be explained by the fact that 
only bright quasars outshining their host galaxies can be identified as point-like sources in the rest-UV quasar selection 
(\citetalias{2018ApJ...869..150M}; \citealt{2020MNRAS.495.2135N,2020MNRAS.494.1771A,2021MNRAS.502.2757O,2021MNRAS.502..662B,2022AJ....164..114K}).
However, at the faint end of $\Muv\gtrsim -24$ mag, extended sources such as the brightest galaxies at $z\sim 6$ 
dominate over quasars in number at the same UV magnitude \citep{2022ApJS..259...20H}.
Thus, the current observations might miss quasars embedded in extended galaxy populations. 
We leave this issue as an important caveat in high-$z$ quasar observations.

\vspace{2mm}
\section{Cosmological evolution of BH mass and quasar luminosity function}\label{sec:cosm}


In previous sections, we calibrate the model for early SMBH evolution based on the observational constraints of the BHMF and QLF at $z\sim6$.
Next, we apply our model to predict the statistical properties of quasars at higher redshifts ($z\gtrsim$ 6)
and provide prediction for future explorations.

\subsection{Black Hole Mass Functions}

In the left panel of Fig.~\ref{fig:BHMF_rhoz}, we show the predicted BHMFs at five different redshifts of $z=6$--$10$
for $\fseed=0.1$ (solid) and $0.01$ (dashed).
As shown in Fig.~\ref{fig:seedmf}, the mass distribution of seed BHs establishes up to $M_\bullet \lesssim 2\times10^5~\Msun$ by $z\gtrsim 17$.
Subsequently, the BHMF develops toward higher mass ranges by the growth of seeds with a fraction of $\fseed$.
In fact, the case with $\fseed=0.1$ yields the number density of those growing seeds ten times higher than that with $\fseed =0.01$,
making the difference in the BH abundance at $M_\bullet \sim 10^{5-7}~\Msun$.
Thus, the information on the seeding fraction still remains in this low mass regime.
Toward lower redshifts, heavier BHs with $\gtrsim 10^7~\Msun$ emerge owing to efficient growth
and thus BHMF in the high-mass tail increases.
Since our model is calibrated against the $z\simeq 6$ BHMF of \citetalias{2010AJ....140..546W},
the two cases with different seeding fractions yield similar shapes of mass distribution at $M_\bullet \simeq 10^8-3\times 10^9~\Msun$ (see Fig.~\ref{fig:fitmf}).
The BHMF at $M_\bullet \gtrsim 10^7~\Msun$, which is potentially accessible by high-$z$ quasar observations,
can be characterized with a double power-law function, 
\begin{equation}
\Phi_{M_\bullet}=\frac{\Phi_{M_\bullet}^\ast}{(M_\bullet/M_{\bullet}^\ast)^{-(\hat \alpha+1)} + (M_\bullet/M_{\bullet}^\ast)^{-(\hat \beta+1)}}.
\end{equation}
This function shape is also used to characterize the BHMF of lower-$z$ quasar populations
\citep[e.g.,][]{2013ApJ...764...45K,2015MNRAS.447.2085S}.
Here, $\Phi_{M_\bullet}^\ast$ (in units of Gpc$^{-3}$ dex$^{-1}$) is the overall normalization, $M_\bullet^\ast$ is the characteristic BH mass,
and $\hat \alpha$ and $\hat \beta$ are the low and high-mass end slopes, respectively.
In Table~\ref{tab:BHMF_fit}, we summarize those parameters at $z=6$--$10$ for the best-fit model in the case of $\fseed=0.01$.
%

\begin{deluxetable}{ccccccc}
\tablecolumns{1}
\renewcommand\thetable{2} 
\tablewidth{0pt}
\tablecaption{Parametric Black Hole Mass Function \label{tab:BHMF_fit}}
\tablehead{
  \colhead{Redshift} & \colhead{$\Phi_{M_\bullet}^\ast$} & \colhead{$M_\bullet^\ast$} & \colhead{$\hat \alpha$} & \colhead{$ \hat \beta$}& \\
   & \colhead{(Gpc$^{-3}$ dex$^{-1}$)} & \colhead{($10^7~\Msun$)} & & &
}
\startdata
$z\sim 6$  & $1310$ & $6.13$ & -1.41 & -2.58    & \\
$z\sim 7$  & $166$   & $8.12$ & -1.81 & -2.88    & \\
$z\sim 8$  & $23.2$  & $9.98$ & -2.07 & -3.12    & \\
$z\sim 9$  & $5.61$  & $9.44$ & -2.24 & -3.29    & \\
$z\sim10$  & $2.13$ & $7.68$  & -2.35 & -3.41    & \\
\enddata
\tablecomments{The fitting parameters of the BHMF with 
a double power-law function for the case with $\fseed=0.01$.
The mass range used for fitting is $10^6~\Msun \leq M_\bullet \leq 10^{10}~\Msun$.
}
\end{deluxetable}

\begin{figure*}
\centering
\includegraphics[width=85mm]{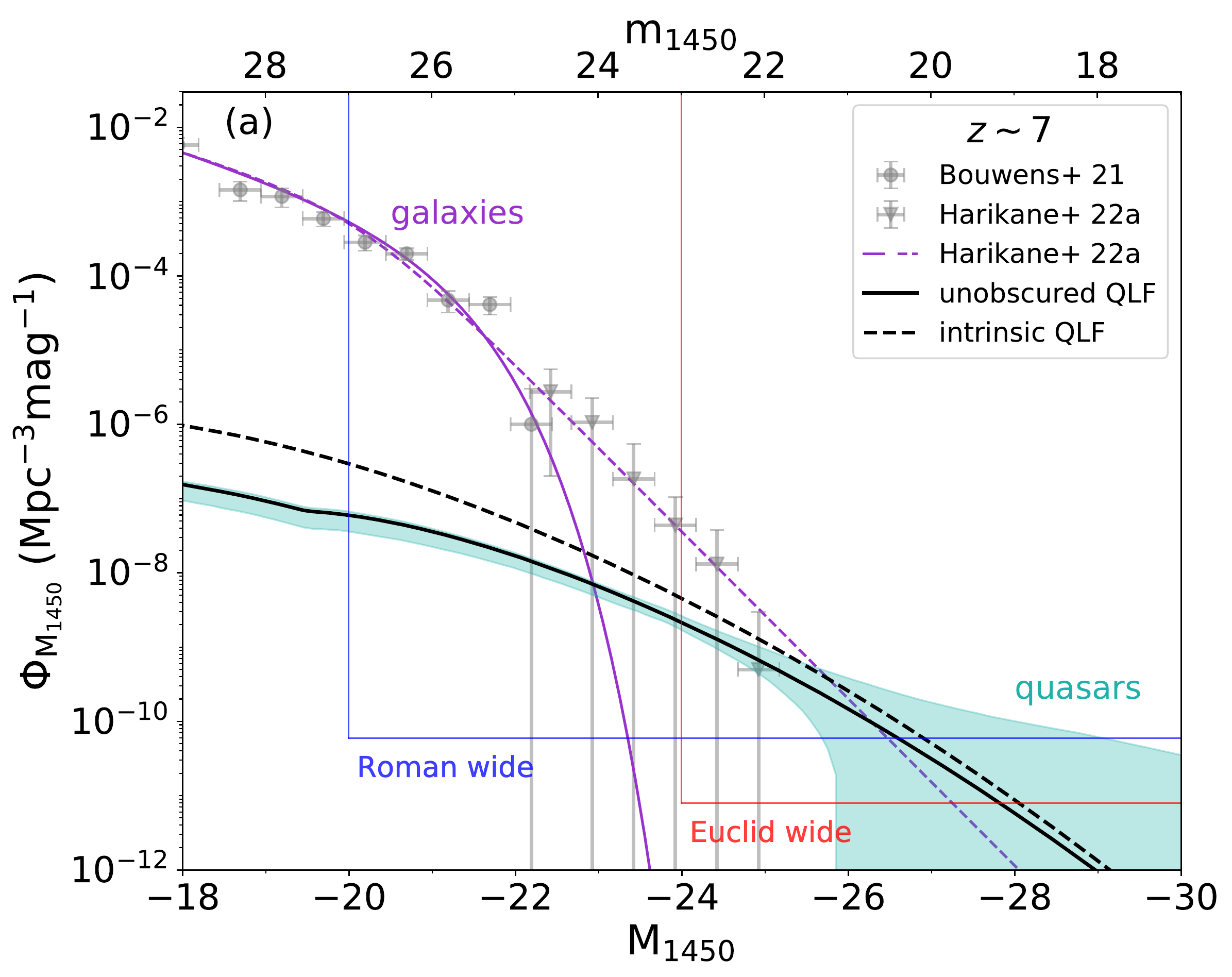}\hspace{2mm}
\includegraphics[width=85mm]{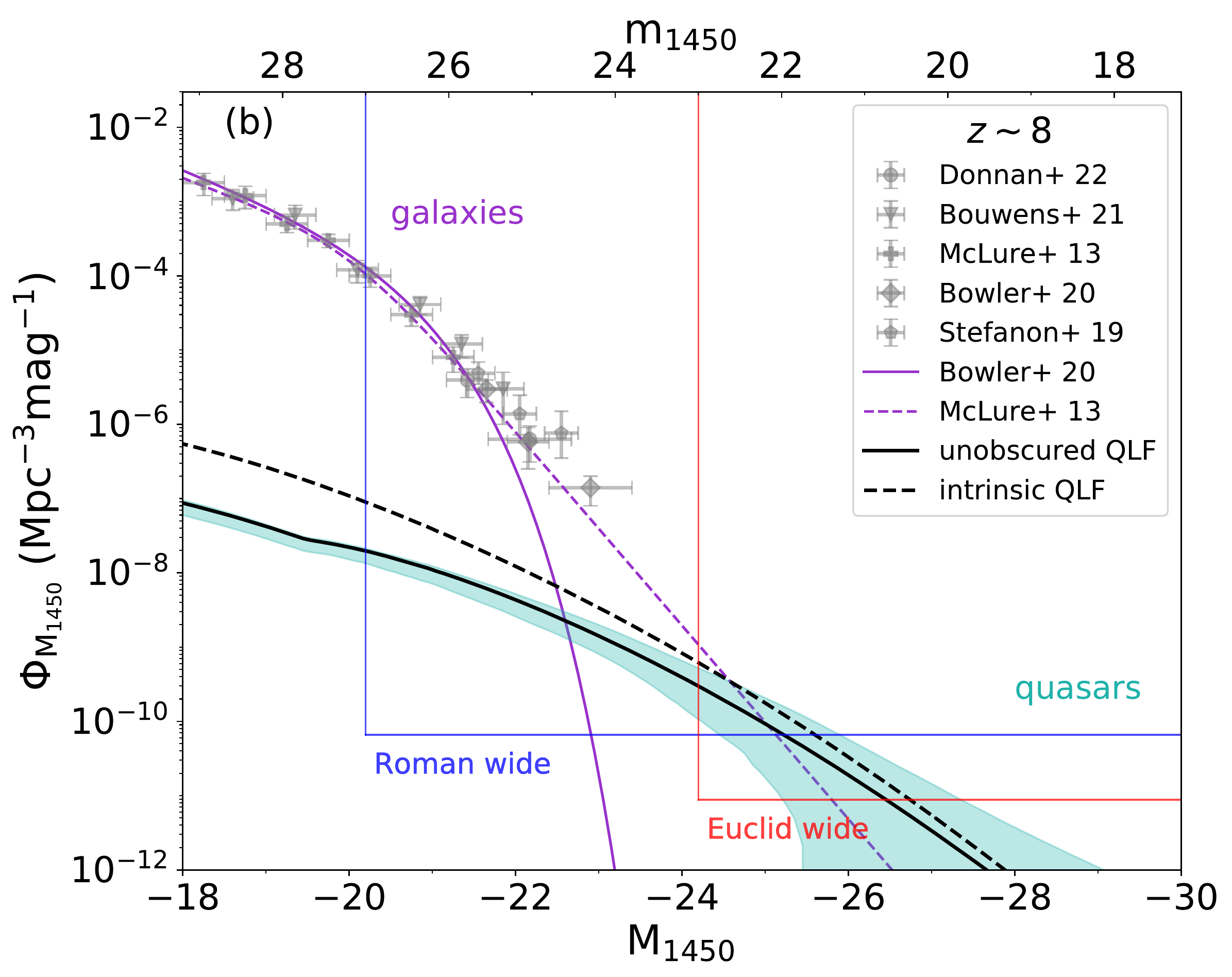}\\\vspace{5mm}
\includegraphics[width=85mm]{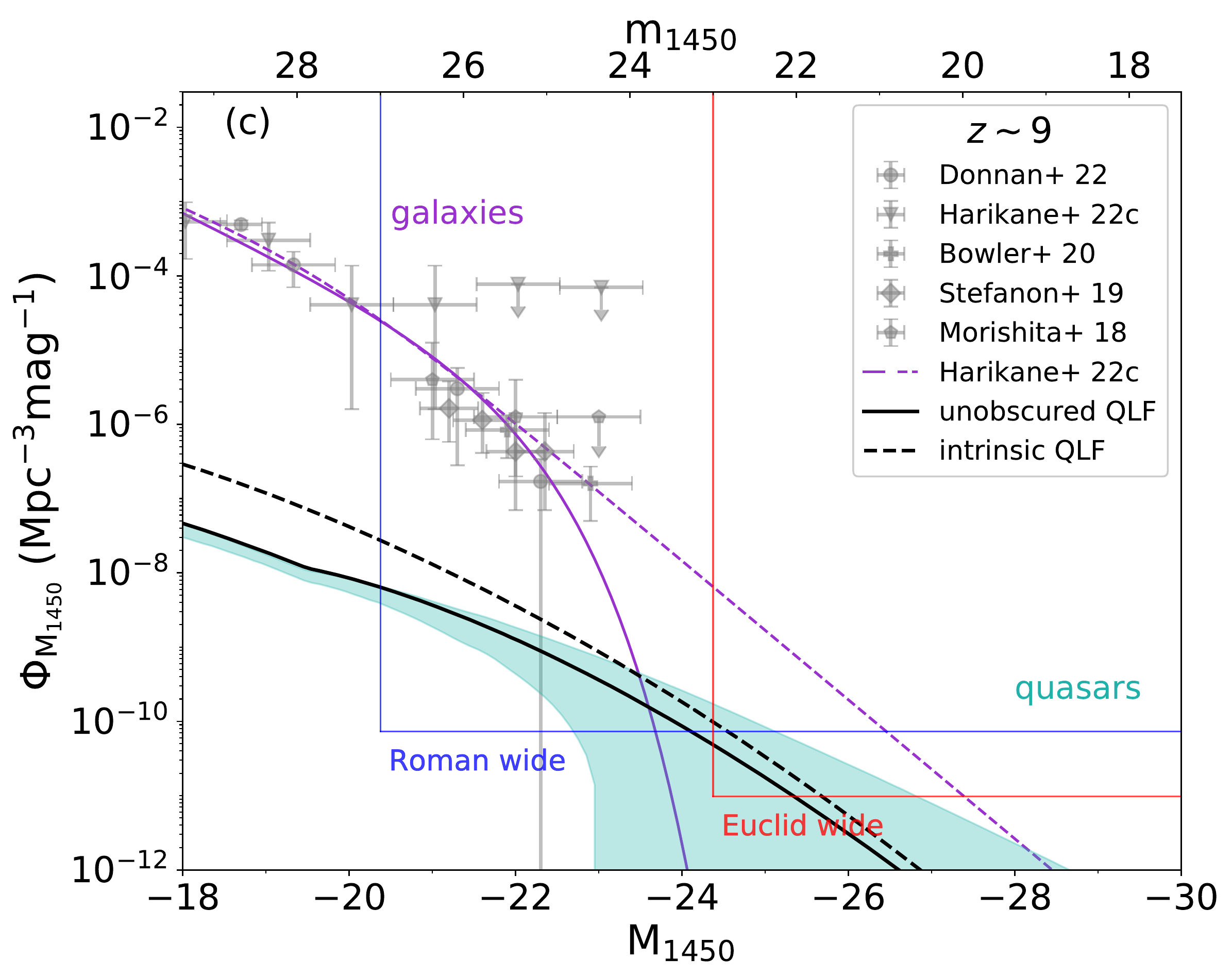}\hspace{2mm}
\includegraphics[width=85mm]{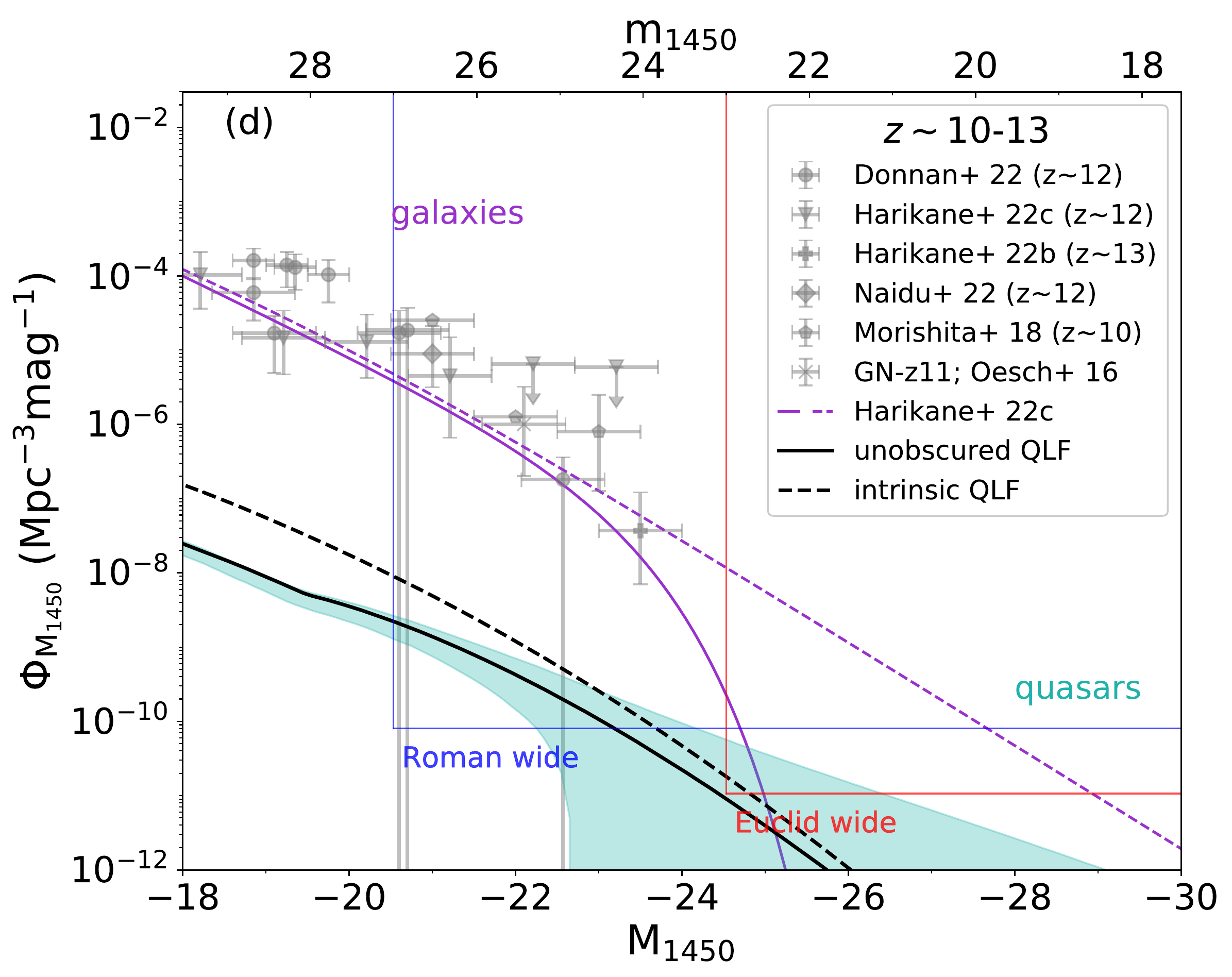}
\caption{
The predicted quasar luminosity functions at (a) $z\sim 7$, (b) $z\sim 8$, (c) $z\sim 9$, and (d) $z\sim 10-13$.
The black solid and dashed curves are the unobscured and intrinsic QLF, respectively, and the shaded region presents the $1\sigma$ statistical error.
The expected depths and volumes of future surveys with Wide Euclid (blue; $m_{1450}= 23$ mag) and Wide RST (red; $m_{1450}= 27$ mag) are overlaid in each panel (see Table~\ref{tab:N_detect}).
For comparison, the galaxy UV luminosity functions are plotted with the data points (grey) and the fitting functions in purple with a Schechter (solid) and double power-law (dashed) shape.
The luminosity function data are taken from \citet{2013MNRAS.432.2696M}, \citet{2016ApJ...819..129O}, \citet{2018ApJ...867..150M},
\citet{2019ApJ...883...99S}, \citet{2020MNRAS.493.2059B}, \citet{2021AJ....162...47B},
\citet{2022ApJS..259...20H,Harikane_2022b,Harikane_2022c}, \citet{2023MNRAS.518.6011D}, \citet{2022ApJ...940L..14N}.
The fitting forms of the galaxy luminosity functions are given by \citet{2022ApJS..259...20H} at $z\sim 7$, 
\citet{2020MNRAS.493.2059B} and \citet{2013MNRAS.432.2696M} at $z\sim 8$, and \citet{Harikane_2022c} at $9\lesssim z\lesssim 13$.}
\label{fig:LFs}
\vspace{5mm}
\end{figure*}

In the right panel of Fig.~\ref{fig:BHMF_rhoz}, we present the cumulative BH mass density $\rho_\bullet(z)$ in a comoving volume
for $\fseed=0.1$ (solid) and $0.01$ (dashed),
which is equivalent to the integration of the BHMF:
\begin{equation}
 \rho_\bullet(z)=\int_{\mathcal{I}} \Phi_{\Mbh} (z) \Mbh ~\D \log \Mbh.
\end{equation}
Here, we consider three mass ranges: $M_\bullet \leq 10^6~\Msun$ (blue), $10^6~\Msun < M_\bullet < 10^8~\Msun$ (orange),
and $M_\bullet \geq 10^8~\Msun$ (green).
Overall, the mass density of the heaviest BHs quickly grows toward lower redshifts,
while for the lightest BHs the mass density barely evolves.
At $z\lesssim 7-8$, heavier BHs with $\gg 10^6~\Msun$ begin to dominate the mass budget of the entire BH population,
promoting the so-called downsizing or anti-hierarchical evolution of massive BHs 
in terms of mass occupation \citep[e.g.,][]{2014ApJ...786..104U}.
We note that our model considers a BH population formed in the overdense region,
and neglects the contribution of BH seeds from ordinary PopIII remnants in lower-mass halos ($\Mh<10^{11}~\Msun$).
The latter population is typically less massive ($M_\star \lesssim 10^3~\Msun$) but more abundant \citep[e.g.,][]{2015MNRAS.448..568H,2023MNRAS.518.1601T}.
Therefore, the BHMF at the low-mass end in our model underestimates the abundance while
the high-mass end at $M_\bullet \gtrsim 10^7~\Msun$ matches the observed distribution well.

For comparison, we overlay in the right panel of Fig.~\ref{fig:BHMF_rhoz} the BH mass density evolution at $10^6~\Msun<M_{\bullet}<10^8~\Msun$
obtained from the cosmological simulations in \citet{2022MNRAS.513..670N} (dashed-dotted curve).
At $z\lesssim 7$, our results are in overall good agreement with those simulation results.
However, our models predict a larger number of $10^{6-8}~\Msun$ BHs emerging at $z\sim 8-10$,
owing to our model prescription that allows BH seed formation earlier on and (modest) super-Eddington episodes.
Furthermore, the unresolved fraction of the cosmic X-ray background at $z\gtrsim 5$ constrains
the global BH accretion history \citep{2012A&A...545L...6S,2013ApJ...778..130T}.
Upper bounds on the accreted-mass density (black arrows) prohibit overproduction of low mass BHs at cosmic dawn.
The mass density of BHs formed in the biased, overdense region of the universe 
is consistent with the constraint, unlike models that require a large number of stellar-mass BH remnants \citep[see also][]{2009ApJ...696.1798T}.

\begin{deluxetable*}{ccccccccccc}
\tablecolumns{1}
\renewcommand\thetable{3} 
\tablewidth{0pt}
\tablecaption{Parametric Quasar Luminosity Function \label{tab:QLF_fit}}
\tablehead{
  & & \multicolumn{4}{c}{Unobscured QLF} && \multicolumn{4}{c}{Intrinsic QLF}  \\
  \cline{3-6} \cline{8-11}
  \colhead{Redshift} && \colhead{$\Phi_{\Muv}^\ast$} & \colhead{$\Muv^\ast$} & \colhead{$\tilde \alpha$} & \colhead{$ \tilde \beta$}& &
  \colhead{$\Phi_{\Muv}^\ast$} & \colhead{$\Muv^\ast$} & \colhead{$\tilde \alpha$} & \colhead{$\tilde \beta$}\\
  & & \colhead{(Gpc$^{-3}$ mag$^{-1}$)} & \colhead{(mag)} & & & &
  \colhead{(Gpc$^{-3}$ mag$^{-1}$)} & \colhead{(mag)} & &
}
\startdata
$z\sim 6$  && 46.5 & -23.62 & -1.33 & -2.57    && 89.2 & -23.73 & -1.59 & -2.67 \\
$z\sim 7$  && 6.98 & -23.70 & -1.61 & -2.82   && 12.4 & -23.85 & -1.87 & -2.92 \\
$z\sim 8$  && 1.18 & -23.80 & -1.83 & -3.02   && 1.91 & -23.98 & -2.08 & -3.12 \\
$z\sim 9$  && 0.268 & -23.80 & -1.98 & -3.18  && 0.429& -23.96 & -2.23 & -3.28 \\
$z\sim 10$ && 0.0841 & -23.68 & -2.01 & -3.30 && 0.144 & -23.82 & -2.35 & -3.40 \\
\enddata
\tablecomments{The fitting parameters of the unobscured and intrinsic QLFs with 
a double power-law function for the case with $\fseed=0.01$.
The fitted magnitude range is $-30~{\rm mag}  \leq \Muv \leq -18~{\rm mag}$.
}
\end{deluxetable*}

\begin{deluxetable*}{cccccccccc}
\tablecolumns{1}
\renewcommand\thetable{4} 
\tablewidth{0pt} 
\tablecaption{Expected Number of Quasars Discovered in Future NIR Surveys \label{tab:N_detect}} 
\tablehead{
\colhead{Survey} & \colhead{Reference}  & \colhead{ Area (\rm{deg}$^2$)} & \colhead{Selection} &  \colhead{$N(z\sim 6)$} & \colhead{$N(z\sim 7)$}& \colhead{$N(z\sim 8)$}&  \colhead{$N(z\sim 9)$}&  \colhead{$N(z\sim 10)$}
}
\startdata
Euclid      & This work             & 15000    & $m_{1450}\leq 24$   & $6654^{+95}_{-2282}$  &  $708^{+213}_{-237}$ &   $93^{+66}_{-56}$ & $14^{+32}_{-14}$ & --  \\
            & This work             &          & $m_{1450}\leq 23$   & $2480^{+188}_{-614}$  &  $207^{+144}_{-114}$ &   $23^{+27}_{-19}$ & $3^{+12}_{-3}$ & --  \\
            & Euclid Collab. (2019) &          & simulation  & --                             &  $204 (117)$                &   $16 (6)$             & $7 (2)$ & --  \\
\hline
RST         & This work       &  2000  & $m_{1450}\leq 27$  & $6759_{-3185}^{+15} $ &  $1514_{-558}^{+169}$ &  $349_{-131}^{53}$ & $85_{-48}^{+23}$ & $25^{+12}_{-15}$  \\
\enddata
\tablecomments{
Prediction of high-$z$ quasar numbers yielded by the Euclid and RST wide-area surveys from $z\sim 6$ to $10$ with $\Delta z=1$.
The QLFs we utilize to estimate the detection numbers are from the BH growth model with $\fseed=0.01$.
The integration of the QLF is conducted assuming 100\% selection completeness for quasar populations brighter than the photometric depths,
which are adopted as the 5$\sigma$ point source detection limits of the $YJH$ bands in each survey.
To mimic realistic quasar identification capability, we also show the detection numbers with $m_{1450}\leq 23$ mag in comparison with
\citet{2019Barnett}.
The latter prediction is migrated from the fourth column of their Table~3 for three redshift ranges: $7.0<z<7.5$, $8.0<z<8.5$, and $8.5<z<9.0$,
with trials on different extrapolations of the quasar number density at higher redshifts ($\propto 10^{k_{\rm L}z}$); $k_{\rm L}=-0.72$ and $-0.92$.
The prediction of no detection or no references in the literature are denoted by the symbol ``--''.
}
\end{deluxetable*}

\subsection{Quasar Luminosity Functions}
\label{sec:qlfglf}

Fig.~\ref{fig:LFs} presents our prediction of QLFs at $z\gtrsim 6$ taken from the BH growth model with $\fseed=0.01$.
We show QLFs at different epochs; (a) $z\sim 7$, (b) $z\sim 8$, (c) $z\sim 9$, and (d) $z\sim 10$--$13$.
The black solid and dashed curves of each panel show the QLF for the unobscured population 
and intrinsic population (including unobscured and obscured quasars), respectively.
The obscured fraction derived in an X-ray quasar sample up to $z\sim 5$ by \citet{2014ApJ...786..104U} is applied to the conversion between the two populations,
while we note that a large uncertainty exists at $\Muv \gtrsim-20$ mag.
Similar to the BHMF, for all the redshift ranges, the shape of the QLF is well approximated with a double power-law function,
\begin{equation}
\Phi_{\Muv} = \frac{\Phi_{\Muv}^\ast}
{10^{0.4(\tilde \alpha+1)\Delta \Muv} + 10^{0.4(\tilde \beta+1)\Delta \Muv}},
\end{equation}
where $\Phi_{\Muv}^\ast$ (in units of Gpc$^{-3}$ mag$^{-1}$) is the overall normalization, $\Muv^\ast$ is the characteristic magnitude,
$\Delta \Muv \equiv \Muv - \Muv^\ast$, and $\tilde \alpha$ and $\tilde \beta$ are the faint and bright-end slopes, respectively.
In Table~\ref{tab:QLF_fit}, we summarize those parameters at $z=6$--$10$ for the best-fit models.
From higher to lower redshifts, the QLF increases and the bright-end slope gets flatter.

We overlay the expected depths and survey volumes of upcoming near-infrared surveys by
Euclid (\citealt{2011arXiv1110.3193L}; \citealt{2019Barnett}) and RST \citep{2019arXiv190205569A}.
We here consider only their wide-area survey layers because narrow-field deep surveys are not optimized for hunting high-$z$ quasars \citep[but see][]{2023ApJ...942L..17O}.
Table~\ref{tab:N_detect} summarizes the expected number yields of quasars from those surveys.
We first integrate the QLFs at the apparent magnitude $m_{1450}\leq$ 24 mag and 27 mag for Euclid and RST, respectively.
These depths correspond to the 5$\sigma$ point source detection limits in the $YJH$ bands of the two facilities.
Euclid will cover $\simeq 100$--$700$ quasars at $z\sim 7$--$8$ down to the limiting magnitude ($\Muv \sim -24$ mag),
and can reach up to $z\sim 9$.
The RST will explore fainter ($\Muv \gtrsim -24$ mag) and more abundant quasar populations up to $z\sim 10$.
However, the estimated values above only give upper bounds of the number of detectable high-$z$ quasars,
because selection completeness is not unity in identifying high-$z$ quasars.
\citet{2019Barnett} investigated the efficiency of quasar selection with Euclid images
and found that combined with ground-based optical observations,
$z\sim 7$--$9$ quasars can be identified down to $J \sim$ 23 mag
due to photometric noise and contaminants in the color selection such as Galactic brown dwarfs and low-$z$ early-type galaxies.
Adopting $m_{1450}=23$ mag as the effective depth, 
our predicted detection numbers are in good agreement with their results but show a better match for the case
where their QLFs are extrapolated from $z\sim 6$ with $10^{-0.72z}$ toward higher redshifts
(see the fourth column of Table~3 in \citealt{2019Barnett}).

\begin{figure*}
\centering
\includegraphics[width=130mm]{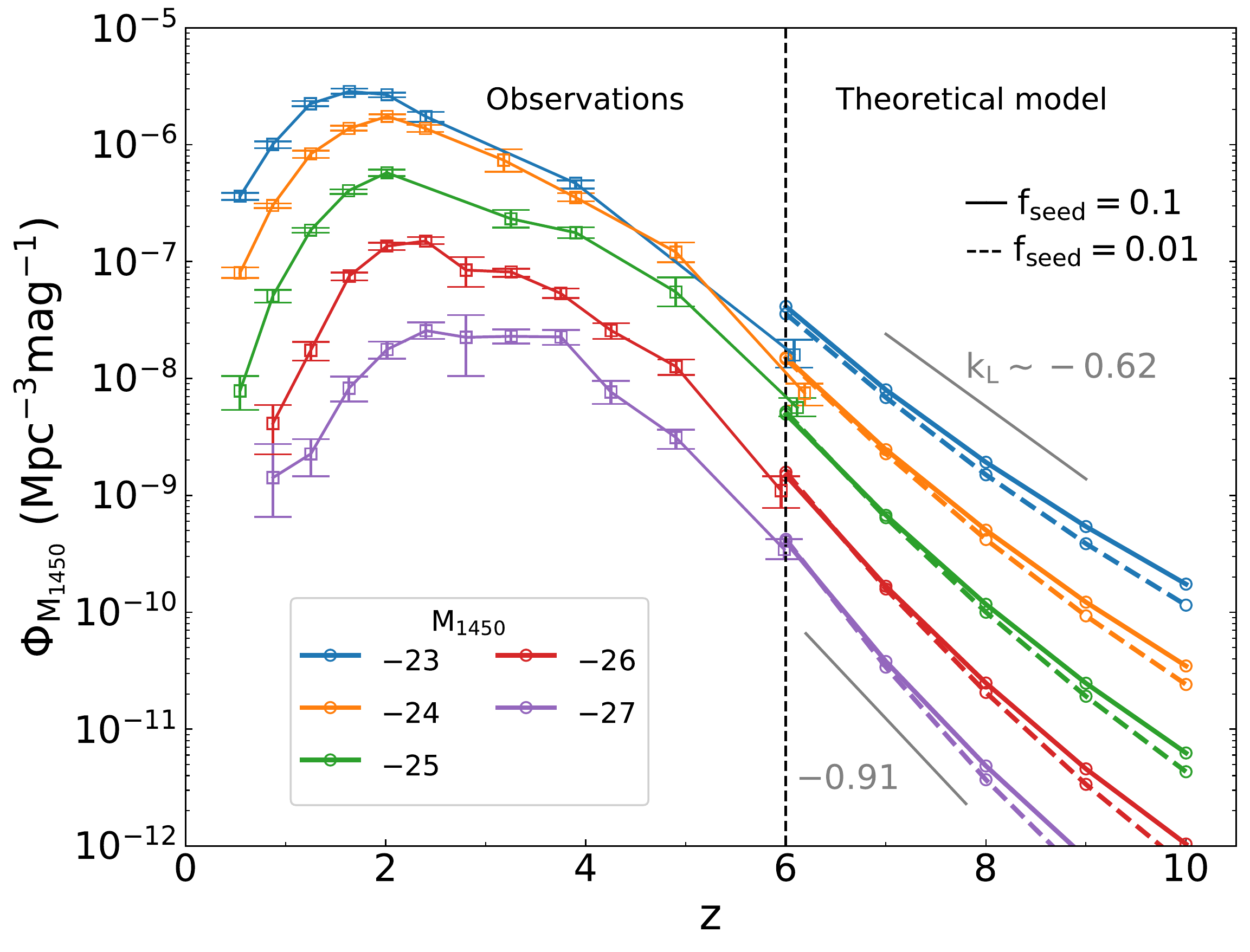}
\caption{
Redshift evolution of the QLF value at different UV magnitudes from $\Muv= -27$ mag to $-23$ mag.
The theoretical model prediction from $z=6$ to $10$ are shown for the cases of $\fseed=0.1$ (solid) and 0.01 (dashed), respectively.
At $z\geq 6$, the quasar number is characterized with $\propto 10^{k_{\rm L}z}$,
where the index is $k_\mathrm{L} \simeq -0.91$ 
and $-0.62$ for quasars at $\Muv=-27$ mag and $-23$ mag, respectively.
The observational data at lower redshifts $0\leq z \leq 6$ are compiled from the Subaru HSC results \citep{2018PASJ...70S..34A,2018ApJ...869..150M,2020ApJ...904...89N},
SDSS \citep{2006AJ....131.2766R,2013ApJ...768..105M},
the 2dFSDSS LRG and Quasar Survey (2SLAQ; \citealt{2009MNRAS.399.1755C}),
and the Spitzer Wide-area Infrared Extragalactic Legacy Survey (SWIRE; \citealt{2008ApJ...675...49S}).
}
\label{fig:QLFmag_z}
\vspace{5mm}
\end{figure*}

Luminous quasars with $\Muv \lesssim -26$ mag are expected to be marginally detected by Euclid and the RST up to $z\sim 7$ and 8, respectively,
with $\sim 1$ detection number predicted by the integration over the model QLFs.
However, \citet{2019BAAS...51c.121F} proposed that quasar detection can reach $z\sim 9$ by the two surveys,
assuming a moderate decline of the quasar number density ($\propto 10^{-0.78z}$) in QLF extrapolations from $z\sim 6$
(see more discussion in Fig.~\ref{fig:QLFmag_z}).
The differences in those predictions will be tested by upcoming high-$z$ quasar observations.

We also show the galaxy UV luminosity function in each panel of Fig.~\ref{fig:LFs}
along with the fitting function in a Schechter (solid) and double power-law (dashed) shape extended to the bright end
\citep{2013MNRAS.432.2696M, 2016ApJ...819..129O, 2018ApJ...867..150M, 2019ApJ...883...99S, 2020MNRAS.493.2059B,
2021AJ....162...47B, 2023MNRAS.518.6011D,
2022ApJS..259...20H, Harikane_2022b, Harikane_2022c, 2022ApJ...940L..14N}.
Galaxies at $\Muv\gtrsim -23$ mag dominate in number at all the redshifts and impede
the identification of relatively rare quasars among those faint objects.
Since the extrapolated galaxy luminosity function with a Schechter shape declines quickly with the luminosity increasing,
the galaxy abundance becomes comparable to that of quasars at $\Muv \sim -23$ mag ($z\sim 7$--$8$), $-24$ mag ($z\sim 9$), and $-25$ mag ($z\sim 10$--$13$).
On the other hand, if the bright end of the galaxy luminosity function is extended with the bright-end power-law slope,
the critical UV magnitude shifts to $\Muv\sim-26$, $-25$, and $\ll -30$ mag for each redshift range.
Future deep and wide surveys will unveil the relative contribution of the two populations and their roles in cosmic reionization
(see also \citetalias{2018ApJ...869..150M}, \citealt{2022NatAs...6..850J}).

In Fig.~\ref{fig:QLFmag_z}, we present the quasar number density evolution as a function of redshift.
The solid curves show the cases for $\fseed = 0.1$ and the dashed curves show the cases for $\fseed = 0.01$, respectively.
Each curve corresponds to the QLF value at a UV magnitude from $\Muv=-27$ mag to $-23$ mag.
Our theoretical predictions are shown at $z\gtrsim 6$ (thick curves),
while observational results compiled in \cite{2020ApJ...904...89N} are shown at lower redshifts (thin curves).
The evolutionary trend can be fitted as $\propto 10^{k_{\rm L}z}$ at high redshifts.
The best-fit indexes are $k_{\rm L}\simeq -0.91$ and $-0.62$ for the brightest and faintest population
at $z\gtrsim 6$, respectively.
For the abundance of luminous quasars with $\Muv \lesssim -26$ mag, a rapid decay toward higher redshifts is suggested by
observational studies \citep[e.g.,][]{2001AJ....122.2833F,2013ApJ...768..105M,2016ApJ...833..222J,2019ApJ...884...30W}.
For instance, \citet{2019ApJ...884...30W} proposed a decay index of $k_{\rm L}=-0.78\pm{0.18}$ for those bright quasars at $6\lesssim z \lesssim 7$.
The slopes in our model prediction of $\Muv\leq -26$ are slightly steeper ($k_\mathrm{L} \simeq -0.84$ at $\Muv=-26$),
but consistent with their finding within the $1\sigma$ error range.

\vspace{2mm}
\section{Discussion}\label{sec:discussion}
\vspace{2mm}
\subsection{Individual BH growth}\label{sec:evol}

From the BH growth model described in Section~\ref{sec:MF}, we further explore individual BH evolutionary tracks starting from their seeding phases.
We generate a sample of 10$^6$ BHs, whose formation time and mass distribution at birth follow the model described in Section~\ref{sec:seed}. 
Adopting the best-fit parameters in the case of $\fseed = 0.01$, in every time interval $\tlife=18.76$ Myr,
we assign a constant Eddington ratio generated from the Schechter-type ERDF characterized with $\lambda_0=0.87$ and $\alpha=0.20$ 
(see Section~\ref{sec:fitting_result}).
With those parameters, we grow the individual BHs until $z=6$ and study their statistical properties.

First, we estimate the duty cycle between at $6\leq z\leq 10$ defined as $f_{\rm duty} \equiv \mathcal{N}\tlife/\Delta t_{\rm H}$, 
where $\mathcal{N}$ is the frequency of super-Eddington accretion bursts with $\lambda \geq 1$ and $\Delta t_{\rm H}\simeq 450$ Myr.
For BH populations grown to $M_\bullet \geq10^7$, $10^8$ and $10^9~\Msun$ at $z=$ 6,
the average duty cycle is found to be $\langle f_{\rm duty}\rangle \simeq 0.12$, 0.15, and 0.19, respectively.
The trend of the duty cycle on the mass suggests that those massive BHs experience multiple active accretion phases to reach the SMBH regime.
On the other hand, the average duty cycle for the entire BH population is calculated as $\langle f_{\rm duty} \rangle =0.075$
adopting this ERDF (see Section~\ref{sec:fitting_result}), and is significantly lower than those for the massive BH sub-samples.
Therefore, we suggest that SMBHs observed in $z\sim 6$ quasars are a biased population
that undergoes longer active accretion episodes.

\begin{figure*}
\centering
\includegraphics[width=125mm]{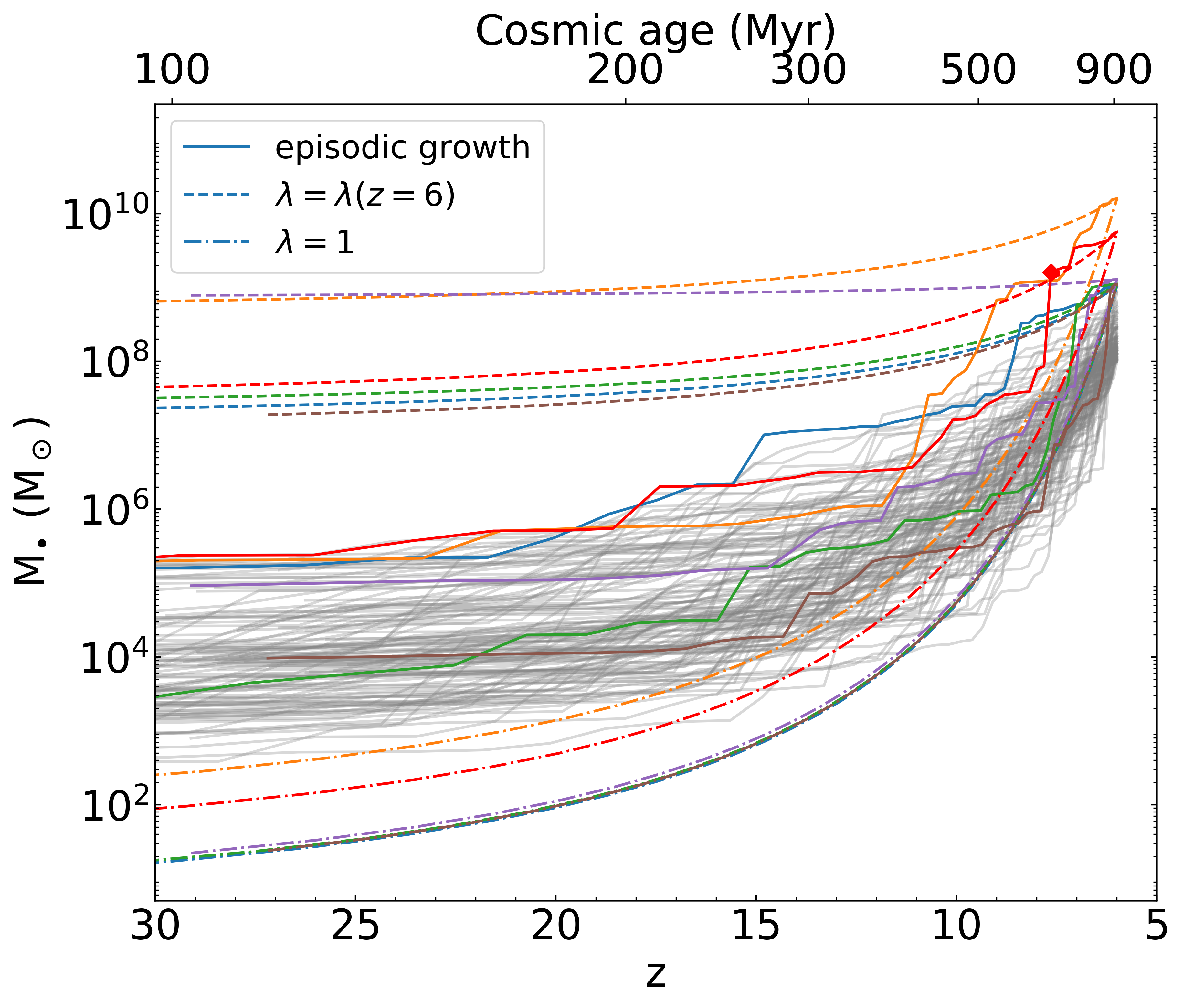}
\caption{
Evolutionary tracks of individual seed BHs with the the best-fit growth model parameters in the $\fseed=0.01$ case.
Among all the samples, we select BHs that reach $M\geq 10^8~\Msun$ at $z=6$ (grey thin curves). 
The six heaviest SMBHs ($M_\bullet \geq 10^9~\Msun$ at $z=6$) are highlighted with colored curves. 
For those BHs, we draw the assembly history assuming an exponential growth with a constant Eddington ratio: 
the $\lambda$-value seen at $z=6$ for each BH (dotted) and $\lambda =1$ (dashed).
Our best-fit model explains the existence of the most massive BHs via multiple accretion episodes with different values of $\lambda$,
and reproduces the mass and Eddington ratio for the highest-$z$ quasar (\citealt{2021ApJ...907L...1W}; red diamond).
}
\label{fig:Mevol}
\vspace{5mm}
\end{figure*}

In Fig.~\ref{fig:Mevol}, we show the BH growth tracks that end up with $M_\bullet>10^8~\Msun$ by $z=6$ (grey curves).
In this sub-sample, we find six BHs heavier than $M=10^9~\Msun$ (highlighted with colored solid curves),
all of which indicate $\lambda \lesssim 1$ at the  $z=6$ snapshot.
Tracing the BH growth curves back with the Eddington ratio observed at $z=6$ (dashed curves),
the extrapolated masses of those massive, sub-Eddington BHs at $z=6$ are as heavy as $\Mbh \gg 10^6~\Msun$ at $z\sim 30$,
which is substantially higher than the mass range for seed BHs.
%
It has been pointed out by \cite{2019ApJ...880...77O} that such simple extrapolation of the growth history for low-$\lambda$ quasars 
does not work properly (see their Figure 10).
For comparison, we show the mass growth tracks assuming continuous Eddington accretion ($\lambda=1$; dashed-dotted curves),
where the extrapolated seed mass is $\Mbh \lesssim 10^2~\Msun$ at $z\sim 30$.
In conclusion, therefore, multiple accretion episodes with different values of $\lambda$ should be taken into account 
to infer the seeding mass for the observed SMBHs as bright quasars at $z\gtrsim 6-7$,
instead of assuming one single number of the Eddington ratio.

\citet{2021ApJ...907L...1W} discovered the currently known most distant quasar at $z=7.642$,
with $M_{\bullet}\simeq 1.6\times 10^9~\Msun$ and $\lambda \simeq 0.67$ (red diamond in Fig.~\ref{fig:Mevol}).
Assuming continuous Eddington accretion, the extrapolated BH is as massive as $10^{4-6}~\Msun$ at $z= 30$. 
Our best-fit model naturally explains the existence of this most extreme BH,
taking into account the episodical accretion nature of quasars (see the red solid curve;
$M_\bullet \simeq 1.7\times 10^9~\Msun$ and $\lambda \simeq 0.6$ at $z=7.6$).
The result demonstrates that
one possible formation channel of the highest-$z$ quasar is through the combination of rare halo environments planting a seed BH 
with $\sim 2\times 10^5~\Msun$ at $z\sim 30$ and several (modest) super-Eddington accretion bursts.
The mass range of the highest-$z$ SMBHs can be achieved by previous cosmological simulations with a certain type of
sub-grid models for BH feeding and feedback \citep{2017MNRAS.467.4243D,2018MNRAS.473.4003B,2019MNRAS.488.4004L}.

Multiple episodes of BH accretion naturally yield variable quasar light curves and leave ionized gas with a complex structure.
The measurement of the ionization state of the intergalactic medium surrounding a quasar is
a powerful tool to constrain the timescale of $t_{\rm Q}$ in a luminous quasar phase
\citep{2007MNRAS.374..493B,2016ApJ...824..133K,2017ApJ...840...24E,2019ApJ...884L..19D}.
Recent observations of the line-of-sight hydrogen Ly$\alpha$ proximity zone sizes indicate
a short quasar age of $t_{\rm Q} \sim 1$ Myr on average, while the ages of some individual quasars are
inferred to be even shorter than $t_{\rm Q} \sim 0.1$ Myr \citep[e.g.,][]{2020ApJ...903...60I,2021ApJ...917...38E}.
However, we note that the line-of-sight proximity effect is only sensitive to the most recent activity of quasars.
Alternatively, the transverse proximity effect is useful to probe a longer quasar activity comparable to a Salpeter timescale \citep{2004ApJ...612..706A,2008ApJ...674..660V,2019ApJ...882..165S}.
Recently, \citet{2022MNRAS.516..582K} proposed a new technique to photometrically map the quasar light echoes using Ly$\alpha$ forest tomography
and discussed an efficient observational strategy.
With our MCMC fitting result reproducing the observed QLF and BHMF, each quasar activity is required to last for $\tau \simeq 20$ Myr on average.
Moreover, a significant fraction ($\sim 10-20\%$) of the massive BHs with $M_\bullet \gtrsim 10^8~\Msun$ in our sample experience
two successive active phases with a total active duration of $\sim 30-40$ Myr at $z\simeq 6$.
Therefore, direct measurements of the distribution of $\tau$ in future observations will test our prediction on quasar activity and
improve our understanding on the BH growth histories through luminous quasar phases.

\vspace{2mm}
\subsection{The distribution of the Eddington ratio: \\observed v.s. underlying populations}\label{sec:ldist}

Even with the current observational facilities, it is challenging to unveil the properties of the underlying quasar population
because of the limitation of the photometric depths of quasar surveys.
In this section, we examine how the unavoidable bias affects the properties of the ERDF for observed quasars,
taking our complete sample of high-$z$ BH populations in the whole luminosity range.

\begin{figure}
\centering
\includegraphics[width=80mm]{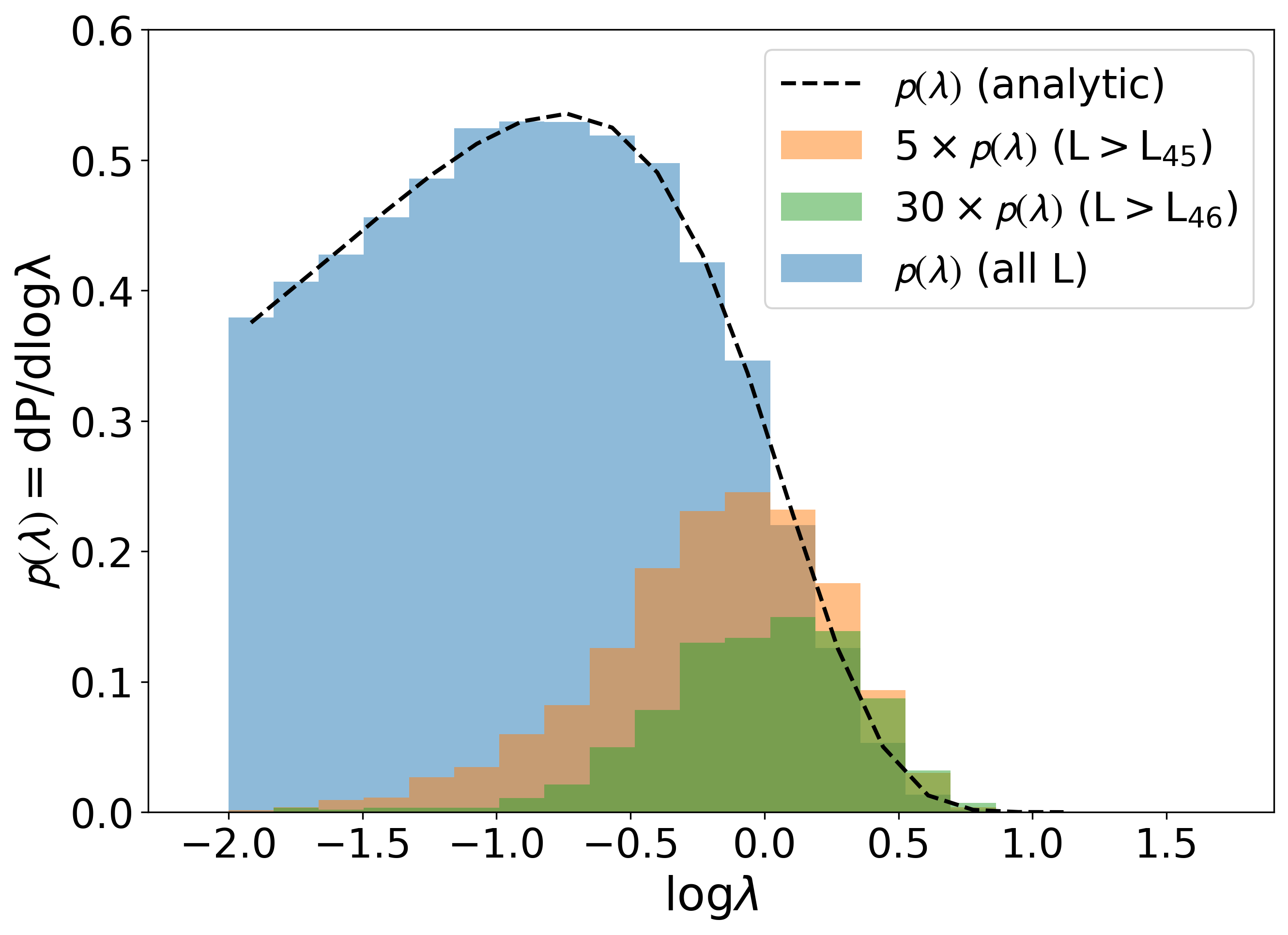}
\caption{
The intrinsic Eddington ratio distribution function for the whole BH sample (blue) and the apparent distribution function for selected samples with
a luminosity cut of $\Lbol \geq 10^{45}~\mathrm{erg~s^{-1}}$ (orange) and $\Lbol \geq 10^{46}~\mathrm{erg~s^{-1}}$ (green), respectively.
The normalization of the ERDFs with the luminosity cuts is scaled by a factor of 5 and 30, respectively, for illustration purposes.
The intrinsic ERDF follows the best-fitted Schechter function (dashed curve),
while the observed ERDF with a detection limit results in a log-normal shape where most sub-Eddington quasars are excluded.
The log-normal shape still holds with the lower luminosity cut, but a lower detection limit alleviates the selection effect against quasar populations with low Eddington ratios.
}
\label{fig:lhist}
\end{figure}

Based on the quasar sample from the best-fit model described in Section~\ref{sec:evol},
we select quasars with bolometric luminosity limits of $L_{45}=10^{45}~\mathrm{erg~s^{-1}}$ and $L_{46}=10^{46}~\mathrm{erg~s^{-1}}$.
In Fig.~\ref{fig:lhist}, we present the intrinsic ERDF for the whole BH sample (blue)
and the apparent ERDFs for selected samples with a luminosity cut of $\Lbol \geq L_{45}$ (orange),
and $\Lbol \geq L_{46}$ (green), respectively.
For illustration purposes, we scale the normalization of the ERDFs with the luminosity cuts, multiplied by a factor of 5 and 30, respectively.
The Eddington ratio for the whole sample follows a Schechter shape with a majority of inactive quasars,
while the observed Eddington ratio is biased toward higher values and the distribution results in a log-normal shape,
consistent with that of the brightest quasars at $z\sim 6$ \citep[e.g.,][]{2010AJ....140..546W,2019ApJ...873...35S,2021ApJ...923..262Y,2022ApJ...941..106F}.
The log-normal shape still holds with the lower luminosity cut, but a lower luminosity threshold allows us to unveil more hidden quasars with low Eddington ratios.
The existence of those sub-Eddington BHs at $z>6$ will be probed by the ongoing survey of the Subaru HSC and JWST
\citep{2019ApJ...880...77O,2021jwst.prop.1967O}, as well as upcoming wide-field surveys such as the Vera C. Rubin Observatory, Euclid, and RST.

\section{Summary}
\label{sec:sum}
In this paper, we propose a theoretical model for the redshift-dependent QLF and BHMF at $z\gtrsim 6$,
applying the mass distribution function of seed BHs over $10^2~\Msun \lesssim M_\bullet \lesssim 10^5~\Msun$
that originate from massive primordial stars formed in quasar host galaxies.
In our accretion model, those early BH populations are assumed to experience multiple accretion bursts,
in each of which a constant Eddington ratio is assigned following a Schechter distribution function.
We further conduct the MCMC fitting to optimize the BH growth parameters
so that the observed QLF and BHMF at $z\simeq 6$ are simultaneously reproduced.
Our major findings are summarized as follows.

\begin{itemize}
\item
The best-fitted model requires the typical duration of accretion bursts to be $\tau \simeq 20-30$ Myr,
which is a fraction of the Salpeter time ($t_{\rm S} \simeq 45$ Myr).
The ERDF with a Schechter-like shape is constrained to produce the observed BHMF and QLF at $z\simeq 6$; namely,
the characteristic Eddington ratio and power-law slope are $\lambda_0 \simeq 0.88$ and $\alpha \simeq 0.12-0.2$,
depending the fraction of seed BHs that participate in the assembly of SMBHs (see Fig.~\ref{fig:contour}).

\item
The cosmological evolution of the BHMF and QLF at $6\leq z \leq 10$ is predicted based on the model parameters
calibrated with the observational data (see Figs.~\ref{fig:BHMF_rhoz}, \ref{fig:LFs}, and \ref{fig:QLFmag_z}).
The fitting parameters of the BHMF and QLF with a double power-law function are summarized in Tables~\ref{tab:BHMF_fit} and \ref{tab:QLF_fit}.
We find that the number density of luminous quasars decays toward higher redshifts as $\propto 10^{k_{\rm L} z}$,
where $k_{\rm L}\simeq -0.91$ and $-0.62$ for the brightest and faintest population at $6<z<10$.
We also show the detection number of high-$z$ quasars expected in Euclid and RST observations (see Table~\ref{tab:N_detect}).
Those results will be tested with upcoming deep and wide-field surveys.


\item
We apply the best-fit model for BH growth to evolution of individual BHs and examine their statistical properties.
We find that the existence of SMBHs hosted in $z\gtrsim 6$ quasars can be explained by
multiple accretion episodes with variable Eddington ratios, instead of assuming continuous Eddington-limited growth (see Fig.~\ref{fig:Mevol}).
In fact, the average duty cycle of active accretion phases with $\lambda \geq 1$ is as high as $\simeq 15~\%$
for massive BHs that reach $\gtrsim 10^8~\Msun$ by $z\simeq 6$.
This result indicates that those biased BH populations undergo longer active accretion episodes to be observed as quasars.
We also find that the observed Eddington-ratio distribution function is skewed to a log-normal shape from the intrinsic Schechter function
owing to detection limits of quasar surveys (see Fig.~\ref{fig:lhist}).
The log-normal shape still holds with a lower luminosity cut of $L_{\rm bol}\gtrsim 10^{45}~{\rm erg~s}^{-1}$.
The existence of those underlying sub-Eddington BHs at $z>6$ will be probed by the ongoing survey of the Subaru HSC and JWST,
as well as upcoming wide-field surveys such as the Vera C. Rubin Observatory, Euclid, and RST.
\end{itemize}

\acknowledgments
We greatly thank the anonymous referee for valuable comments that have helped improve the quality of this paper.
We are grateful to Yuichi Harikane, Nobunari Kashikawa, Yuanqi Liu, Alessandro Lupi, Yoshiki Matsuoka, Tohru Nagao, Zhiwei Pan, and Jin Wu for constructive discussions. 
We acknowledge support from the National Natural Science Foundation of China 
(12073003, 12003003, 11721303, 11991052, 11950410493, 1215041030, and 12150410307), 
and the China Manned Space Project Nos. CMS-CSST-2021-A04 and CMS-CSST-2021-A06. 
The numerical simulations were performed with the Cray XC50 at the Center for Computational Astrophysics (CfCA) 
of the National Astronomical Observatory of Japan and with the High-performance Computing Platform of Peking University.

\appendix

\begin{figure*}
\centering
\includegraphics[width=80mm]{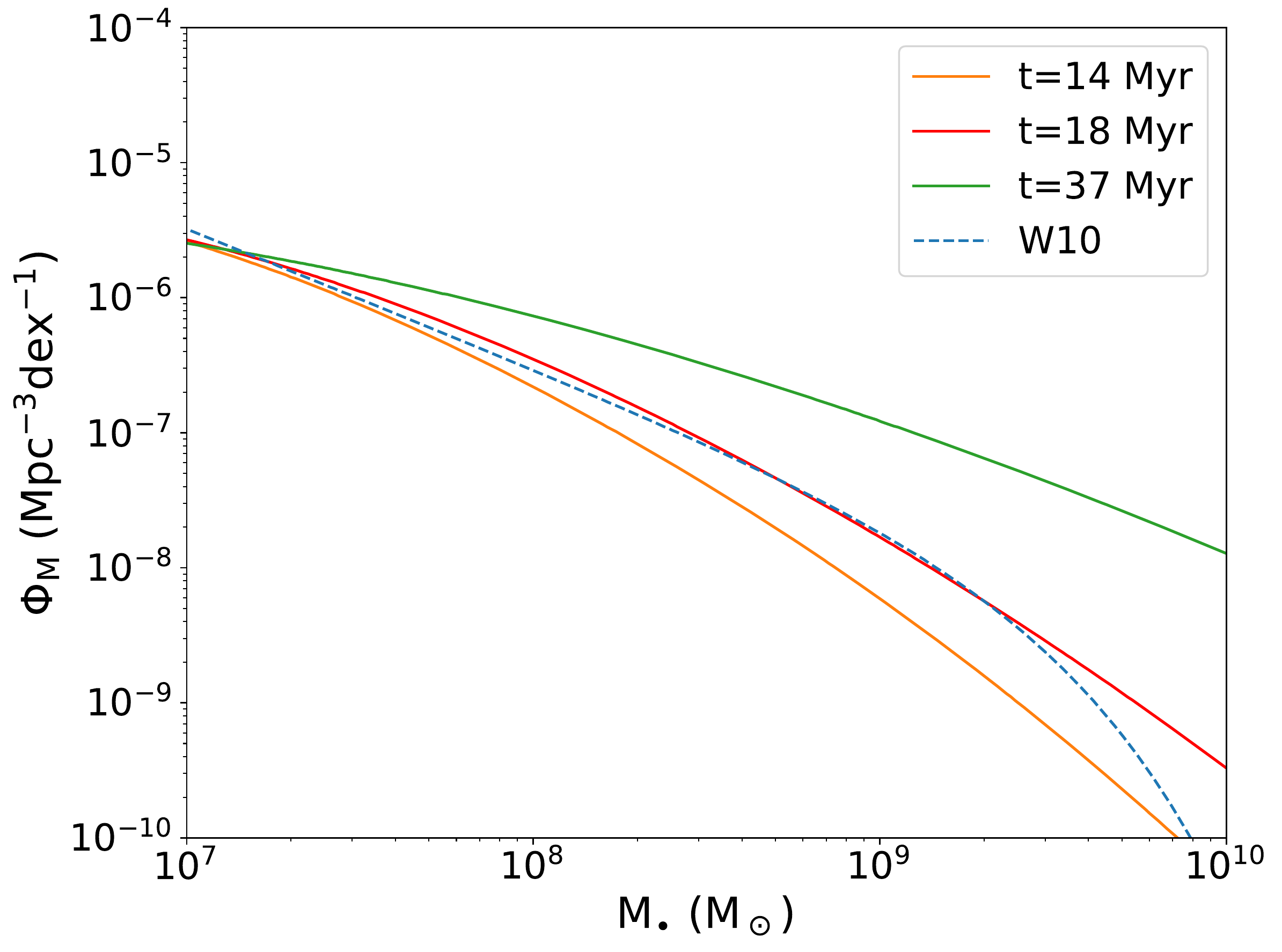}
\includegraphics[width=80mm]{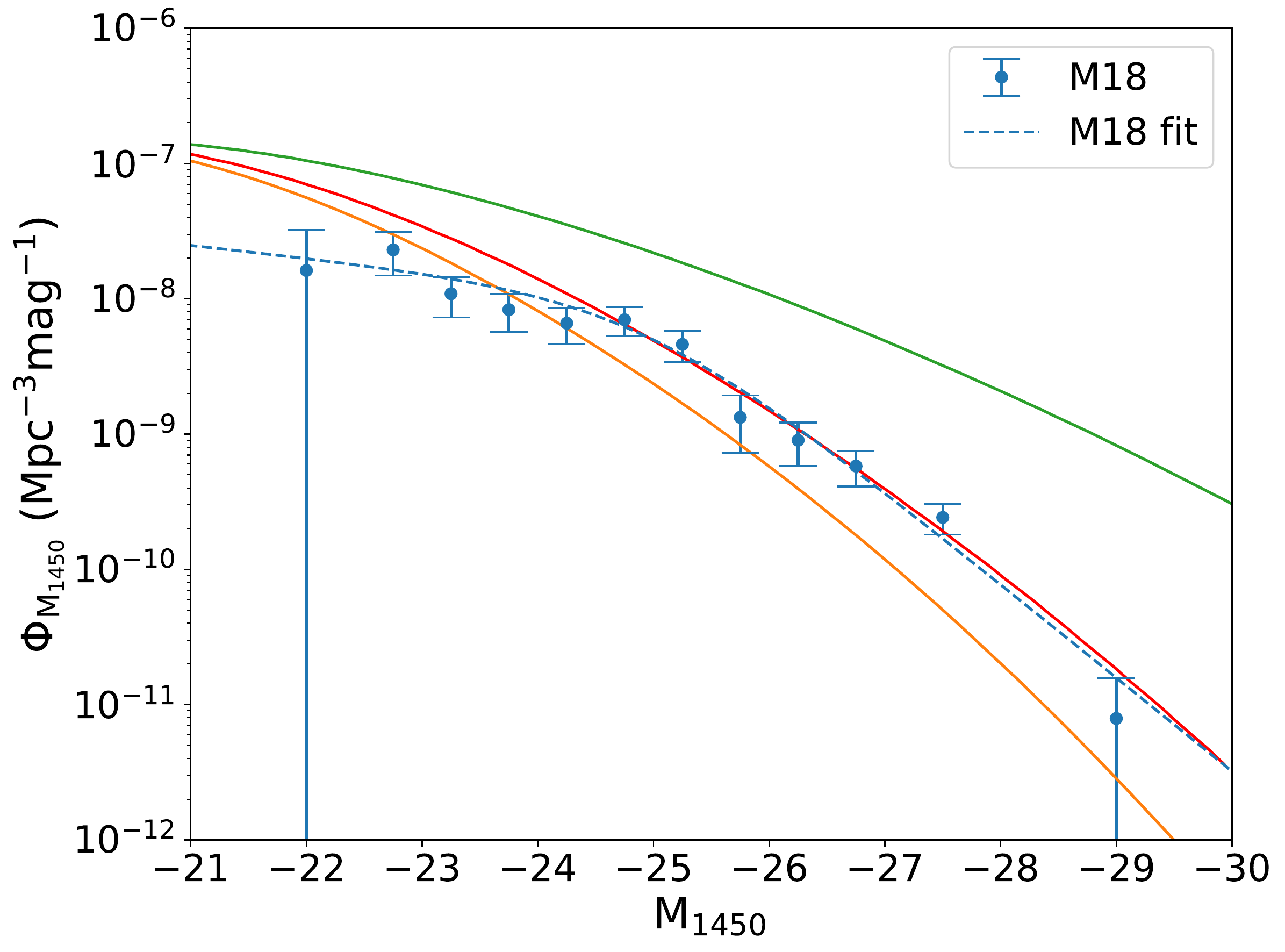}
\caption{
The BH mass function (left) and quasar luminosity function (right) at $z=$ 6 in the case of $\fseed=0.01$,
with different values of $\tlife=$ 14, 18 (the best-fit value), and 37 Myr, respectively.
Note that the other parameters are fixed to the best-fit values ($\delta=0.055$, $\lambda_0=0.87$, and $\alpha=0.20$).
For comparison, we overlay the BHMF (\citetalias{2010AJ....140..546W}) and QLF (\citetalias{2018ApJ...869..150M}) at $z\simeq 6$ in blue.
With the shorter (longer) timescales of each accretion episode $\tlife$, the number of BHs at the high mass end and bright end is under(over)-produced.
Multiple accretion bursts with each duration of $\tau \simeq 18$ Myr, which is nearly half of the $e$-folding time of the exponential BH mass growth,
reproduce the observed BHMF and QLF at $z\simeq 6$.
}
\label{fig:tau}
\vspace{4mm}
\end{figure*}

In our modeling for BH growth, we introduce a time duration $\tau$ of individual quasar activity,
where a single value of the Eddington ratio $\lambda$ generated from the ERDF is assigned to all the BHs (see Section~\ref{sec:model}).
Multiple sampling of $\lambda$ between the seeding epoch and $z=6$ reduces the probability that one BH
keeps a fast growth speed at $\lambda \gtrsim \lambda_0$ all the way to $z=6$, compared to the case with a single sampling of $\lambda$.
In what follows, we demonstrate the role of $\tlife$ in shaping the BHMF and QLF.

In the left panel of Fig.~\ref{fig:tau}, we present the BHMF at $z=6$ for three different values of $\tlife=14$, $18$, and $37$ Myr,
but keeping the other parameters to the best-fit values ($\delta=0.055$, $\lambda_0=0.87$, and $\alpha=0.20$) for the case with $\fseed=0.01$.
We note that $\tlife \simeq$ 18 Myr is the best-fit value reproducing the observed BHMF and QLF at $z=6$,
and the other two choices correspond to the $16\%$ and $84\%$ quantiles in the cumulative one-dimensional posterior distribution of $\tau$, respectively (see Fig.~\ref{fig:contour}).
With the shorter quasar activity of $\tlife=14$ Myr, the frequency of Eddington ratio assignment increases by a factor of 1.3 until $z=6$, compared to the best-fit case.
Therefore, continuous rapid growth of BHs with $\lambda \gtrsim \lambda_0$ is less likely to take place,
preventing heavy BHs from forming.
Moreover, even if a high Eddington ratio is given with a certain but low probability, the BH hardly grows in mass within such a short time $\tlife=14$ Myr,
which is $\simeq 30\%$ of the $e$-folding timescale.
On the other hand, with the longer quasar activity of $\tlife=$ 37 Myr, a substantial fraction of BHs experience high accretion rates with $\lambda\gtrsim \lambda_0$
through their assembly history.
Therefore, the high mass end of the BHMF is extended and massive BH populations with $\gg 10^8~\Msun$ are overproduced compared to the observed abundance.

In the right panel of Fig.~\ref{fig:tau}, we show the QLF at $z=6$ with the three different values of $\tau$.
Overall, the dependence of the QLF on the choice of $\tau$ shows the same trend as in the BHMF,
i.e., a longer (shorter) $\tlife$ leads to under(over)-production of the luminous quasar population.
This result demonstrates the importance of multiple accretion episodes for the earliest BHs to
reproduce the observed BHMF and QLF at $z\sim$ 6.



\bibliography{ref}{}

\begin{thebibliography}{}
\expandafter\ifx\csname natexlab\endcsname\relax\def\natexlab#1{#1}\fi
\providecommand{\url}[1]{\href{#1}{#1}}
\providecommand{\dodoi}[1]{doi:~\href{http://doi.org/#1}{\nolinkurl{#1}}}
\providecommand{\doeprint}[1]{\href{http://ascl.net/#1}{\nolinkurl{http://ascl.net/#1}}}
\providecommand{\doarXiv}[1]{\href{https://arxiv.org/abs/#1}{\nolinkurl{https://arxiv.org/abs/#1}}}

\bibitem[{{Adams} {et~al.}(2020){Adams}, {Bowler}, {Jarvis}, {H{\"a}u{\ss}ler},
  {McLure}, {Bunker}, {Dunlop}, \& {Verma}}]{2020MNRAS.494.1771A}
{Adams}, N.~J., {Bowler}, R.~A.~A., {Jarvis}, M.~J., {et~al.} 2020, \mnras,
  494, 1771, \dodoi{10.1093/mnras/staa687}

\bibitem[{{Adelberger}(2004)}]{2004ApJ...612..706A}
{Adelberger}, K.~L. 2004, \apj, 612, 706, \dodoi{10.1086/422804}

\bibitem[{{Ahn} {et~al.}(2009){Ahn}, {Shapiro}, {Iliev}, {Mellema}, \&
  {Pen}}]{2009ApJ...695.1430A}
{Ahn}, K., {Shapiro}, P.~R., {Iliev}, I.~T., {Mellema}, G., \& {Pen}, U.-L.
  2009, \apj, 695, 1430, \dodoi{10.1088/0004-637X/695/2/1430}

\bibitem[{{Aird} {et~al.}(2018){Aird}, {Coil}, \&
  {Georgakakis}}]{2018MNRAS.474.1225A}
{Aird}, J., {Coil}, A.~L., \& {Georgakakis}, A. 2018, \mnras, 474, 1225,
  \dodoi{10.1093/mnras/stx2700}

\bibitem[{{Akeson} {et~al.}(2019){Akeson}, {Armus}, {Bachelet}, {Bailey},
  {Bartusek}, {Bellini}, {Benford}, {Bennett}, {Bhattacharya}, {Bohlin},
  {Boyer}, {Bozza}, {Bryden}, {Calchi Novati}, {Carpenter}, {Casertano},
  {Choi}, {Content}, {Dayal}, {Dressler}, {Dor{\'e}}, {Fall}, {Fan}, {Fang},
  {Filippenko}, {Finkelstein}, {Foley}, {Furlanetto}, {Kalirai}, {Gaudi},
  {Gilbert}, {Girard}, {Grady}, {Greene}, {Guhathakurta}, {Heinrich},
  {Hemmati}, {Hendel}, {Henderson}, {Henning}, {Hirata}, {Ho}, {Huff},
  {Hutter}, {Jansen}, {Jha}, {Johnson}, {Jones}, {Kasdin}, {Kelly}, {Kirshner},
  {Koekemoer}, {Kruk}, {Lewis}, {Macintosh}, {Madau}, {Malhotra}, {Mandel},
  {Massara}, {Masters}, {McEnery}, {McQuinn}, {Melchior}, {Melton},
  {Mennesson}, {Peeples}, {Penny}, {Perlmutter}, {Pisani}, {Plazas}, {Poleski},
  {Postman}, {Ranc}, {Rauscher}, {Rest}, {Roberge}, {Robertson}, {Rodney},
  {Rhoads}, {Rhodes}, {Ryan}, {Sahu}, {Sand}, {Scolnic}, {Seth}, {Shvartzvald},
  {Siellez}, {Smith}, {Spergel}, {Stassun}, {Street}, {Strolger}, {Szalay},
  {Trauger}, {Troxel}, {Turnbull}, {van der Marel}, {von der Linden}, {Wang},
  {Weinberg}, {Williams}, {Windhorst}, {Wollack}, {Wu}, {Yee}, \&
  {Zimmerman}}]{2019arXiv190205569A}
{Akeson}, R., {Armus}, L., {Bachelet}, E., {et~al.} 2019, arXiv e-prints,
  arXiv:1902.05569.
\newblock \doarXiv{1902.05569}

\bibitem[{{Akiyama} {et~al.}(2018){Akiyama}, {He}, {Ikeda}, {Niida}, {Nagao},
  {Bosch}, {Coupon}, {Enoki}, {Imanishi}, {Kashikawa}, {Kawaguchi}, {Komiyama},
  {Lee}, {Matsuoka}, {Miyazaki}, {Nishizawa}, {Oguri}, {Ono}, {Onoue}, {Ouchi},
  {Schulze}, {Silverman}, {Tanaka}, {Tanaka}, {Terashima}, {Toba}, \&
  {Ueda}}]{2018PASJ...70S..34A}
{Akiyama}, M., {He}, W., {Ikeda}, H., {et~al.} 2018, \pasj, 70, S34,
  \dodoi{10.1093/pasj/psx091}

\bibitem[{{Amaro-Seoane} {et~al.}(2017){Amaro-Seoane}, {Audley}, {Babak},
  {Baker}, {Barausse}, {Bender}, {Berti}, {Binetruy}, {Born}, {Bortoluzzi},
  {Camp}, {Caprini}, {Cardoso}, {Colpi}, {Conklin}, {Cornish}, {Cutler},
  {Danzmann}, {Dolesi}, {Ferraioli}, {Ferroni}, {Fitzsimons}, {Gair}, {Gesa
  Bote}, {Giardini}, {Gibert}, {Grimani}, {Halloin}, {Heinzel}, {Hertog},
  {Hewitson}, {Holley-Bockelmann}, {Hollington}, {Hueller}, {Inchauspe},
  {Jetzer}, {Karnesis}, {Killow}, {Klein}, {Klipstein}, {Korsakova}, {Larson},
  {Livas}, {Lloro}, {Man}, {Mance}, {Martino}, {Mateos}, {McKenzie},
  {McWilliams}, {Miller}, {Mueller}, {Nardini}, {Nelemans}, {Nofrarias},
  {Petiteau}, {Pivato}, {Plagnol}, {Porter}, {Reiche}, {Robertson},
  {Robertson}, {Rossi}, {Russano}, {Schutz}, {Sesana}, {Shoemaker}, {Slutsky},
  {Sopuerta}, {Sumner}, {Tamanini}, {Thorpe}, {Troebs}, {Vallisneri},
  {Vecchio}, {Vetrugno}, {Vitale}, {Volonteri}, {Wanner}, {Ward}, {Wass},
  {Weber}, {Ziemer}, \& {Zweifel}}]{LISA_2017}
{Amaro-Seoane}, P., {Audley}, H., {Babak}, S., {et~al.} 2017, arXiv e-prints,
  arXiv:1702.00786.
\newblock \doarXiv{1702.00786}

\bibitem[{{Ba{\~n}ados} {et~al.}(2018){Ba{\~n}ados}, {Venemans},
  {Mazzucchelli}, {Farina}, {Walter}, {Wang}, {Decarli}, {Stern}, {Fan},
  {Davies}, {Hennawi}, {Simcoe}, {Turner}, {Rix}, {Yang}, {Kelson}, {Rudie}, \&
  {Winters}}]{2018Natur.553..473B}
{Ba{\~n}ados}, E., {Venemans}, B.~P., {Mazzucchelli}, C., {et~al.} 2018, \nat,
  553, 473, \dodoi{10.1038/nature25180}

\bibitem[{{Barai} {et~al.}(2018){Barai}, {Gallerani}, {Pallottini}, {Ferrara},
  {Marconi}, {Cicone}, {Maiolino}, \& {Carniani}}]{2018MNRAS.473.4003B}
{Barai}, P., {Gallerani}, S., {Pallottini}, A., {et~al.} 2018, \mnras, 473,
  4003, \dodoi{10.1093/mnras/stx2563}

\bibitem[{{Becerra} {et~al.}(2015){Becerra}, {Greif}, {Springel}, \&
  {Hernquist}}]{2015MNRAS.446.2380B}
{Becerra}, F., {Greif}, T.~H., {Springel}, V., \& {Hernquist}, L.~E. 2015,
  \mnras, 446, 2380, \dodoi{10.1093/mnras/stu2284}

\bibitem[{{Begelman} {et~al.}(2006){Begelman}, {Volonteri}, \&
  {Rees}}]{2006MNRAS.370..289B}
{Begelman}, M.~C., {Volonteri}, M., \& {Rees}, M.~J. 2006, \mnras, 370, 289,
  \dodoi{10.1111/j.1365-2966.2006.10467.x}

\bibitem[{{Bhowmick} {et~al.}(2022){Bhowmick}, {Blecha}, {Ni}, {Di Matteo},
  {Torrey}, {Kelley}, {Vogelsberger}, {Weinberger}, \&
  {Hernquist}}]{2022MNRAS.516..138B}
{Bhowmick}, A.~K., {Blecha}, L., {Ni}, Y., {et~al.} 2022, \mnras, 516, 138,
  \dodoi{10.1093/mnras/stac2238}

\bibitem[{{Bolton} \& {Haehnelt}(2007)}]{2007MNRAS.374..493B}
{Bolton}, J.~S., \& {Haehnelt}, M.~G. 2007, \mnras, 374, 493,
  \dodoi{10.1111/j.1365-2966.2006.11176.x}

\bibitem[{{Bouwens} {et~al.}(2021){Bouwens}, {Oesch}, {Stefanon},
  {Illingworth}, {Labb{\'e}}, {Reddy}, {Atek}, {Montes}, {Naidu},
  {Nanayakkara}, {Nelson}, \& {Wilkins}}]{2021AJ....162...47B}
{Bouwens}, R.~J., {Oesch}, P.~A., {Stefanon}, M., {et~al.} 2021, \aj, 162, 47,
  \dodoi{10.3847/1538-3881/abf83e}

\bibitem[{{Bowler} {et~al.}(2021){Bowler}, {Adams}, {Jarvis}, \&
  {H{\"a}u{\ss}ler}}]{2021MNRAS.502..662B}
{Bowler}, R.~A.~A., {Adams}, N.~J., {Jarvis}, M.~J., \& {H{\"a}u{\ss}ler}, B.
  2021, \mnras, 502, 662, \dodoi{10.1093/mnras/stab038}

\bibitem[{{Bowler} {et~al.}(2020){Bowler}, {Jarvis}, {Dunlop}, {McLure},
  {McLeod}, {Adams}, {Milvang-Jensen}, \& {McCracken}}]{2020MNRAS.493.2059B}
{Bowler}, R.~A.~A., {Jarvis}, M.~J., {Dunlop}, J.~S., {et~al.} 2020, \mnras,
  493, 2059, \dodoi{10.1093/mnras/staa313}

\bibitem[{{Bromm} \& {Loeb}(2003)}]{2003Natur.425..812B}
{Bromm}, V., \& {Loeb}, A. 2003, \nat, 425, 812, \dodoi{10.1038/nature02071}

\bibitem[{{Cao}(2010)}]{2010ApJ...725..388C}
{Cao}, X. 2010, \apj, 725, 388, \dodoi{10.1088/0004-637X/725/1/388}

\bibitem[{{Cao} \& {Li}(2008)}]{2008MNRAS.390..561C}
{Cao}, X., \& {Li}, F. 2008, \mnras, 390, 561,
  \dodoi{10.1111/j.1365-2966.2008.13800.x}

\bibitem[{{Cavaliere} {et~al.}(1971){Cavaliere}, {Morrison}, \&
  {Wood}}]{1971ApJ...170..223C}
{Cavaliere}, A., {Morrison}, P., \& {Wood}, K. 1971, \apj, 170, 223,
  \dodoi{10.1086/151206}

\bibitem[{{Chambers} {et~al.}(2016){Chambers}, {Magnier}, {Metcalfe},
  {Flewelling}, {Huber}, {Waters}, {Denneau}, {Draper}, {Farrow}, {Finkbeiner},
  {Holmberg}, {Koppenhoefer}, {Price}, {Rest}, {Saglia}, {Schlafly}, {Smartt},
  {Sweeney}, {Wainscoat}, {Burgett}, {Chastel}, {Grav}, {Heasley}, {Hodapp},
  {Jedicke}, {Kaiser}, {Kudritzki}, {Luppino}, {Lupton}, {Monet}, {Morgan},
  {Onaka}, {Shiao}, {Stubbs}, {Tonry}, {White}, {Ba{\~n}ados}, {Bell},
  {Bender}, {Bernard}, {Boegner}, {Boffi}, {Botticella}, {Calamida},
  {Casertano}, {Chen}, {Chen}, {Cole}, {Deacon}, {Frenk}, {Fitzsimmons},
  {Gezari}, {Gibbs}, {Goessl}, {Goggia}, {Gourgue}, {Goldman}, {Grant},
  {Grebel}, {Hambly}, {Hasinger}, {Heavens}, {Heckman}, {Henderson}, {Henning},
  {Holman}, {Hopp}, {Ip}, {Isani}, {Jackson}, {Keyes}, {Koekemoer}, {Kotak},
  {Le}, {Liska}, {Long}, {Lucey}, {Liu}, {Martin}, {Masci}, {McLean}, {Mindel},
  {Misra}, {Morganson}, {Murphy}, {Obaika}, {Narayan}, {Nieto-Santisteban},
  {Norberg}, {Peacock}, {Pier}, {Postman}, {Primak}, {Rae}, {Rai}, {Riess},
  {Riffeser}, {Rix}, {R{\"o}ser}, {Russel}, {Rutz}, {Schilbach}, {Schultz},
  {Scolnic}, {Strolger}, {Szalay}, {Seitz}, {Small}, {Smith}, {Soderblom},
  {Taylor}, {Thomson}, {Taylor}, {Thakar}, {Thiel}, {Thilker}, {Unger},
  {Urata}, {Valenti}, {Wagner}, {Walder}, {Walter}, {Watters}, {Werner},
  {Wood-Vasey}, \& {Wyse}}]{2016arXiv161205560C}
{Chambers}, K.~C., {Magnier}, E.~A., {Metcalfe}, N., {et~al.} 2016, arXiv
  e-prints, arXiv:1612.05560.
\newblock \doarXiv{1612.05560}

\bibitem[{{Chiaki} {et~al.}(2018){Chiaki}, {Susa}, \&
  {Hirano}}]{2018MNRAS.475.4378C}
{Chiaki}, G., {Susa}, H., \& {Hirano}, S. 2018, \mnras, 475, 4378,
  \dodoi{10.1093/mnras/sty040}

\bibitem[{{Chon} {et~al.}(2016){Chon}, {Hirano}, {Hosokawa}, \&
  {Yoshida}}]{2016ApJ...832..134C}
{Chon}, S., {Hirano}, S., {Hosokawa}, T., \& {Yoshida}, N. 2016, \apj, 832,
  134, \dodoi{10.3847/0004-637X/832/2/134}

\bibitem[{{Clark} {et~al.}(2011){Clark}, {Glover}, {Smith}, {Greif}, {Klessen},
  \& {Bromm}}]{2011Sci...331.1040C}
{Clark}, P.~C., {Glover}, S. C.~O., {Smith}, R.~J., {et~al.} 2011, Science,
  331, 1040, \dodoi{10.1126/science.1198027}

\bibitem[{{Cole} {et~al.}(2000){Cole}, {Lacey}, {Baugh}, \&
  {Frenk}}]{2000MNRAS.319..168C}
{Cole}, S., {Lacey}, C.~G., {Baugh}, C.~M., \& {Frenk}, C.~S. 2000, \mnras,
  319, 168, \dodoi{10.1046/j.1365-8711.2000.03879.x}

\bibitem[{{Croom} {et~al.}(2009){Croom}, {Richards}, {Shanks}, {Boyle},
  {Strauss}, {Myers}, {Nichol}, {Pimbblet}, {Ross}, {Schneider}, {Sharp}, \&
  {Wake}}]{2009MNRAS.399.1755C}
{Croom}, S.~M., {Richards}, G.~T., {Shanks}, T., {et~al.} 2009, \mnras, 399,
  1755, \dodoi{10.1111/j.1365-2966.2009.15398.x}

\bibitem[{{Davies} {et~al.}(2019){Davies}, {Hennawi}, \&
  {Eilers}}]{2019ApJ...884L..19D}
{Davies}, F.~B., {Hennawi}, J.~F., \& {Eilers}, A.-C. 2019, \apjl, 884, L19,
  \dodoi{10.3847/2041-8213/ab42e3}

\bibitem[{{Dayal} {et~al.}(2019){Dayal}, {Rossi}, {Shiralilou}, {Piana},
  {Choudhury}, \& {Volonteri}}]{2019MNRAS.486.2336D}
{Dayal}, P., {Rossi}, E.~M., {Shiralilou}, B., {et~al.} 2019, \mnras, 486,
  2336, \dodoi{10.1093/mnras/stz897}

\bibitem[{{Dekel} {et~al.}(2009){Dekel}, {Birnboim}, {Engel}, {Freundlich},
  {Goerdt}, {Mumcuoglu}, {Neistein}, {Pichon}, {Teyssier}, \&
  {Zinger}}]{2009Natur.457..451D}
{Dekel}, A., {Birnboim}, Y., {Engel}, G., {et~al.} 2009, \nat, 457, 451,
  \dodoi{10.1038/nature07648}

\bibitem[{{Delvecchio} {et~al.}(2014){Delvecchio}, {Gruppioni}, {Pozzi},
  {Berta}, {Zamorani}, {Cimatti}, {Lutz}, {Scott}, {Vignali}, {Cresci},
  {Feltre}, {Cooray}, {Vaccari}, {Fritz}, {Le Floc'h}, {Magnelli}, {Popesso},
  {Oliver}, {Bock}, {Carollo}, {Contini}, {Le F{\'e}vre}, {Lilly}, {Mainieri},
  {Renzini}, \& {Scodeggio}}]{2014MNRAS.439.2736D}
{Delvecchio}, I., {Gruppioni}, C., {Pozzi}, F., {et~al.} 2014, \mnras, 439,
  2736, \dodoi{10.1093/mnras/stu130}

\bibitem[{{Dey} {et~al.}(2019){Dey}, {Schlegel}, {Lang}, {Blum}, {Burleigh},
  {Fan}, {Findlay}, {Finkbeiner}, {Herrera}, {Juneau}, {Landriau}, {Levi},
  {McGreer}, {Meisner}, {Myers}, {Moustakas}, {Nugent}, {Patej}, {Schlafly},
  {Walker}, {Valdes}, {Weaver}, {Y{\`e}che}, {Zou}, {Zhou}, {Abareshi},
  {Abbott}, {Abolfathi}, {Aguilera}, {Alam}, {Allen}, {Alvarez}, {Annis},
  {Ansarinejad}, {Aubert}, {Beechert}, {Bell}, {BenZvi}, {Beutler}, {Bielby},
  {Bolton}, {Brice{\~n}o}, {Buckley-Geer}, {Butler}, {Calamida}, {Carlberg},
  {Carter}, {Casas}, {Castander}, {Choi}, {Comparat}, {Cukanovaite}, {Delubac},
  {DeVries}, {Dey}, {Dhungana}, {Dickinson}, {Ding}, {Donaldson}, {Duan},
  {Duckworth}, {Eftekharzadeh}, {Eisenstein}, {Etourneau}, {Fagrelius},
  {Farihi}, {Fitzpatrick}, {Font-Ribera}, {Fulmer}, {G{\"a}nsicke},
  {Gaztanaga}, {George}, {Gerdes}, {Gontcho}, {Gorgoni}, {Green}, {Guy},
  {Harmer}, {Hernandez}, {Honscheid}, {Huang}, {James}, {Jannuzi}, {Jiang},
  {Joyce}, {Karcher}, {Karkar}, {Kehoe}, {Kneib}, {Kueter-Young}, {Lan},
  {Lauer}, {Le Guillou}, {Le Van Suu}, {Lee}, {Lesser}, {Perreault Levasseur},
  {Li}, {Mann}, {Marshall}, {Mart{\'\i}nez-V{\'a}zquez}, {Martini}, {du Mas des
  Bourboux}, {McManus}, {Meier}, {M{\'e}nard}, {Metcalfe},
  {Mu{\~n}oz-Guti{\'e}rrez}, {Najita}, {Napier}, {Narayan}, {Newman}, {Nie},
  {Nord}, {Norman}, {Olsen}, {Paat}, {Palanque-Delabrouille}, {Peng},
  {Poppett}, {Poremba}, {Prakash}, {Rabinowitz}, {Raichoor}, {Rezaie},
  {Robertson}, {Roe}, {Ross}, {Ross}, {Rudnick}, {Safonova}, {Saha},
  {S{\'a}nchez}, {Savary}, {Schweiker}, {Scott}, {Seo}, {Shan}, {Silva},
  {Slepian}, {Soto}, {Sprayberry}, {Staten}, {Stillman}, {Stupak}, {Summers},
  {Sien Tie}, {Tirado}, {Vargas-Maga{\~n}a}, {Vivas}, {Wechsler}, {Williams},
  {Yang}, {Yang}, {Yapici}, {Zaritsky}, {Zenteno}, {Zhang}, {Zhang}, {Zhou}, \&
  {Zhou}}]{2019AJ....157..168D}
{Dey}, A., {Schlegel}, D.~J., {Lang}, D., {et~al.} 2019, \aj, 157, 168,
  \dodoi{10.3847/1538-3881/ab089d}

\bibitem[{{Di Matteo} {et~al.}(2017){Di Matteo}, {Croft}, {Feng}, {Waters}, \&
  {Wilkins}}]{2017MNRAS.467.4243D}
{Di Matteo}, T., {Croft}, R. A.~C., {Feng}, Y., {Waters}, D., \& {Wilkins}, S.
  2017, \mnras, 467, 4243, \dodoi{10.1093/mnras/stx319}

\bibitem[{{Di Matteo} {et~al.}(2005){Di Matteo}, {Springel}, \&
  {Hernquist}}]{2005Natur.433..604D}
{Di Matteo}, T., {Springel}, V., \& {Hernquist}, L. 2005, \nat, 433, 604,
  \dodoi{10.1038/nature03335}

\bibitem[{{Dijkstra} {et~al.}(2014){Dijkstra}, {Ferrara}, \&
  {Mesinger}}]{2014MNRAS.442.2036D}
{Dijkstra}, M., {Ferrara}, A., \& {Mesinger}, A. 2014, \mnras, 442, 2036,
  \dodoi{10.1093/mnras/stu1007}

\bibitem[{{Dijkstra} {et~al.}(2008){Dijkstra}, {Haiman}, {Mesinger}, \&
  {Wyithe}}]{2008MNRAS.391.1961D}
{Dijkstra}, M., {Haiman}, Z., {Mesinger}, A., \& {Wyithe}, J. S.~B. 2008,
  \mnras, 391, 1961, \dodoi{10.1111/j.1365-2966.2008.14031.x}

\bibitem[{{Donnan} {et~al.}(2023){Donnan}, {McLeod}, {Dunlop}, {McLure},
  {Carnall}, {Begley}, {Cullen}, {Hamadouche}, {Bowler}, {Magee}, {McCracken},
  {Milvang-Jensen}, {Moneti}, \& {Targett}}]{2023MNRAS.518.6011D}
{Donnan}, C.~T., {McLeod}, D.~J., {Dunlop}, J.~S., {et~al.} 2023, \mnras, 518,
  6011, \dodoi{10.1093/mnras/stac3472}

\bibitem[{{Duras} {et~al.}(2020){Duras}, {Bongiorno}, {Ricci}, {Piconcelli},
  {Shankar}, {Lusso}, {Bianchi}, {Fiore}, {Maiolino}, {Marconi}, {Onori},
  {Sani}, {Schneider}, {Vignali}, \& {La Franca}}]{2020A&A...636A..73D}
{Duras}, F., {Bongiorno}, A., {Ricci}, F., {et~al.} 2020, \aap, 636, A73,
  \dodoi{10.1051/0004-6361/201936817}

\bibitem[{{Eilers} {et~al.}(2017){Eilers}, {Davies}, {Hennawi}, {Prochaska},
  {Luki{\'c}}, \& {Mazzucchelli}}]{2017ApJ...840...24E}
{Eilers}, A.-C., {Davies}, F.~B., {Hennawi}, J.~F., {et~al.} 2017, \apj, 840,
  24, \dodoi{10.3847/1538-4357/aa6c60}

\bibitem[{{Eilers} {et~al.}(2018){Eilers}, {Hennawi}, \&
  {Davies}}]{2018ApJ...867...30E}
{Eilers}, A.-C., {Hennawi}, J.~F., \& {Davies}, F.~B. 2018, \apj, 867, 30,
  \dodoi{10.3847/1538-4357/aae081}

\bibitem[{{Eilers} {et~al.}(2021){Eilers}, {Hennawi}, {Davies}, \&
  {Simcoe}}]{2021ApJ...917...38E}
{Eilers}, A.-C., {Hennawi}, J.~F., {Davies}, F.~B., \& {Simcoe}, R.~A. 2021,
  \apj, 917, 38, \dodoi{10.3847/1538-4357/ac0a76}

\bibitem[{{Ekstr{\"o}m} {et~al.}(2008){Ekstr{\"o}m}, {Meynet}, {Chiappini},
  {Hirschi}, \& {Maeder}}]{2008A&A...489..685E}
{Ekstr{\"o}m}, S., {Meynet}, G., {Chiappini}, C., {Hirschi}, R., \& {Maeder},
  A. 2008, \aap, 489, 685, \dodoi{10.1051/0004-6361:200809633}

\bibitem[{{Euclid Collaboration} {et~al.}(2019){Euclid Collaboration},
  {Barnett}, {Warren}, {Mortlock}, {Cuby}, {Conselice}, {Hewett}, {Willott},
  {Auricchio}, {Balaguera-Antol{\'\i}nez}, {Baldi}, {Bardelli}, {Bellagamba},
  {Bender}, {Biviano}, {Bonino}, {Bozzo}, {Branchini}, {Brescia}, {Brinchmann},
  {Burigana}, {Camera}, {Capobianco}, {Carbone}, {Carretero}, {Carvalho},
  {Castander}, {Castellano}, {Cavuoti}, {Cimatti}, {Cl{\'e}dassou}, {Congedo},
  {Conversi}, {Copin}, {Corcione}, {Coupon}, {Courtois}, {Cropper}, {Da Silva},
  {Duncan}, {Dusini}, {Ealet}, {Farrens}, {Fosalba}, {Fotopoulou},
  {Fourmanoit}, {Frailis}, {Fumana}, {Galeotta}, {Garilli}, {Gillard},
  {Gillis}, {Graci{\'a}-Carpio}, {Grupp}, {Hoekstra}, {Hormuth}, {Israel},
  {Jahnke}, {Kermiche}, {Kilbinger}, {Kirkpatrick}, {Kitching}, {Kohley},
  {Kubik}, {Kunz}, {Kurki-Suonio}, {Laureijs}, {Ligori}, {Lilje}, {Lloro},
  {Maiorano}, {Mansutti}, {Marggraf}, {Martinet}, {Marulli}, {Massey}, {Mauri},
  {Medinaceli}, {Mei}, {Mellier}, {Metcalf}, {Metge}, {Meylan}, {Moresco},
  {Moscardini}, {Munari}, {Neissner}, {Niemi}, {Nutma}, {Padilla}, {Paltani},
  {Pasian}, {Paykari}, {Percival}, {Pettorino}, {Polenta}, {Poncet},
  {Pozzetti}, {Raison}, {Renzi}, {Rhodes}, {Rix}, {Romelli}, {Roncarelli},
  {Rossetti}, {Saglia}, {Sapone}, {Scaramella}, {Schneider}, {Scottez},
  {Secroun}, {Serrano}, {Sirri}, {Stanco}, {Sureau}, {Tallada-Cresp{\'\i}},
  {Tavagnacco}, {Taylor}, {Tenti}, {Tereno}, {Toledo-Moreo}, {Torradeflot},
  {Valenziano}, {Vassallo}, {Wang}, {Zacchei}, {Zamorani}, {Zoubian}, \&
  {Zucca}}]{2019Barnett}
{Euclid Collaboration}, {Barnett}, R., {Warren}, S.~J., {et~al.} 2019, \aap,
  631, A85, \dodoi{10.1051/0004-6361/201936427}

\bibitem[{{Fan} {et~al.}(2001){Fan}, {Narayanan}, {Lupton}, {Strauss}, {Knapp},
  {Becker}, {White}, {Pentericci}, {Leggett}, {Haiman}, {Gunn}, {Ivezi{\'c}},
  {Schneider}, {Anderson}, {Brinkmann}, {Bahcall}, {Connolly}, {Csabai}, {Doi},
  {Fukugita}, {Geballe}, {Grebel}, {Harbeck}, {Hennessy}, {Lamb}, {Miknaitis},
  {Munn}, {Nichol}, {Okamura}, {Pier}, {Prada}, {Richards}, {Szalay}, \&
  {York}}]{2001AJ....122.2833F}
{Fan}, X., {Narayanan}, V.~K., {Lupton}, R.~H., {et~al.} 2001, \aj, 122, 2833,
  \dodoi{10.1086/324111}

\bibitem[{{Fan} {et~al.}(2019){Fan}, {Barth}, {Banados}, {De Rosa}, {Decarli},
  {Eilers}, {Farina}, {Greene}, {Habouzit}, {Jiang}, {Jun}, {Koekemoer},
  {Malhotra}, {Mazzucchelli}, {Pacucci}, {Rhoads}, {Riechers}, {Rigby}, {Shen},
  {Simcoe}, {Stern}, {Strauss}, {Treu}, {Venemans}, {Vestergaard}, {Volonteri},
  {Walter}, {Yang}, \& {Wang}}]{2019BAAS...51c.121F}
{Fan}, X., {Barth}, A., {Banados}, E., {et~al.} 2019, \baas, 51, 121.
\newblock \doarXiv{1903.04078}

\bibitem[{{Farina} {et~al.}(2022){Farina}, {Schindler}, {Walter},
  {Ba{\~n}ados}, {Davies}, {Decarli}, {Eilers}, {Fan}, {Hennawi},
  {Mazzucchelli}, {Meyer}, {Trakhtenbrot}, {Volonteri}, {Wang}, {Worseck},
  {Yang}, {Gutcke}, {Venemans}, {Bosman}, {Costa}, {Rosa}, {Drake}, \&
  {Onoue}}]{2022ApJ...941..106F}
{Farina}, E.~P., {Schindler}, J.-T., {Walter}, F., {et~al.} 2022, \apj, 941,
  106, \dodoi{10.3847/1538-4357/ac9626}

\bibitem[{{Ferrarese}(2002)}]{2002ApJ...578...90F}
{Ferrarese}, L. 2002, \apj, 578, 90, \dodoi{10.1086/342308}

\bibitem[{{Fialkov} {et~al.}(2012){Fialkov}, {Barkana}, {Tseliakhovich}, \&
  {Hirata}}]{2012MNRAS.424.1335F}
{Fialkov}, A., {Barkana}, R., {Tseliakhovich}, D., \& {Hirata}, C.~M. 2012,
  \mnras, 424, 1335, \dodoi{10.1111/j.1365-2966.2012.21318.x}

\bibitem[{{Finkelstein} \& {Bagley}(2022)}]{2022ApJ...938...25F}
{Finkelstein}, S.~L., \& {Bagley}, M.~B. 2022, \apj, 938, 25,
  \dodoi{10.3847/1538-4357/ac89eb}

\bibitem[{{Foreman-Mackey} {et~al.}(2013){Foreman-Mackey}, {Hogg}, {Lang}, \&
  {Goodman}}]{2013PASP..125..306F}
{Foreman-Mackey}, D., {Hogg}, D.~W., {Lang}, D., \& {Goodman}, J. 2013, \pasp,
  125, 306, \dodoi{10.1086/670067}

\bibitem[{{Giallongo} {et~al.}(2019){Giallongo}, {Grazian}, {Fiore}, {Kodra},
  {Urrutia}, {Castellano}, {Cristiani}, {Dickinson}, {Fontana}, {Menci},
  {Pentericci}, {Boutsia}, {Newman}, \& {Puccetti}}]{2019ApJ...884...19G}
{Giallongo}, E., {Grazian}, A., {Fiore}, F., {et~al.} 2019, \apj, 884, 19,
  \dodoi{10.3847/1538-4357/ab39e1}

\bibitem[{{Gilli} {et~al.}(2007){Gilli}, {Comastri}, \&
  {Hasinger}}]{2007A&A...463...79G}
{Gilli}, R., {Comastri}, A., \& {Hasinger}, G. 2007, \aap, 463, 79,
  \dodoi{10.1051/0004-6361:20066334}

\bibitem[{{Gilli} {et~al.}(2022){Gilli}, {Norman}, {Calura}, {Vito}, {Decarli},
  {Marchesi}, {Iwasawa}, {Comastri}, {Lanzuisi}, {Pozzi}, {D'Amato}, {Vignali},
  {Brusa}, {Mignoli}, \& {Cox}}]{2022A&A...666A..17G}
{Gilli}, R., {Norman}, C., {Calura}, F., {et~al.} 2022, \aap, 666, A17,
  \dodoi{10.1051/0004-6361/202243708}

\bibitem[{{Greif} {et~al.}(2011){Greif}, {Springel}, {White}, {Glover},
  {Clark}, {Smith}, {Klessen}, \& {Bromm}}]{2011ApJ...737...75G}
{Greif}, T.~H., {Springel}, V., {White}, S. D.~M., {et~al.} 2011, \apj, 737,
  75, \dodoi{10.1088/0004-637X/737/2/75}

\bibitem[{{Habouzit} {et~al.}(2022){Habouzit}, {Onoue}, {Ba{\~n}ados},
  {Neeleman}, {Angl{\'e}s-Alc{\'a}zar}, {Walter}, {Pillepich}, {Dav{\'e}},
  {Jahnke}, \& {Dubois}}]{2022MNRAS.511.3751H}
{Habouzit}, M., {Onoue}, M., {Ba{\~n}ados}, E., {et~al.} 2022, \mnras, 511,
  3751, \dodoi{10.1093/mnras/stac225}

\bibitem[{{Haemmerl{\'e}} {et~al.}(2018){Haemmerl{\'e}}, {Woods}, {Klessen},
  {Heger}, \& {Whalen}}]{2018MNRAS.474.2757H}
{Haemmerl{\'e}}, L., {Woods}, T.~E., {Klessen}, R.~S., {Heger}, A., \&
  {Whalen}, D.~J. 2018, \mnras, 474, 2757, \dodoi{10.1093/mnras/stx2919}

\bibitem[{{Haiman}(2013)}]{2013ASSL..396..293H}
{Haiman}, Z. 2013, in Astrophysics and Space Science Library, Vol. 396, The
  First Galaxies, ed. T.~{Wiklind}, B.~{Mobasher}, \& V.~{Bromm}, 293,
  \dodoi{10.1007/978-3-642-32362-1\_6}

\bibitem[{{Haiman} {et~al.}(2009){Haiman}, {Kocsis}, \&
  {Menou}}]{2009ApJ...700.1952H}
{Haiman}, Z., {Kocsis}, B., \& {Menou}, K. 2009, \apj, 700, 1952,
  \dodoi{10.1088/0004-637X/700/2/1952}

\bibitem[{{Haiman} \& {Loeb}(1998)}]{1998ApJ...503..505H}
{Haiman}, Z., \& {Loeb}, A. 1998, \apj, 503, 505, \dodoi{10.1086/306017}

\bibitem[{{Haiman} {et~al.}(1996){Haiman}, {Thoul}, \&
  {Loeb}}]{1996ApJ...464..523H}
{Haiman}, Z., {Thoul}, A.~A., \& {Loeb}, A. 1996, \apj, 464, 523,
  \dodoi{10.1086/177343}

\bibitem[{{Harikane} {et~al.}(2022{\natexlab{a}}){Harikane}, {Ono}, {Ouchi},
  {Liu}, {Sawicki}, {Shibuya}, {Behroozi}, {He}, {Shimasaku}, {Arnouts},
  {Coupon}, {Fujimoto}, {Gwyn}, {Huang}, {Inoue}, {Kashikawa}, {Komiyama},
  {Matsuoka}, \& {Willott}}]{2022ApJS..259...20H}
{Harikane}, Y., {Ono}, Y., {Ouchi}, M., {et~al.} 2022{\natexlab{a}}, \apjs,
  259, 20, \dodoi{10.3847/1538-4365/ac3dfc}

\bibitem[{{Harikane} {et~al.}(2022{\natexlab{b}}){Harikane}, {Inoue},
  {Mawatari}, {Hashimoto}, {Yamanaka}, {Fudamoto}, {Matsuo}, {Tamura}, {Dayal},
  {Yung}, {Hutter}, {Pacucci}, {Sugahara}, \& {Koekemoer}}]{Harikane_2022b}
{Harikane}, Y., {Inoue}, A.~K., {Mawatari}, K., {et~al.} 2022{\natexlab{b}},
  \apj, 929, 1, \dodoi{10.3847/1538-4357/ac53a9}

\bibitem[{{Harikane} {et~al.}(2023){Harikane}, {Ouchi}, {Oguri}, {Ono},
  {Nakajima}, {Isobe}, {Umeda}, {Mawatari}, \& {Zhang}}]{Harikane_2022c}
{Harikane}, Y., {Ouchi}, M., {Oguri}, M., {et~al.} 2023, \apjs, 265, 5,
  \dodoi{10.3847/1538-4365/acaaa9}

\bibitem[{{Hasinger}(2008)}]{2008A&A...490..905H}
{Hasinger}, G. 2008, \aap, 490, 905, \dodoi{10.1051/0004-6361:200809839}

\bibitem[{{Heger} {et~al.}(2003){Heger}, {Fryer}, {Woosley}, {Langer}, \&
  {Hartmann}}]{2003ApJ...591..288H}
{Heger}, A., {Fryer}, C.~L., {Woosley}, S.~E., {Langer}, N., \& {Hartmann},
  D.~H. 2003, \apj, 591, 288, \dodoi{10.1086/375341}

\bibitem[{{Hickox} \& {Alexander}(2018)}]{2018ARA&A..56..625H}
{Hickox}, R.~C., \& {Alexander}, D.~M. 2018, \araa, 56, 625,
  \dodoi{10.1146/annurev-astro-081817-051803}

\bibitem[{{Hirano} {et~al.}(2017){Hirano}, {Hosokawa}, {Yoshida}, \&
  {Kuiper}}]{2017Sci...357.1375H}
{Hirano}, S., {Hosokawa}, T., {Yoshida}, N., \& {Kuiper}, R. 2017, Science,
  357, 1375, \dodoi{10.1126/science.aai9119}

\bibitem[{{Hirano} {et~al.}(2015){Hirano}, {Hosokawa}, {Yoshida}, {Omukai}, \&
  {Yorke}}]{2015MNRAS.448..568H}
{Hirano}, S., {Hosokawa}, T., {Yoshida}, N., {Omukai}, K., \& {Yorke}, H.~W.
  2015, \mnras, 448, 568, \dodoi{10.1093/mnras/stv044}

\bibitem[{{Hirano} {et~al.}(2014){Hirano}, {Hosokawa}, {Yoshida}, {Umeda},
  {Omukai}, {Chiaki}, \& {Yorke}}]{2014ApJ...781...60H}
{Hirano}, S., {Hosokawa}, T., {Yoshida}, N., {et~al.} 2014, \apj, 781, 60,
  \dodoi{10.1088/0004-637X/781/2/60}

\bibitem[{{Hirano} {et~al.}(2018){Hirano}, {Yoshida}, {Sakurai}, \&
  {Fujii}}]{2018ApJ...855...17H}
{Hirano}, S., {Yoshida}, N., {Sakurai}, Y., \& {Fujii}, M.~S. 2018, \apj, 855,
  17, \dodoi{10.3847/1538-4357/aaaaba}

\bibitem[{{Hopkins} \& {Hernquist}(2009)}]{2009ApJ...698.1550H}
{Hopkins}, P.~F., \& {Hernquist}, L. 2009, \apj, 698, 1550,
  \dodoi{10.1088/0004-637X/698/2/1550}

\bibitem[{{Hopkins} {et~al.}(2005){Hopkins}, {Hernquist}, {Cox}, {Di Matteo},
  {Martini}, {Robertson}, \& {Springel}}]{2005ApJ...630..705H}
{Hopkins}, P.~F., {Hernquist}, L., {Cox}, T.~J., {et~al.} 2005, \apj, 630, 705,
  \dodoi{10.1086/432438}

\bibitem[{{Hopkins} {et~al.}(2006){Hopkins}, {Hernquist}, {Cox}, {Robertson},
  {Di Matteo}, \& {Springel}}]{2006ApJ...639..700H}
---. 2006, \apj, 639, 700, \dodoi{10.1086/499351}

\bibitem[{{Hopkins} {et~al.}(2007){Hopkins}, {Hernquist}, {Cox}, {Robertson},
  \& {Krause}}]{2007ApJ...669...45H}
{Hopkins}, P.~F., {Hernquist}, L., {Cox}, T.~J., {Robertson}, B., \& {Krause},
  E. 2007, \apj, 669, 45, \dodoi{10.1086/521590}

\bibitem[{{Hosokawa} {et~al.}(2011){Hosokawa}, {Omukai}, {Yoshida}, \&
  {Yorke}}]{2011Sci...334.1250H}
{Hosokawa}, T., {Omukai}, K., {Yoshida}, N., \& {Yorke}, H.~W. 2011, Science,
  334, 1250, \dodoi{10.1126/science.1207433}

\bibitem[{{Hosokawa} {et~al.}(2013){Hosokawa}, {Yorke}, {Inayoshi}, {Omukai},
  \& {Yoshida}}]{2013ApJ...778..178H}
{Hosokawa}, T., {Yorke}, H.~W., {Inayoshi}, K., {Omukai}, K., \& {Yoshida}, N.
  2013, \apj, 778, 178, \dodoi{10.1088/0004-637X/778/2/178}

\bibitem[{{Inayoshi} {et~al.}(2018){Inayoshi}, {Li}, \&
  {Haiman}}]{2018MNRAS.479.4017I}
{Inayoshi}, K., {Li}, M., \& {Haiman}, Z. 2018, \mnras, 479, 4017,
  \dodoi{10.1093/mnras/sty1720}

\bibitem[{{Inayoshi} {et~al.}(2022){Inayoshi}, {Nakatani}, {Toyouchi},
  {Hosokawa}, {Kuiper}, \& {Onoue}}]{2022ApJ...927..237I}
{Inayoshi}, K., {Nakatani}, R., {Toyouchi}, D., {et~al.} 2022, \apj, 927, 237,
  \dodoi{10.3847/1538-4357/ac4751}

\bibitem[{{Inayoshi} \& {Omukai}(2012)}]{2012MNRAS.422.2539I}
{Inayoshi}, K., \& {Omukai}, K. 2012, \mnras, 422, 2539,
  \dodoi{10.1111/j.1365-2966.2012.20812.x}

\bibitem[{{Inayoshi} {et~al.}(2014){Inayoshi}, {Omukai}, \&
  {Tasker}}]{2014MNRAS.445L.109I}
{Inayoshi}, K., {Omukai}, K., \& {Tasker}, E. 2014, \mnras, 445, L109,
  \dodoi{10.1093/mnrasl/slu151}

\bibitem[{{Inayoshi} \& {Tanaka}(2015)}]{2015MNRAS.450.4350I}
{Inayoshi}, K., \& {Tanaka}, T.~L. 2015, \mnras, 450, 4350,
  \dodoi{10.1093/mnras/stv871}

\bibitem[{{Inayoshi} {et~al.}(2020){Inayoshi}, {Visbal}, \&
  {Haiman}}]{2020ARA&A..58...27I}
{Inayoshi}, K., {Visbal}, E., \& {Haiman}, Z. 2020, \araa, 58, 27,
  \dodoi{10.1146/annurev-astro-120419-014455}

\bibitem[{{Ishimoto} {et~al.}(2020){Ishimoto}, {Kashikawa}, {Onoue},
  {Matsuoka}, {Izumi}, {Strauss}, {Fujimoto}, {Imanishi}, {Ito}, {Iwasawa},
  {Kawaguchi}, {Lee}, {Liang}, {Lu}, {Momose}, {Toba}, \&
  {Uchiyama}}]{2020ApJ...903...60I}
{Ishimoto}, R., {Kashikawa}, N., {Onoue}, M., {et~al.} 2020, \apj, 903, 60,
  \dodoi{10.3847/1538-4357/abb80b}

\bibitem[{{Izumi} {et~al.}(2021){Izumi}, {Matsuoka}, {Fujimoto}, {Onoue},
  {Strauss}, {Umehata}, {Imanishi}, {Kohno}, {Kawaguchi}, {Kawamuro}, {Baba},
  {Nagao}, {Toba}, {Inayoshi}, {Silverman}, {Inoue}, {Ikarashi}, {Iwasawa},
  {Kashikawa}, {Hashimoto}, {Nakanishi}, {Ueda}, {Schramm}, {Lee}, \&
  {Suh}}]{2021ApJ...914...36I}
{Izumi}, T., {Matsuoka}, Y., {Fujimoto}, S., {et~al.} 2021, \apj, 914, 36,
  \dodoi{10.3847/1538-4357/abf6dc}

\bibitem[{{Jiang} {et~al.}(2007){Jiang}, {Fan}, {Vestergaard}, {Kurk},
  {Walter}, {Kelly}, \& {Strauss}}]{2007AJ....134.1150J}
{Jiang}, L., {Fan}, X., {Vestergaard}, M., {et~al.} 2007, \aj, 134, 1150,
  \dodoi{10.1086/520811}

\bibitem[{{Jiang} {et~al.}(2008){Jiang}, {Fan}, {Annis}, {Becker}, {White},
  {Chiu}, {Lin}, {Lupton}, {Richards}, {Strauss}, {Jester}, \&
  {Schneider}}]{2008AJ....135.1057J}
{Jiang}, L., {Fan}, X., {Annis}, J., {et~al.} 2008, \aj, 135, 1057,
  \dodoi{10.1088/0004-6256/135/3/1057}

\bibitem[{{Jiang} {et~al.}(2016){Jiang}, {McGreer}, {Fan}, {Strauss},
  {Ba{\~n}ados}, {Becker}, {Bian}, {Farnsworth}, {Shen}, {Wang}, {Wang},
  {Wang}, {White}, {Wu}, {Wu}, {Yang}, \& {Yang}}]{2016ApJ...833..222J}
{Jiang}, L., {McGreer}, I.~D., {Fan}, X., {et~al.} 2016, \apj, 833, 222,
  \dodoi{10.3847/1538-4357/833/2/222}

\bibitem[{{Jiang} {et~al.}(2022){Jiang}, {Ning}, {Fan}, {Ho}, {Luo}, {Wang},
  {Wu}, {Wu}, {Yang}, \& {Zheng}}]{2022NatAs...6..850J}
{Jiang}, L., {Ning}, Y., {Fan}, X., {et~al.} 2022, Nature Astronomy, 6, 850,
  \dodoi{10.1038/s41550-022-01708-w}

\bibitem[{{Jones} {et~al.}(2016){Jones}, {Hickox}, {Black}, {Hainline},
  {DiPompeo}, \& {Goulding}}]{2016ApJ...826...12J}
{Jones}, M.~L., {Hickox}, R.~C., {Black}, C.~S., {et~al.} 2016, \apj, 826, 12,
  \dodoi{10.3847/0004-637X/826/1/12}

\bibitem[{{Kakiichi} {et~al.}(2022){Kakiichi}, {Schmidt}, \&
  {Hennawi}}]{2022MNRAS.516..582K}
{Kakiichi}, K., {Schmidt}, T., \& {Hennawi}, J. 2022, \mnras, 516, 582,
  \dodoi{10.1093/mnras/stac2026}

\bibitem[{{Kashikawa} {et~al.}(2015){Kashikawa}, {Ishizaki}, {Willott},
  {Onoue}, {Im}, {Furusawa}, {Toshikawa}, {Ishikawa}, {Niino}, {Shimasaku},
  {Ouchi}, \& {Hibon}}]{2015ApJ...798...28K}
{Kashikawa}, N., {Ishizaki}, Y., {Willott}, C.~J., {et~al.} 2015, \apj, 798,
  28, \dodoi{10.1088/0004-637X/798/1/28}

\bibitem[{{Kelly} \& {Shen}(2013)}]{2013ApJ...764...45K}
{Kelly}, B.~C., \& {Shen}, Y. 2013, \apj, 764, 45,
  \dodoi{10.1088/0004-637X/764/1/45}

\bibitem[{{Khrykin} {et~al.}(2016){Khrykin}, {Hennawi}, {McQuinn}, \&
  {Worseck}}]{2016ApJ...824..133K}
{Khrykin}, I.~S., {Hennawi}, J.~F., {McQuinn}, M., \& {Worseck}, G. 2016, \apj,
  824, 133, \dodoi{10.3847/0004-637X/824/2/133}

\bibitem[{{Kim} \& {Im}(2021)}]{2021ApJ...910L..11K}
{Kim}, Y., \& {Im}, M. 2021, \apjl, 910, L11, \dodoi{10.3847/2041-8213/abed58}

\bibitem[{{Kim} {et~al.}(2022){Kim}, {Im}, {Jeon}, {Kim}, {Jiang}, {Shin},
  {Choi}, {Hyun}, {Jun}, {Kim}, {Kim}, {Kim}, {Kim}, {Lee}, {Lee}, {Molina},
  {Pak}, {Park}, {Taak}, \& {Yoon}}]{2022AJ....164..114K}
{Kim}, Y., {Im}, M., {Jeon}, Y., {et~al.} 2022, \aj, 164, 114,
  \dodoi{10.3847/1538-3881/ac81c8}

\bibitem[{{King} \& {Pringle}(2007)}]{2007MNRAS.377L..25K}
{King}, A.~R., \& {Pringle}, J.~E. 2007, \mnras, 377, L25,
  \dodoi{10.1111/j.1745-3933.2007.00296.x}

\bibitem[{{Kocsis} {et~al.}(2006){Kocsis}, {Frei}, {Haiman}, \&
  {Menou}}]{2006ApJ...637...27K}
{Kocsis}, B., {Frei}, Z., {Haiman}, Z., \& {Menou}, K. 2006, \apj, 637, 27,
  \dodoi{10.1086/498236}

\bibitem[{{Kormendy} \& {Ho}(2013)}]{2013ARA&A..51..511K}
{Kormendy}, J., \& {Ho}, L.~C. 2013, \araa, 51, 511,
  \dodoi{10.1146/annurev-astro-082708-101811}

\bibitem[{{Kulkarni} {et~al.}(2019){Kulkarni}, {Worseck}, \&
  {Hennawi}}]{2019MNRAS.488.1035K}
{Kulkarni}, G., {Worseck}, G., \& {Hennawi}, J.~F. 2019, \mnras, 488, 1035,
  \dodoi{10.1093/mnras/stz1493}

\bibitem[{{Kurk} {et~al.}(2007){Kurk}, {Walter}, {Fan}, {Jiang}, {Riechers},
  {Rix}, {Pentericci}, {Strauss}, {Carilli}, \& {Wagner}}]{2007ApJ...669...32K}
{Kurk}, J.~D., {Walter}, F., {Fan}, X., {et~al.} 2007, \apj, 669, 32,
  \dodoi{10.1086/521596}

\bibitem[{{Larson}(1969)}]{1969MNRAS.145..271L}
{Larson}, R.~B. 1969, \mnras, 145, 271, \dodoi{10.1093/mnras/145.3.271}

\bibitem[{{Latif} \& {Ferrara}(2016)}]{2016PASA...33...51L}
{Latif}, M.~A., \& {Ferrara}, A. 2016, \pasa, 33, e051,
  \dodoi{10.1017/pasa.2016.41}

\bibitem[{{Latif} {et~al.}(2022){Latif}, {Whalen}, {Khochfar}, {Herrington}, \&
  {Woods}}]{2022Natur.607...48L}
{Latif}, M.~A., {Whalen}, D.~J., {Khochfar}, S., {Herrington}, N.~P., \&
  {Woods}, T.~E. 2022, \nat, 607, 48, \dodoi{10.1038/s41586-022-04813-y}

\bibitem[{{Laureijs} {et~al.}(2011){Laureijs}, {Amiaux}, {Arduini},
  {Augu{\`e}res}, {Brinchmann}, {Cole}, {Cropper}, {Dabin}, {Duvet}, {Ealet},
  {Garilli}, {Gondoin}, {Guzzo}, {Hoar}, {Hoekstra}, {Holmes}, {Kitching},
  {Maciaszek}, {Mellier}, {Pasian}, {Percival}, {Rhodes}, {Saavedra Criado},
  {Sauvage}, {Scaramella}, {Valenziano}, {Warren}, {Bender}, {Castander},
  {Cimatti}, {Le F{\`e}vre}, {Kurki-Suonio}, {Levi}, {Lilje}, {Meylan},
  {Nichol}, {Pedersen}, {Popa}, {Rebolo Lopez}, {Rix}, {Rottgering},
  {Zeilinger}, {Grupp}, {Hudelot}, {Massey}, {Meneghetti}, {Miller}, {Paltani},
  {Paulin-Henriksson}, {Pires}, {Saxton}, {Schrabback}, {Seidel}, {Walsh},
  {Aghanim}, {Amendola}, {Bartlett}, {Baccigalupi}, {Beaulieu}, {Benabed},
  {Cuby}, {Elbaz}, {Fosalba}, {Gavazzi}, {Helmi}, {Hook}, {Irwin}, {Kneib},
  {Kunz}, {Mannucci}, {Moscardini}, {Tao}, {Teyssier}, {Weller}, {Zamorani},
  {Zapatero Osorio}, {Boulade}, {Foumond}, {Di Giorgio}, {Guttridge}, {James},
  {Kemp}, {Martignac}, {Spencer}, {Walton}, {Bl{\"u}mchen}, {Bonoli},
  {Bortoletto}, {Cerna}, {Corcione}, {Fabron}, {Jahnke}, {Ligori}, {Madrid},
  {Martin}, {Morgante}, {Pamplona}, {Prieto}, {Riva}, {Toledo}, {Trifoglio},
  {Zerbi}, {Abdalla}, {Douspis}, {Grenet}, {Borgani}, {Bouwens}, {Courbin},
  {Delouis}, {Dubath}, {Fontana}, {Frailis}, {Grazian}, {Koppenh{\"o}fer},
  {Mansutti}, {Melchior}, {Mignoli}, {Mohr}, {Neissner}, {Noddle}, {Poncet},
  {Scodeggio}, {Serrano}, {Shane}, {Starck}, {Surace}, {Taylor},
  {Verdoes-Kleijn}, {Vuerli}, {Williams}, {Zacchei}, {Altieri}, {Escudero
  Sanz}, {Kohley}, {Oosterbroek}, {Astier}, {Bacon}, {Bardelli}, {Baugh},
  {Bellagamba}, {Benoist}, {Bianchi}, {Biviano}, {Branchini}, {Carbone},
  {Cardone}, {Clements}, {Colombi}, {Conselice}, {Cresci}, {Deacon}, {Dunlop},
  {Fedeli}, {Fontanot}, {Franzetti}, {Giocoli}, {Garcia-Bellido}, {Gow},
  {Heavens}, {Hewett}, {Heymans}, {Holland}, {Huang}, {Ilbert}, {Joachimi},
  {Jennins}, {Kerins}, {Kiessling}, {Kirk}, {Kotak}, {Krause}, {Lahav}, {van
  Leeuwen}, {Lesgourgues}, {Lombardi}, {Magliocchetti}, {Maguire}, {Majerotto},
  {Maoli}, {Marulli}, {Maurogordato}, {McCracken}, {McLure}, {Melchiorri},
  {Merson}, {Moresco}, {Nonino}, {Norberg}, {Peacock}, {Pello}, {Penny},
  {Pettorino}, {Di Porto}, {Pozzetti}, {Quercellini}, {Radovich}, {Rassat},
  {Roche}, {Ronayette}, {Rossetti}, {Sartoris}, {Schneider}, {Semboloni},
  {Serjeant}, {Simpson}, {Skordis}, {Smadja}, {Smartt}, {Spano}, {Spiro},
  {Sullivan}, {Tilquin}, {Trotta}, {Verde}, {Wang}, {Williger}, {Zhao},
  {Zoubian}, \& {Zucca}}]{2011arXiv1110.3193L}
{Laureijs}, R., {Amiaux}, J., {Arduini}, S., {et~al.} 2011, arXiv e-prints,
  arXiv:1110.3193.
\newblock \doarXiv{1110.3193}

\bibitem[{{Li} {et~al.}(2021){Li}, {Inayoshi}, \& {Qiu}}]{2021ApJ...917...60L}
{Li}, W., {Inayoshi}, K., \& {Qiu}, Y. 2021, \apj, 917, 60,
  \dodoi{10.3847/1538-4357/ac0adc}

\bibitem[{{Li} {et~al.}(2012){Li}, {Wang}, \& {Ho}}]{2012ApJ...749..187L}
{Li}, Y.-R., {Wang}, J.-M., \& {Ho}, L.~C. 2012, \apj, 749, 187,
  \dodoi{10.1088/0004-637X/749/2/187}

\bibitem[{{Luo} {et~al.}(2016){Luo}, {Chen}, {Duan}, {Gong}, {Hu}, {Ji}, {Liu},
  {Mei}, {Milyukov}, {Sazhin}, {Shao}, {Toth}, {Tu}, {Wang}, {Wang}, {Yeh},
  {Zhan}, {Zhang}, {Zharov}, \& {Zhou}}]{Tianqin_2016}
{Luo}, J., {Chen}, L.-S., {Duan}, H.-Z., {et~al.} 2016, Classical and Quantum
  Gravity, 33, 035010, \dodoi{10.1088/0264-9381/33/3/035010}

\bibitem[{{Lupi} {et~al.}(2021){Lupi}, {Haiman}, \&
  {Volonteri}}]{2021MNRAS.503.5046L}
{Lupi}, A., {Haiman}, Z., \& {Volonteri}, M. 2021, \mnras, 503, 5046,
  \dodoi{10.1093/mnras/stab692}

\bibitem[{{Lupi} {et~al.}(2019){Lupi}, {Volonteri}, {Decarli}, {Bovino},
  {Silk}, \& {Bergeron}}]{2019MNRAS.488.4004L}
{Lupi}, A., {Volonteri}, M., {Decarli}, R., {et~al.} 2019, \mnras, 488, 4004,
  \dodoi{10.1093/mnras/stz1959}

\bibitem[{{Marconi} {et~al.}(2004){Marconi}, {Risaliti}, {Gilli}, {Hunt},
  {Maiolino}, \& {Salvati}}]{2004MNRAS.351..169M}
{Marconi}, A., {Risaliti}, G., {Gilli}, R., {et~al.} 2004, \mnras, 351, 169,
  \dodoi{10.1111/j.1365-2966.2004.07765.x}

\bibitem[{{Martini}(2004)}]{2004cbhg.symp..169M}
{Martini}, P. 2004, in Coevolution of Black Holes and Galaxies, ed. L.~C. {Ho},
  169.
\newblock \doarXiv{astro-ph/0304009}

\bibitem[{{Matsuoka} {et~al.}(2018{\natexlab{a}}){Matsuoka}, {Onoue},
  {Kashikawa}, {Iwasawa}, {Strauss}, {Nagao}, {Imanishi}, {Lee}, {Akiyama},
  {Asami}, {Bosch}, {Foucaud}, {Furusawa}, {Goto}, {Gunn}, {Harikane}, {Ikeda},
  {Izumi}, {Kawaguchi}, {Kikuta}, {Kohno}, {Komiyama}, {Lupton}, {Minezaki},
  {Miyazaki}, {Morokuma}, {Murayama}, {Niida}, {Nishizawa}, {Oguri}, {Ono},
  {Ouchi}, {Price}, {Sameshima}, {Schulze}, {Shirakata}, {Silverman},
  {Sugiyama}, {Tait}, {Takada}, {Takata}, {Tanaka}, {Tang}, {Toba}, {Utsumi},
  \& {Wang}}]{2018PASJ...70S..35M}
{Matsuoka}, Y., {Onoue}, M., {Kashikawa}, N., {et~al.} 2018{\natexlab{a}},
  \pasj, 70, S35, \dodoi{10.1093/pasj/psx046}

\bibitem[{{Matsuoka} {et~al.}(2018{\natexlab{b}}){Matsuoka}, {Strauss},
  {Kashikawa}, {Onoue}, {Iwasawa}, {Tang}, {Lee}, {Imanishi}, {Nagao},
  {Akiyama}, {Asami}, {Bosch}, {Furusawa}, {Goto}, {Gunn}, {Harikane}, {Ikeda},
  {Izumi}, {Kawaguchi}, {Kato}, {Kikuta}, {Kohno}, {Komiyama}, {Lupton},
  {Minezaki}, {Miyazaki}, {Murayama}, {Niida}, {Nishizawa}, {Noboriguchi},
  {Oguri}, {Ono}, {Ouchi}, {Price}, {Sameshima}, {Schulze}, {Shirakata},
  {Silverman}, {Sugiyama}, {Tait}, {Takada}, {Takata}, {Tanaka}, {Toba},
  {Utsumi}, {Wang}, \& {Yamashita}}]{2018ApJ...869..150M}
{Matsuoka}, Y., {Strauss}, M.~A., {Kashikawa}, N., {et~al.} 2018{\natexlab{b}},
  \apj, 869, 150, \dodoi{10.3847/1538-4357/aaee7a}

\bibitem[{{Matsuoka} {et~al.}(2019){Matsuoka}, {Iwasawa}, {Onoue}, {Kashikawa},
  {Strauss}, {Lee}, {Imanishi}, {Nagao}, {Akiyama}, {Asami}, {Bosch},
  {Furusawa}, {Goto}, {Gunn}, {Harikane}, {Ikeda}, {Izumi}, {Kawaguchi},
  {Kato}, {Kikuta}, {Kohno}, {Komiyama}, {Koyama}, {Lupton}, {Minezaki},
  {Miyazaki}, {Murayama}, {Niida}, {Nishizawa}, {Noboriguchi}, {Oguri}, {Ono},
  {Ouchi}, {Price}, {Sameshima}, {Schulze}, {Silverman}, {Sugiyama}, {Tait},
  {Takada}, {Takata}, {Tanaka}, {Tang}, {Toba}, {Utsumi}, {Wang}, \&
  {Yamashita}}]{2019ApJ...883..183M}
{Matsuoka}, Y., {Iwasawa}, K., {Onoue}, M., {et~al.} 2019, \apj, 883, 183,
  \dodoi{10.3847/1538-4357/ab3c60}

\bibitem[{{Mayer} {et~al.}(2015){Mayer}, {Fiacconi}, {Bonoli}, {Quinn},
  {Ro{\v{s}}kar}, {Shen}, \& {Wadsley}}]{2015ApJ...810...51M}
{Mayer}, L., {Fiacconi}, D., {Bonoli}, S., {et~al.} 2015, \apj, 810, 51,
  \dodoi{10.1088/0004-637X/810/1/51}

\bibitem[{{Mayer} {et~al.}(2010){Mayer}, {Kazantzidis}, {Escala}, \&
  {Callegari}}]{2010Natur.466.1082M}
{Mayer}, L., {Kazantzidis}, S., {Escala}, A., \& {Callegari}, S. 2010, \nat,
  466, 1082, \dodoi{10.1038/nature09294}

\bibitem[{{McGreer} {et~al.}(2013){McGreer}, {Jiang}, {Fan}, {Richards},
  {Strauss}, {Ross}, {White}, {Shen}, {Schneider}, {Myers}, {Brandt}, {DeGraf},
  {Glikman}, {Ge}, \& {Streblyanska}}]{2013ApJ...768..105M}
{McGreer}, I.~D., {Jiang}, L., {Fan}, X., {et~al.} 2013, \apj, 768, 105,
  \dodoi{10.1088/0004-637X/768/2/105}

\bibitem[{{McKee} \& {Tan}(2008)}]{2008ApJ...681..771M}
{McKee}, C.~F., \& {Tan}, J.~C. 2008, \apj, 681, 771, \dodoi{10.1086/587434}

\bibitem[{{McLure} {et~al.}(2013){McLure}, {Dunlop}, {Bowler}, {Curtis-Lake},
  {Schenker}, {Ellis}, {Robertson}, {Koekemoer}, {Rogers}, {Ono}, {Ouchi},
  {Charlot}, {Wild}, {Stark}, {Furlanetto}, {Cirasuolo}, \&
  {Targett}}]{2013MNRAS.432.2696M}
{McLure}, R.~J., {Dunlop}, J.~S., {Bowler}, R.~A.~A., {et~al.} 2013, \mnras,
  432, 2696, \dodoi{10.1093/mnras/stt627}

\bibitem[{{Merloni} \& {Heinz}(2008)}]{2008MNRAS.388.1011M}
{Merloni}, A., \& {Heinz}, S. 2008, \mnras, 388, 1011,
  \dodoi{10.1111/j.1365-2966.2008.13472.x}

\bibitem[{{Merloni} {et~al.}(2014){Merloni}, {Bongiorno}, {Brusa}, {Iwasawa},
  {Mainieri}, {Magnelli}, {Salvato}, {Berta}, {Cappelluti}, {Comastri},
  {Fiore}, {Gilli}, {Koekemoer}, {Le Floc'h}, {Lusso}, {Lutz}, {Miyaji},
  {Pozzi}, {Riguccini}, {Rosario}, {Silverman}, {Symeonidis}, {Treister},
  {Vignali}, \& {Zamorani}}]{2014MNRAS.437.3550M}
{Merloni}, A., {Bongiorno}, A., {Brusa}, M., {et~al.} 2014, \mnras, 437, 3550,
  \dodoi{10.1093/mnras/stt2149}

\bibitem[{{Morishita} {et~al.}(2018){Morishita}, {Trenti}, {Stiavelli},
  {Bradley}, {Coe}, {Oesch}, {Mason}, {Bridge}, {Holwerda}, {Livermore},
  {Salmon}, {Schmidt}, {Shull}, \& {Treu}}]{2018ApJ...867..150M}
{Morishita}, T., {Trenti}, M., {Stiavelli}, M., {et~al.} 2018, \apj, 867, 150,
  \dodoi{10.3847/1538-4357/aae68c}

\bibitem[{{Naidu} {et~al.}(2022){Naidu}, {Oesch}, {van Dokkum}, {Nelson},
  {Suess}, {Brammer}, {Whitaker}, {Illingworth}, {Bouwens}, {Tacchella},
  {Matthee}, {Allen}, {Bezanson}, {Conroy}, {Labbe}, {Leja}, {Leonova},
  {Magee}, {Price}, {Setton}, {Strait}, {Stefanon}, {Toft}, {Weaver}, \&
  {Weibel}}]{2022ApJ...940L..14N}
{Naidu}, R.~P., {Oesch}, P.~A., {van Dokkum}, P., {et~al.} 2022, \apjl, 940,
  L14, \dodoi{10.3847/2041-8213/ac9b22}

\bibitem[{{Natarajan} \& {Volonteri}(2012)}]{2012MNRAS.422.2051N}
{Natarajan}, P., \& {Volonteri}, M. 2012, \mnras, 422, 2051,
  \dodoi{10.1111/j.1365-2966.2012.20708.x}

\bibitem[{{Ni} {et~al.}(2020){Ni}, {Di Matteo}, {Gilli}, {Croft}, {Feng}, \&
  {Norman}}]{2020MNRAS.495.2135N}
{Ni}, Y., {Di Matteo}, T., {Gilli}, R., {et~al.} 2020, \mnras, 495, 2135,
  \dodoi{10.1093/mnras/staa1313}

\bibitem[{{Ni} {et~al.}(2022){Ni}, {Di Matteo}, {Bird}, {Croft}, {Feng},
  {Chen}, {Tremmel}, {DeGraf}, \& {Li}}]{2022MNRAS.513..670N}
{Ni}, Y., {Di Matteo}, T., {Bird}, S., {et~al.} 2022, \mnras, 513, 670,
  \dodoi{10.1093/mnras/stac351}

\bibitem[{{Niida} {et~al.}(2020){Niida}, {Nagao}, {Ikeda}, {Akiyama},
  {Matsuoka}, {He}, {Matsuoka}, {Toba}, {Onoue}, {Kobayashi}, {Taniguchi},
  {Furusawa}, {Harikane}, {Imanishi}, {Kashikawa}, {Kawaguchi}, {Komiyama},
  {Shirakata}, {Terashima}, \& {Ueda}}]{2020ApJ...904...89N}
{Niida}, M., {Nagao}, T., {Ikeda}, H., {et~al.} 2020, \apj, 904, 89,
  \dodoi{10.3847/1538-4357/abbe11}

\bibitem[{{Novak} {et~al.}(2011){Novak}, {Ostriker}, \&
  {Ciotti}}]{2011ApJ...737...26N}
{Novak}, G.~S., {Ostriker}, J.~P., \& {Ciotti}, L. 2011, \apj, 737, 26,
  \dodoi{10.1088/0004-637X/737/1/26}

\bibitem[{{Oesch} {et~al.}(2016){Oesch}, {Brammer}, {van Dokkum},
  {Illingworth}, {Bouwens}, {Labb{\'e}}, {Franx}, {Momcheva}, {Ashby}, {Fazio},
  {Gonzalez}, {Holden}, {Magee}, {Skelton}, {Smit}, {Spitler}, {Trenti}, \&
  {Willner}}]{2016ApJ...819..129O}
{Oesch}, P.~A., {Brammer}, G., {van Dokkum}, P.~G., {et~al.} 2016, \apj, 819,
  129, \dodoi{10.3847/0004-637X/819/2/129}

\bibitem[{{Oh} \& {Haiman}(2002)}]{2002ApJ...569..558O}
{Oh}, S.~P., \& {Haiman}, Z. 2002, \apj, 569, 558, \dodoi{10.1086/339393}

\bibitem[{{Omukai}(2001)}]{2001ApJ...546..635O}
{Omukai}, K. 2001, \apj, 546, 635, \dodoi{10.1086/318296}

\bibitem[{{Omukai} \& {Palla}(2001)}]{2001ApJ...561L..55O}
{Omukai}, K., \& {Palla}, F. 2001, \apjl, 561, L55, \dodoi{10.1086/324410}

\bibitem[{{Onoue} {et~al.}(2017){Onoue}, {Kashikawa}, {Willott}, {Hibon}, {Im},
  {Furusawa}, {Harikane}, {Imanishi}, {Ishikawa}, {Kikuta}, {Matsuoka},
  {Nagao}, {Niino}, {Ono}, {Ouchi}, {Tanaka}, {Tang}, {Toshikawa}, \&
  {Uchiyama}}]{2017ApJ...847L..15O}
{Onoue}, M., {Kashikawa}, N., {Willott}, C.~J., {et~al.} 2017, \apjl, 847, L15,
  \dodoi{10.3847/2041-8213/aa8cc6}

\bibitem[{{Onoue} {et~al.}(2019){Onoue}, {Kashikawa}, {Matsuoka}, {Kato},
  {Izumi}, {Nagao}, {Strauss}, {Harikane}, {Imanishi}, {Ito}, {Iwasawa},
  {Kawaguchi}, {Lee}, {Noboriguchi}, {Suh}, {Tanaka}, \&
  {Toba}}]{2019ApJ...880...77O}
{Onoue}, M., {Kashikawa}, N., {Matsuoka}, Y., {et~al.} 2019, \apj, 880, 77,
  \dodoi{10.3847/1538-4357/ab29e9}

\bibitem[{{Onoue} {et~al.}(2021){Onoue}, {Ding}, {Izumi}, {Matsuoka},
  {Silverman}, {Akiyama}, {Andika}, {Aoki}, {Baba}, {Bieri}, {Bosman},
  {Eilers}, {Fujimoto}, {Habouzit}, {Haiman}, {Imanishi}, {Inayoshi}, {Jahnke},
  {Kashikawa}, {Kawaguchi}, {Kikuta}, {Kohno}, {Lee}, {Lupi}, {Lyu}, {Marian},
  {Nagao}, {Overzier}, {Schindler}, {Schramm}, {Shimasaku}, {Strauss},
  {Trakhtenbrot}, {Trebitsch}, {Umehata}, {Venemans}, {Vestergaard},
  {Volonteri}, {Walter}, {Wang}, {Yang}, \& {he}}]{2021jwst.prop.1967O}
{Onoue}, M., {Ding}, X., {Izumi}, T., {et~al.} 2021, {A Complete Census of
  Supermassive Black Holes and Host Galaxies at z=6}, JWST Proposal. Cycle 1,
  ID. \#1967

\bibitem[{{Onoue} {et~al.}(2023){Onoue}, {Inayoshi}, {Ding}, {Li}, {Li},
  {Molina}, {Inoue}, {Jiang}, \& {Ho}}]{2023ApJ...942L..17O}
{Onoue}, M., {Inayoshi}, K., {Ding}, X., {et~al.} 2023, \apjl, 942, L17,
  \dodoi{10.3847/2041-8213/aca9d3}

\bibitem[{{Oogi} {et~al.}(2022){Oogi}, {Ishiyama}, {Prada}, {Sinha}, {Croton},
  {Cora}, {Jullo}, {Klypin}, {Nagashima}, {L{\'o}pez Cacheiro}, {Ruedas},
  {Kobayashi}, \& {Makiya}}]{2022arXiv220714689O}
{Oogi}, T., {Ishiyama}, T., {Prada}, F., {et~al.} 2022, arXiv e-prints,
  arXiv:2207.14689.
\newblock \doarXiv{2207.14689}

\bibitem[{{Orofino} {et~al.}(2021){Orofino}, {Ferrara}, \&
  {Gallerani}}]{2021MNRAS.502.2757O}
{Orofino}, M.~C., {Ferrara}, A., \& {Gallerani}, S. 2021, \mnras, 502, 2757,
  \dodoi{10.1093/mnras/stab160}

\bibitem[{{Padmanabhan} \& {Loeb}(2022)}]{2022arXiv220714309P}
{Padmanabhan}, H., \& {Loeb}, A. 2022, arXiv e-prints, arXiv:2207.14309.
\newblock \doarXiv{2207.14309}

\bibitem[{{Parkinson} {et~al.}(2008){Parkinson}, {Cole}, \&
  {Helly}}]{2008MNRAS.383..557P}
{Parkinson}, H., {Cole}, S., \& {Helly}, J. 2008, \mnras, 383, 557,
  \dodoi{10.1111/j.1365-2966.2007.12517.x}

\bibitem[{{Penston}(1969)}]{1969MNRAS.144..425P}
{Penston}, M.~V. 1969, \mnras, 144, 425, \dodoi{10.1093/mnras/144.4.425}

\bibitem[{{Piana} {et~al.}(2021){Piana}, {Dayal}, {Volonteri}, \&
  {Choudhury}}]{2021MNRAS.500.2146P}
{Piana}, O., {Dayal}, P., {Volonteri}, M., \& {Choudhury}, T.~R. 2021, \mnras,
  500, 2146, \dodoi{10.1093/mnras/staa3363}

\bibitem[{{Planck Collaboration} {et~al.}(2016){Planck Collaboration}, {Ade},
  {Aghanim}, {Arnaud}, {Ashdown}, {Aumont}, {Baccigalupi}, {Banday},
  {Barreiro}, {Bartlett}, {Bartolo}, {Battaner}, {Battye}, {Benabed},
  {Beno{\^\i}t}, {Benoit-L{\'e}vy}, {Bernard}, {Bersanelli}, {Bielewicz},
  {Bock}, {Bonaldi}, {Bonavera}, {Bond}, {Borrill}, {Bouchet}, {Boulanger},
  {Bucher}, {Burigana}, {Butler}, {Calabrese}, {Cardoso}, {Catalano},
  {Challinor}, {Chamballu}, {Chary}, {Chiang}, {Chluba}, {Christensen},
  {Church}, {Clements}, {Colombi}, {Colombo}, {Combet}, {Coulais}, {Crill},
  {Curto}, {Cuttaia}, {Danese}, {Davies}, {Davis}, {de Bernardis}, {de Rosa},
  {de Zotti}, {Delabrouille}, {D{\'e}sert}, {Di Valentino}, {Dickinson},
  {Diego}, {Dolag}, {Dole}, {Donzelli}, {Dor{\'e}}, {Douspis}, {Ducout},
  {Dunkley}, {Dupac}, {Efstathiou}, {Elsner}, {En{\ss}lin}, {Eriksen},
  {Farhang}, {Fergusson}, {Finelli}, {Forni}, {Frailis}, {Fraisse},
  {Franceschi}, {Frejsel}, {Galeotta}, {Galli}, {Ganga}, {Gauthier}, {Gerbino},
  {Ghosh}, {Giard}, {Giraud-H{\'e}raud}, {Giusarma}, {Gjerl{\o}w},
  {Gonz{\'a}lez-Nuevo}, {G{\'o}rski}, {Gratton}, {Gregorio}, {Gruppuso},
  {Gudmundsson}, {Hamann}, {Hansen}, {Hanson}, {Harrison}, {Helou},
  {Henrot-Versill{\'e}}, {Hern{\'a}ndez-Monteagudo}, {Herranz}, {Hildebrandt},
  {Hivon}, {Hobson}, {Holmes}, {Hornstrup}, {Hovest}, {Huang}, {Huffenberger},
  {Hurier}, {Jaffe}, {Jaffe}, {Jones}, {Juvela}, {Keih{\"a}nen}, {Keskitalo},
  {Kisner}, {Kneissl}, {Knoche}, {Knox}, {Kunz}, {Kurki-Suonio}, {Lagache},
  {L{\"a}hteenm{\"a}ki}, {Lamarre}, {Lasenby}, {Lattanzi}, {Lawrence}, {Leahy},
  {Leonardi}, {Lesgourgues}, {Levrier}, {Lewis}, {Liguori}, {Lilje},
  {Linden-V{\o}rnle}, {L{\'o}pez-Caniego}, {Lubin}, {Mac{\'\i}as-P{\'e}rez},
  {Maggio}, {Maino}, {Mandolesi}, {Mangilli}, {Marchini}, {Maris}, {Martin},
  {Martinelli}, {Mart{\'\i}nez-Gonz{\'a}lez}, {Masi}, {Matarrese}, {McGehee},
  {Meinhold}, {Melchiorri}, {Melin}, {Mendes}, {Mennella}, {Migliaccio},
  {Millea}, {Mitra}, {Miville-Desch{\^e}nes}, {Moneti}, {Montier}, {Morgante},
  {Mortlock}, {Moss}, {Munshi}, {Murphy}, {Naselsky}, {Nati}, {Natoli},
  {Netterfield}, {N{\o}rgaard-Nielsen}, {Noviello}, {Novikov}, {Novikov},
  {Oxborrow}, {Paci}, {Pagano}, {Pajot}, {Paladini}, {Paoletti}, {Partridge},
  {Pasian}, {Patanchon}, {Pearson}, {Perdereau}, {Perotto}, {Perrotta},
  {Pettorino}, {Piacentini}, {Piat}, {Pierpaoli}, {Pietrobon}, {Plaszczynski},
  {Pointecouteau}, {Polenta}, {Popa}, {Pratt}, {Pr{\'e}zeau}, {Prunet},
  {Puget}, {Rachen}, {Reach}, {Rebolo}, {Reinecke}, {Remazeilles}, {Renault},
  {Renzi}, {Ristorcelli}, {Rocha}, {Rosset}, {Rossetti}, {Roudier},
  {Rouill{\'e} d'Orfeuil}, {Rowan-Robinson}, {Rubi{\~n}o-Mart{\'\i}n},
  {Rusholme}, {Said}, {Salvatelli}, {Salvati}, {Sandri}, {Santos},
  {Savelainen}, {Savini}, {Scott}, {Seiffert}, {Serra}, {Shellard}, {Spencer},
  {Spinelli}, {Stolyarov}, {Stompor}, {Sudiwala}, {Sunyaev}, {Sutton},
  {Suur-Uski}, {Sygnet}, {Tauber}, {Terenzi}, {Toffolatti}, {Tomasi},
  {Tristram}, {Trombetti}, {Tucci}, {Tuovinen}, {T{\"u}rler}, {Umana},
  {Valenziano}, {Valiviita}, {Van Tent}, {Vielva}, {Villa}, {Wade}, {Wandelt},
  {Wehus}, {White}, {White}, {Wilkinson}, {Yvon}, {Zacchei}, \&
  {Zonca}}]{2016A&A...594A..13P}
{Planck Collaboration}, {Ade}, P.~A.~R., {Aghanim}, N., {et~al.} 2016, \aap,
  594, A13, \dodoi{10.1051/0004-6361/201525830}

\bibitem[{{Press} \& {Schechter}(1974)}]{1974ApJ...187..425P}
{Press}, W.~H., \& {Schechter}, P. 1974, \apj, 187, 425, \dodoi{10.1086/152650}

\bibitem[{{Pringle}(1991)}]{1991MNRAS.248..754P}
{Pringle}, J.~E. 1991, \mnras, 248, 754, \dodoi{10.1093/mnras/248.4.754}

\bibitem[{{Regan} \& {Haehnelt}(2009)}]{2009MNRAS.396..343R}
{Regan}, J.~A., \& {Haehnelt}, M.~G. 2009, \mnras, 396, 343,
  \dodoi{10.1111/j.1365-2966.2009.14579.x}

\bibitem[{{Ricarte} \& {Natarajan}(2018{\natexlab{a}})}]{2018MNRAS.474.1995R}
{Ricarte}, A., \& {Natarajan}, P. 2018{\natexlab{a}}, \mnras, 474, 1995,
  \dodoi{10.1093/mnras/stx2851}

\bibitem[{{Ricarte} \& {Natarajan}(2018{\natexlab{b}})}]{2018MNRAS.481.3278R}
---. 2018{\natexlab{b}}, \mnras, 481, 3278, \dodoi{10.1093/mnras/sty2448}

\bibitem[{{Richards} {et~al.}(2006{\natexlab{a}}){Richards}, {Lacy},
  {Storrie-Lombardi}, {Hall}, {Gallagher}, {Hines}, {Fan}, {Papovich}, {Vanden
  Berk}, {Trammell}, {Schneider}, {Vestergaard}, {York}, {Jester}, {Anderson},
  {Budav{\'a}ri}, \& {Szalay}}]{2006ApJS..166..470R}
{Richards}, G.~T., {Lacy}, M., {Storrie-Lombardi}, L.~J., {et~al.}
  2006{\natexlab{a}}, \apjs, 166, 470, \dodoi{10.1086/506525}

\bibitem[{{Richards} {et~al.}(2006{\natexlab{b}}){Richards}, {Strauss}, {Fan},
  {Hall}, {Jester}, {Schneider}, {Vanden Berk}, {Stoughton}, {Anderson},
  {Brunner}, {Gray}, {Gunn}, {Ivezi{\'c}}, {Kirkland}, {Knapp}, {Loveday},
  {Meiksin}, {Pope}, {Szalay}, {Thakar}, {Yanny}, {York}, {Barentine},
  {Brewington}, {Brinkmann}, {Fukugita}, {Harvanek}, {Kent}, {Kleinman},
  {Krzesi{\'n}ski}, {Long}, {Lupton}, {Nash}, {Neilsen}, {Nitta}, {Schlegel},
  \& {Snedden}}]{2006AJ....131.2766R}
{Richards}, G.~T., {Strauss}, M.~A., {Fan}, X., {et~al.} 2006{\natexlab{b}},
  \aj, 131, 2766, \dodoi{10.1086/503559}

\bibitem[{{Rieke} {et~al.}(2019){Rieke}, {Arribas}, {Bunker}, {Charlot},
  {Finkelstein}, {Maiolino}, {Robertson}, {Willott}, {Windhorst}, {Eisenstein},
  {Nelson}, {Tacchella}, {Egami}, {Endsley}, {Frye}, {Hainline}, {Hviding},
  {Rieke}, {Williams}, {Willmer}, \& {Woodrum}}]{2019BAAS...51c..45R}
{Rieke}, M., {Arribas}, S., {Bunker}, A., {et~al.} 2019, \baas, 51, 45

\bibitem[{{Sakurai} {et~al.}(2015){Sakurai}, {Hosokawa}, {Yoshida}, \&
  {Yorke}}]{2015MNRAS.452..755S}
{Sakurai}, Y., {Hosokawa}, T., {Yoshida}, N., \& {Yorke}, H.~W. 2015, \mnras,
  452, 755, \dodoi{10.1093/mnras/stv1346}

\bibitem[{{Sakurai} {et~al.}(2016){Sakurai}, {Vorobyov}, {Hosokawa}, {Yoshida},
  {Omukai}, \& {Yorke}}]{2016MNRAS.459.1137S}
{Sakurai}, Y., {Vorobyov}, E.~I., {Hosokawa}, T., {et~al.} 2016, \mnras, 459,
  1137, \dodoi{10.1093/mnras/stw637}

\bibitem[{{Salpeter}(1964)}]{1964ApJ...140..796S}
{Salpeter}, E.~E. 1964, \apj, 140, 796, \dodoi{10.1086/147973}

\bibitem[{{Salvaterra} {et~al.}(2012){Salvaterra}, {Haardt}, {Volonteri}, \&
  {Moretti}}]{2012A&A...545L...6S}
{Salvaterra}, R., {Haardt}, F., {Volonteri}, M., \& {Moretti}, A. 2012, \aap,
  545, L6, \dodoi{10.1051/0004-6361/201219965}

\bibitem[{{Sassano} {et~al.}(2021){Sassano}, {Schneider}, {Valiante},
  {Inayoshi}, {Chon}, {Omukai}, {Mayer}, \& {Capelo}}]{2021MNRAS.506..613S}
{Sassano}, F., {Schneider}, R., {Valiante}, R., {et~al.} 2021, \mnras, 506,
  613, \dodoi{10.1093/mnras/stab1737}

\bibitem[{{Schauer} {et~al.}(2019){Schauer}, {Glover}, {Klessen}, \&
  {Ceverino}}]{2019MNRAS.484.3510S}
{Schauer}, A. T.~P., {Glover}, S. C.~O., {Klessen}, R.~S., \& {Ceverino}, D.
  2019, \mnras, 484, 3510, \dodoi{10.1093/mnras/stz013}

\bibitem[{{Schauer} {et~al.}(2017){Schauer}, {Regan}, {Glover}, \&
  {Klessen}}]{2017MNRAS.471.4878S}
{Schauer}, A. T.~P., {Regan}, J., {Glover}, S. C.~O., \& {Klessen}, R.~S. 2017,
  \mnras, 471, 4878, \dodoi{10.1093/mnras/stx1915}

\bibitem[{{Schleicher} {et~al.}(2013){Schleicher}, {Palla}, {Ferrara}, {Galli},
  \& {Latif}}]{2013A&A...558A..59S}
{Schleicher}, D. R.~G., {Palla}, F., {Ferrara}, A., {Galli}, D., \& {Latif}, M.
  2013, \aap, 558, A59, \dodoi{10.1051/0004-6361/201321949}

\bibitem[{{Schmidt} {et~al.}(2019){Schmidt}, {Hennawi}, {Lee}, {Luki{\'c}},
  {O{\~n}orbe}, \& {White}}]{2019ApJ...882..165S}
{Schmidt}, T.~M., {Hennawi}, J.~F., {Lee}, K.-G., {et~al.} 2019, \apj, 882,
  165, \dodoi{10.3847/1538-4357/ab2fcb}

\bibitem[{{Schulze} {et~al.}(2015){Schulze}, {Bongiorno}, {Gavignaud},
  {Schramm}, {Silverman}, {Merloni}, {Zamorani}, {Hirschmann}, {Mainieri},
  {Wisotzki}, {Shankar}, {Fiore}, {Koekemoer}, \&
  {Temporin}}]{2015MNRAS.447.2085S}
{Schulze}, A., {Bongiorno}, A., {Gavignaud}, I., {et~al.} 2015, \mnras, 447,
  2085, \dodoi{10.1093/mnras/stu2549}

\bibitem[{{Sesana} {et~al.}(2008){Sesana}, {Vecchio}, \&
  {Colacino}}]{2008MNRAS.390..192S}
{Sesana}, A., {Vecchio}, A., \& {Colacino}, C.~N. 2008, \mnras, 390, 192,
  \dodoi{10.1111/j.1365-2966.2008.13682.x}

\bibitem[{{Shakura} \& {Sunyaev}(1973)}]{1973A&A....24..337S}
{Shakura}, N.~I., \& {Sunyaev}, R.~A. 1973, \aap, 24, 337

\bibitem[{{Shang} {et~al.}(2010){Shang}, {Bryan}, \&
  {Haiman}}]{2010MNRAS.402.1249S}
{Shang}, C., {Bryan}, G.~L., \& {Haiman}, Z. 2010, \mnras, 402, 1249,
  \dodoi{10.1111/j.1365-2966.2009.15960.x}

\bibitem[{{Shankar} {et~al.}(2010){Shankar}, {Crocce}, {Miralda-Escud{\'e}},
  {Fosalba}, \& {Weinberg}}]{2010ApJ...718..231S}
{Shankar}, F., {Crocce}, M., {Miralda-Escud{\'e}}, J., {Fosalba}, P., \&
  {Weinberg}, D.~H. 2010, \apj, 718, 231, \dodoi{10.1088/0004-637X/718/1/231}

\bibitem[{{Shankar} {et~al.}(2004){Shankar}, {Salucci}, {Granato}, {De Zotti},
  \& {Danese}}]{2004MNRAS.354.1020S}
{Shankar}, F., {Salucci}, P., {Granato}, G.~L., {De Zotti}, G., \& {Danese}, L.
  2004, \mnras, 354, 1020, \dodoi{10.1111/j.1365-2966.2004.08261.x}

\bibitem[{{Shankar} {et~al.}(2009){Shankar}, {Weinberg}, \&
  {Miralda-Escud{\'e}}}]{2009ApJ...690...20S}
{Shankar}, F., {Weinberg}, D.~H., \& {Miralda-Escud{\'e}}, J. 2009, \apj, 690,
  20, \dodoi{10.1088/0004-637X/690/1/20}

\bibitem[{{Shen} {et~al.}(2020){Shen}, {Hopkins}, {Faucher-Gigu{\`e}re},
  {Alexander}, {Richards}, {Ross}, \& {Hickox}}]{2020MNRAS.495.3252S}
{Shen}, X., {Hopkins}, P.~F., {Faucher-Gigu{\`e}re}, C.-A., {et~al.} 2020,
  \mnras, 495, 3252, \dodoi{10.1093/mnras/staa1381}

\bibitem[{{Shen} {et~al.}(2019){Shen}, {Wu}, {Jiang}, {Ba{\~n}ados}, {Fan},
  {Ho}, {Riechers}, {Strauss}, {Venemans}, {Vestergaard}, {Walter}, {Wang},
  {Willott}, {Wu}, \& {Yang}}]{2019ApJ...873...35S}
{Shen}, Y., {Wu}, J., {Jiang}, L., {et~al.} 2019, \apj, 873, 35,
  \dodoi{10.3847/1538-4357/ab03d9}

\bibitem[{{Sheth} {et~al.}(2001){Sheth}, {Mo}, \&
  {Tormen}}]{2001MNRAS.323....1S}
{Sheth}, R.~K., {Mo}, H.~J., \& {Tormen}, G. 2001, \mnras, 323, 1,
  \dodoi{10.1046/j.1365-8711.2001.04006.x}

\bibitem[{{Shibata} {et~al.}(2016){Shibata}, {Sekiguchi}, {Uchida}, \&
  {Umeda}}]{2016PhRvD..94b1501S}
{Shibata}, M., {Sekiguchi}, Y., {Uchida}, H., \& {Umeda}, H. 2016, \prd, 94,
  021501, \dodoi{10.1103/PhysRevD.94.021501}

\bibitem[{{Shibata} \& {Shapiro}(2002)}]{2002ApJ...572L..39S}
{Shibata}, M., \& {Shapiro}, S.~L. 2002, \apjl, 572, L39,
  \dodoi{10.1086/341516}

\bibitem[{{Shimasaku} \& {Izumi}(2019)}]{2019ApJ...872L..29S}
{Shimasaku}, K., \& {Izumi}, T. 2019, \apjl, 872, L29,
  \dodoi{10.3847/2041-8213/ab053f}

\bibitem[{{Siana} {et~al.}(2008){Siana}, {Polletta}, {Smith}, {Lonsdale},
  {Gonzalez-Solares}, {Farrah}, {Babbedge}, {Rowan-Robinson}, {Surace},
  {Shupe}, {Fang}, {Franceschini}, \& {Oliver}}]{2008ApJ...675...49S}
{Siana}, B., {Polletta}, M. d.~C., {Smith}, H.~E., {et~al.} 2008, \apj, 675,
  49, \dodoi{10.1086/527025}

\bibitem[{{Sicilia} {et~al.}(2022){Sicilia}, {Lapi}, {Boco}, {Shankar},
  {Alexander}, {Allevato}, {Villforth}, {Massardi}, {Spera}, {Bressan}, \&
  {Danese}}]{2022ApJ...934...66S}
{Sicilia}, A., {Lapi}, A., {Boco}, L., {et~al.} 2022, \apj, 934, 66,
  \dodoi{10.3847/1538-4357/ac7873}

\bibitem[{{Small} \& {Blandford}(1992)}]{1992MNRAS.259..725S}
{Small}, T.~A., \& {Blandford}, R.~D. 1992, \mnras, 259, 725,
  \dodoi{10.1093/mnras/259.4.725}

\bibitem[{{Smits} {et~al.}(2009){Smits}, {Kramer}, {Stappers}, {Lorimer},
  {Cordes}, \& {Faulkner}}]{SKA_09}
{Smits}, R., {Kramer}, M., {Stappers}, B., {et~al.} 2009, \aap, 493, 1161,
  \dodoi{10.1051/0004-6361:200810383}

\bibitem[{{Soltan}(1982)}]{1982MNRAS.200..115S}
{Soltan}, A. 1982, \mnras, 200, 115, \dodoi{10.1093/mnras/200.1.115}

\bibitem[{{Spera} {et~al.}(2015){Spera}, {Mapelli}, \&
  {Bressan}}]{2015MNRAS.451.4086S}
{Spera}, M., {Mapelli}, M., \& {Bressan}, A. 2015, \mnras, 451, 4086,
  \dodoi{10.1093/mnras/stv1161}

\bibitem[{{Stacy} {et~al.}(2010){Stacy}, {Greif}, \&
  {Bromm}}]{2010MNRAS.403...45S}
{Stacy}, A., {Greif}, T.~H., \& {Bromm}, V. 2010, \mnras, 403, 45,
  \dodoi{10.1111/j.1365-2966.2009.16113.x}

\bibitem[{{Stefanon} {et~al.}(2019){Stefanon}, {Labb{\'e}}, {Bouwens}, {Oesch},
  {Ashby}, {Caputi}, {Franx}, {Fynbo}, {Illingworth}, {Le F{\`e}vre},
  {Marchesini}, {McCracken}, {Milvang-Jensen}, {Muzzin}, \& {van
  Dokkum}}]{2019ApJ...883...99S}
{Stefanon}, M., {Labb{\'e}}, I., {Bouwens}, R.~J., {et~al.} 2019, \apj, 883,
  99, \dodoi{10.3847/1538-4357/ab3792}

\bibitem[{{Sugimura} {et~al.}(2014){Sugimura}, {Omukai}, \&
  {Inoue}}]{2014MNRAS.445..544S}
{Sugimura}, K., {Omukai}, K., \& {Inoue}, A.~K. 2014, \mnras, 445, 544,
  \dodoi{10.1093/mnras/stu1778}

\bibitem[{{Tanaka} {et~al.}(2013){Tanaka}, {Nakamoto}, \&
  {Omukai}}]{2013ApJ...773..155T}
{Tanaka}, K. E.~I., {Nakamoto}, T., \& {Omukai}, K. 2013, \apj, 773, 155,
  \dodoi{10.1088/0004-637X/773/2/155}

\bibitem[{{Tanaka} \& {Haiman}(2009)}]{2009ApJ...696.1798T}
{Tanaka}, T., \& {Haiman}, Z. 2009, \apj, 696, 1798,
  \dodoi{10.1088/0004-637X/696/2/1798}

\bibitem[{{Tanaka} \& {Li}(2014)}]{2014MNRAS.439.1092T}
{Tanaka}, T.~L., \& {Li}, M. 2014, \mnras, 439, 1092,
  \dodoi{10.1093/mnras/stu042}

\bibitem[{{Tegmark} {et~al.}(1997){Tegmark}, {Silk}, {Rees}, {Blanchard},
  {Abel}, \& {Palla}}]{1997ApJ...474....1T}
{Tegmark}, M., {Silk}, J., {Rees}, M.~J., {et~al.} 1997, \apj, 474, 1,
  \dodoi{10.1086/303434}

\bibitem[{{Toyouchi} {et~al.}(2021){Toyouchi}, {Inayoshi}, {Hosokawa}, \&
  {Kuiper}}]{2021ApJ...907...74T}
{Toyouchi}, D., {Inayoshi}, K., {Hosokawa}, T., \& {Kuiper}, R. 2021, \apj,
  907, 74, \dodoi{10.3847/1538-4357/abcfc2}

\bibitem[{{Toyouchi} {et~al.}(2023){Toyouchi}, {Inayoshi}, {Li}, {Haiman}, \&
  {Kuiper}}]{2023MNRAS.518.1601T}
{Toyouchi}, D., {Inayoshi}, K., {Li}, W., {Haiman}, Z., \& {Kuiper}, R. 2023,
  \mnras, 518, 1601, \dodoi{10.1093/mnras/stac3191}

\bibitem[{{Treister} {et~al.}(2013){Treister}, {Schawinski}, {Volonteri}, \&
  {Natarajan}}]{2013ApJ...778..130T}
{Treister}, E., {Schawinski}, K., {Volonteri}, M., \& {Natarajan}, P. 2013,
  \apj, 778, 130, \dodoi{10.1088/0004-637X/778/2/130}

\bibitem[{{Trinca} {et~al.}(2022){Trinca}, {Schneider}, {Valiante}, {Graziani},
  {Zappacosta}, \& {Shankar}}]{2022MNRAS.511..616T}
{Trinca}, A., {Schneider}, R., {Valiante}, R., {et~al.} 2022, \mnras, 511, 616,
  \dodoi{10.1093/mnras/stac062}

\bibitem[{{Tseliakhovich} \& {Hirata}(2010)}]{2010PhRvD..82h3520T}
{Tseliakhovich}, D., \& {Hirata}, C. 2010, \prd, 82, 083520,
  \dodoi{10.1103/PhysRevD.82.083520}

\bibitem[{{Tucci} \& {Volonteri}(2017)}]{2017A&A...600A..64T}
{Tucci}, M., \& {Volonteri}, M. 2017, \aap, 600, A64,
  \dodoi{10.1051/0004-6361/201628419}

\bibitem[{{Ueda} {et~al.}(2014){Ueda}, {Akiyama}, {Hasinger}, {Miyaji}, \&
  {Watson}}]{2014ApJ...786..104U}
{Ueda}, Y., {Akiyama}, M., {Hasinger}, G., {Miyaji}, T., \& {Watson}, M.~G.
  2014, \apj, 786, 104, \dodoi{10.1088/0004-637X/786/2/104}

\bibitem[{{Ueda} {et~al.}(2003){Ueda}, {Akiyama}, {Ohta}, \&
  {Miyaji}}]{2003ApJ...598..886U}
{Ueda}, Y., {Akiyama}, M., {Ohta}, K., \& {Miyaji}, T. 2003, \apj, 598, 886,
  \dodoi{10.1086/378940}

\bibitem[{{Valiante} {et~al.}(2018){Valiante}, {Schneider}, {Graziani}, \&
  {Zappacosta}}]{2018MNRAS.474.3825V}
{Valiante}, R., {Schneider}, R., {Graziani}, L., \& {Zappacosta}, L. 2018,
  \mnras, 474, 3825, \dodoi{10.1093/mnras/stx3028}

\bibitem[{{Venemans} {et~al.}(2017){Venemans}, {Walter}, {Decarli},
  {Ba{\~n}ados}, {Carilli}, {Winters}, {Schuster}, {da Cunha}, {Fan}, {Farina},
  {Mazzucchelli}, {Rix}, \& {Weiss}}]{2017ApJ...851L...8V}
{Venemans}, B.~P., {Walter}, F., {Decarli}, R., {et~al.} 2017, \apjl, 851, L8,
  \dodoi{10.3847/2041-8213/aa943a}

\bibitem[{{Visbal} \& {Croft}(2008)}]{2008ApJ...674..660V}
{Visbal}, E., \& {Croft}, R. A.~C. 2008, \apj, 674, 660, \dodoi{10.1086/523843}

\bibitem[{{Visbal} {et~al.}(2014){Visbal}, {Haiman}, {Terrazas}, {Bryan}, \&
  {Barkana}}]{2014MNRAS.445..107V}
{Visbal}, E., {Haiman}, Z., {Terrazas}, B., {Bryan}, G.~L., \& {Barkana}, R.
  2014, \mnras, 445, 107, \dodoi{10.1093/mnras/stu1710}

\bibitem[{{Vito} {et~al.}(2018){Vito}, {Brandt}, {Yang}, {Gilli}, {Luo},
  {Vignali}, {Xue}, {Comastri}, {Koekemoer}, {Lehmer}, {Liu}, {Paolillo},
  {Ranalli}, {Schneider}, {Shemmer}, {Volonteri}, \&
  {Wang}}]{2018MNRAS.473.2378V}
{Vito}, F., {Brandt}, W.~N., {Yang}, G., {et~al.} 2018, \mnras, 473, 2378,
  \dodoi{10.1093/mnras/stx2486}

\bibitem[{{Volonteri}(2012)}]{2012Sci...337..544V}
{Volonteri}, M. 2012, Science, 337, 544, \dodoi{10.1126/science.1220843}

\bibitem[{{Wang} {et~al.}(2019){Wang}, {Yang}, {Fan}, {Wu}, {Yue}, {Li},
  {Bian}, {Jiang}, {Ba{\~n}ados}, {Schindler}, {Findlay}, {Davies}, {Decarli},
  {Farina}, {Green}, {Hennawi}, {Huang}, {Mazzuccheli}, {McGreer}, {Venemans},
  {Walter}, {Dye}, {Lyke}, {Myers}, \& {Nunez}}]{2019ApJ...884...30W}
{Wang}, F., {Yang}, J., {Fan}, X., {et~al.} 2019, \apj, 884, 30,
  \dodoi{10.3847/1538-4357/ab2be5}

\bibitem[{{Wang} {et~al.}(2021){Wang}, {Yang}, {Fan}, {Hennawi}, {Barth},
  {Banados}, {Bian}, {Boutsia}, {Connor}, {Davies}, {Decarli}, {Eilers},
  {Farina}, {Green}, {Jiang}, {Li}, {Mazzucchelli}, {Nanni}, {Schindler},
  {Venemans}, {Walter}, {Wu}, \& {Yue}}]{2021ApJ...907L...1W}
---. 2021, \apjl, 907, L1, \dodoi{10.3847/2041-8213/abd8c6}

\bibitem[{{Wang} {et~al.}(2013){Wang}, {Wagg}, {Carilli}, {Walter}, {Lentati},
  {Fan}, {Riechers}, {Bertoldi}, {Narayanan}, {Strauss}, {Cox}, {Omont},
  {Menten}, {Knudsen}, {Neri}, \& {Jiang}}]{2013ApJ...773...44W}
{Wang}, R., {Wagg}, J., {Carilli}, C.~L., {et~al.} 2013, \apj, 773, 44,
  \dodoi{10.1088/0004-637X/773/1/44}

\bibitem[{{Willott} {et~al.}(2010{\natexlab{a}}){Willott}, {Delorme},
  {Reyl{\'e}}, {Albert}, {Bergeron}, {Crampton}, {Delfosse}, {Forveille},
  {Hutchings}, {McLure}, {Omont}, \& {Schade}}]{2010AJ....139..906W}
{Willott}, C.~J., {Delorme}, P., {Reyl{\'e}}, C., {et~al.} 2010{\natexlab{a}},
  \aj, 139, 906, \dodoi{10.1088/0004-6256/139/3/906}

\bibitem[{{Willott} {et~al.}(2010{\natexlab{b}}){Willott}, {Albert},
  {Arzoumanian}, {Bergeron}, {Crampton}, {Delorme}, {Hutchings}, {Omont},
  {Reyl{\'e}}, \& {Schade}}]{2010AJ....140..546W}
{Willott}, C.~J., {Albert}, L., {Arzoumanian}, D., {et~al.} 2010{\natexlab{b}},
  \aj, 140, 546, \dodoi{10.1088/0004-6256/140/2/546}

\bibitem[{{Wise} {et~al.}(2019){Wise}, {Regan}, {O'Shea}, {Norman}, {Downes},
  \& {Xu}}]{2019Natur.566...85W}
{Wise}, J.~H., {Regan}, J.~A., {O'Shea}, B.~W., {et~al.} 2019, \nat, 566, 85,
  \dodoi{10.1038/s41586-019-0873-4}

\bibitem[{{Woods} {et~al.}(2017){Woods}, {Heger}, {Whalen}, {Haemmerl{\'e}}, \&
  {Klessen}}]{2017ApJ...842L...6W}
{Woods}, T.~E., {Heger}, A., {Whalen}, D.~J., {Haemmerl{\'e}}, L., \&
  {Klessen}, R.~S. 2017, \apjl, 842, L6, \dodoi{10.3847/2041-8213/aa7412}

\bibitem[{{Woods} {et~al.}(2019){Woods}, {Agarwal}, {Bromm}, {Bunker}, {Chen},
  {Chon}, {Ferrara}, {Glover}, {Haemmerl{\'e}}, {Haiman}, {Hartwig}, {Heger},
  {Hirano}, {Hosokawa}, {Inayoshi}, {Klessen}, {Kobayashi}, {Koliopanos},
  {Latif}, {Li}, {Mayer}, {Mezcua}, {Natarajan}, {Pacucci}, {Rees}, {Regan},
  {Sakurai}, {Salvadori}, {Schneider}, {Surace}, {Tanaka}, {Whalen}, \&
  {Yoshida}}]{2019PASA...36...27W}
{Woods}, T.~E., {Agarwal}, B., {Bromm}, V., {et~al.} 2019, \pasa, 36, e027,
  \dodoi{10.1017/pasa.2019.14}

\bibitem[{{Wu} {et~al.}(2022){Wu}, {Shen}, {Jiang}, {Ba{\~n}ados}, {Fan}, {Ho},
  {Vestergaard}, {Wang}, {Wang}, {Wu}, \& {Yang}}]{2022MNRAS.517.2659W}
{Wu}, J., {Shen}, Y., {Jiang}, L., {et~al.} 2022, \mnras, 517, 2659,
  \dodoi{10.1093/mnras/stac2833}

\bibitem[{{Yang} {et~al.}(2016){Yang}, {Wang}, {Wu}, {Fan}, {McGreer}, {Bian},
  {Yi}, {Yang}, {Ai}, {Dong}, {Zuo}, {Green}, {Jiang}, {Wang}, {Wang}, \&
  {Yue}}]{2016ApJ...829...33Y}
{Yang}, J., {Wang}, F., {Wu}, X.-B., {et~al.} 2016, \apj, 829, 33,
  \dodoi{10.3847/0004-637X/829/1/33}

\bibitem[{{Yang} {et~al.}(2021){Yang}, {Wang}, {Fan}, {Barth}, {Hennawi},
  {Nanni}, {Bian}, {Davies}, {Farina}, {Schindler}, {Ba{\~n}ados}, {Decarli},
  {Eilers}, {Green}, {Guo}, {Jiang}, {Li}, {Venemans}, {Walter}, {Wu}, \&
  {Yue}}]{2021ApJ...923..262Y}
{Yang}, J., {Wang}, F., {Fan}, X., {et~al.} 2021, \apj, 923, 262,
  \dodoi{10.3847/1538-4357/ac2b32}

\bibitem[{{Yoon} {et~al.}(2012){Yoon}, {Dierks}, \&
  {Langer}}]{2012A&A...542A.113Y}
{Yoon}, S.~C., {Dierks}, A., \& {Langer}, N. 2012, \aap, 542, A113,
  \dodoi{10.1051/0004-6361/201117769}

\bibitem[{{Yoshida} {et~al.}(2003){Yoshida}, {Abel}, {Hernquist}, \&
  {Sugiyama}}]{2003ApJ...592..645Y}
{Yoshida}, N., {Abel}, T., {Hernquist}, L., \& {Sugiyama}, N. 2003, \apj, 592,
  645, \dodoi{10.1086/375810}

\bibitem[{{Younger} {et~al.}(2008){Younger}, {Hopkins}, {Cox}, \&
  {Hernquist}}]{2008ApJ...686..815Y}
{Younger}, J.~D., {Hopkins}, P.~F., {Cox}, T.~J., \& {Hernquist}, L. 2008,
  \apj, 686, 815, \dodoi{10.1086/591639}

\bibitem[{{Yu} {et~al.}(2005){Yu}, {Lu}, \& {Kauffmann}}]{2005ApJ...634..901Y}
{Yu}, Q., {Lu}, Y., \& {Kauffmann}, G. 2005, \apj, 634, 901,
  \dodoi{10.1086/433166}

\bibitem[{{Yu} \& {Tremaine}(2002)}]{2002MNRAS.335..965Y}
{Yu}, Q., \& {Tremaine}, S. 2002, \mnras, 335, 965,
  \dodoi{10.1046/j.1365-8711.2002.05532.x}

\bibitem[{{Yung} {et~al.}(2021){Yung}, {Somerville}, {Finkelstein},
  {Hirschmann}, {Dav{\'e}}, {Popping}, {Gardner}, \&
  {Venkatesan}}]{2021MNRAS.508.2706Y}
{Yung}, L.~Y.~A., {Somerville}, R.~S., {Finkelstein}, S.~L., {et~al.} 2021,
  \mnras, 508, 2706, \dodoi{10.1093/mnras/stab2761}

\end{thebibliography}
\bibliographystyle{aasjournal}



\end{CJK*}
\end{document}